\pdfoutput=1
\documentclass[cernpreprint, atlasdraft=true, texlive=2016, UKenglish, subcaption, texmf]{atlasdoc}
\usepackage[subcaption]{atlaspackage}
\usepackage{atlasbiblatex}
\usepackage{atlasphysics}
\graphicspath{{logos/}{figures/}}
\usepackage{ANA-STDM-2017-30-PAPER-defs}
 
\PreprintIdNumber{CERN-EP-2021-105}
 
\AtlasTitle{Measurement of the production cross section of pairs of isolated photons in \textit{pp} collisions at 13~TeV with the ATLAS detector}
 
\AtlasJournal{JHEP}
\AtlasJournalRef{JHEP 11 (2021) 169}
\AtlasDOI{10.1007/JHEP11(2021)169}
\AtlasAbstract{
A measurement of prompt photon-pair production in proton--proton collisions at $\sqrt{s}=13~\mathrm{TeV}$\ is presented. The data were recorded by the ATLAS detector at the LHC with an integrated luminosity of 139~fb$^{-1}$.
Events with two photons in the well-instrumented region of the detector are selected. The photons are required to be isolated and have a transverse momentum of $p_\mathrm{T,\gamma_{1(2)}} > 40\:(30)~\mathrm{GeV}$\ for the leading (sub-leading) photon.
The differential cross sections as functions of several observables for the diphoton system are measured and compared with theoretical predictions from state-of-the-art Monte Carlo and fixed-order calculations.
The QCD predictions from next-to-next-to-leading-order calculations and multi-leg merged calculations are able to describe the measured integrated and differential cross sections within uncertainties, whereas lower-order calculations show significant deviations, demonstrating that higher-order perturbative QCD corrections are crucial for this process. The resummed predictions with parton showers additionally provide an excellent description of the low transverse-momentum regime of the diphoton system.
}
 
\author{The ATLAS Collaboration}
 
\AtlasRefCode{STDM-2017-30}
 
\AtlasDate{\today}

\hypersetup{pdftitle={Diphotons at 13TeV with ATLAS},pdfauthor={The ATLAS Collaboration}}
 
\begin{document}
\maketitle
\tableofcontents

\section{Introduction}
The production of a prompt photon pair in proton--proton collisions is one of the cornerstone processes of the LHC physics programme. Most prominently, photon pairs are of vital importance in searching for or studying the properties of particles that decay into photon pairs, such as the Higgs boson or other neutral particles.
The main background to resonant diphoton production originates from the continuum production of such pairs, which is the topic of this publication. Despite the electromagnetic nature of this process, diphoton production in hadron colliders involves intricate strong-interaction dynamics. Theoretical predictions are therefore highly non-trivial and measurements are necessary to scrutinise and validate such predictions.
 
\begin{figure}[tp]
\centering
\begin{subfigure}[b]{.25\linewidth}
\centering
\includegraphics[width=0.8\textwidth]{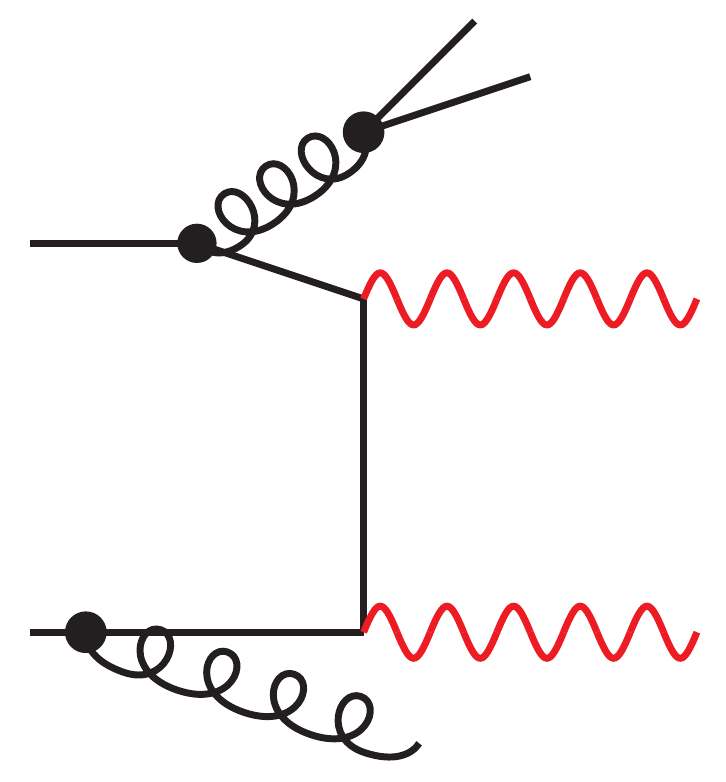}
\caption{Direct photons}
\label{fig:psdirect}
\end{subfigure}
\begin{subfigure}[b]{.43\linewidth}
\centering
\includegraphics[width=0.8\textwidth]{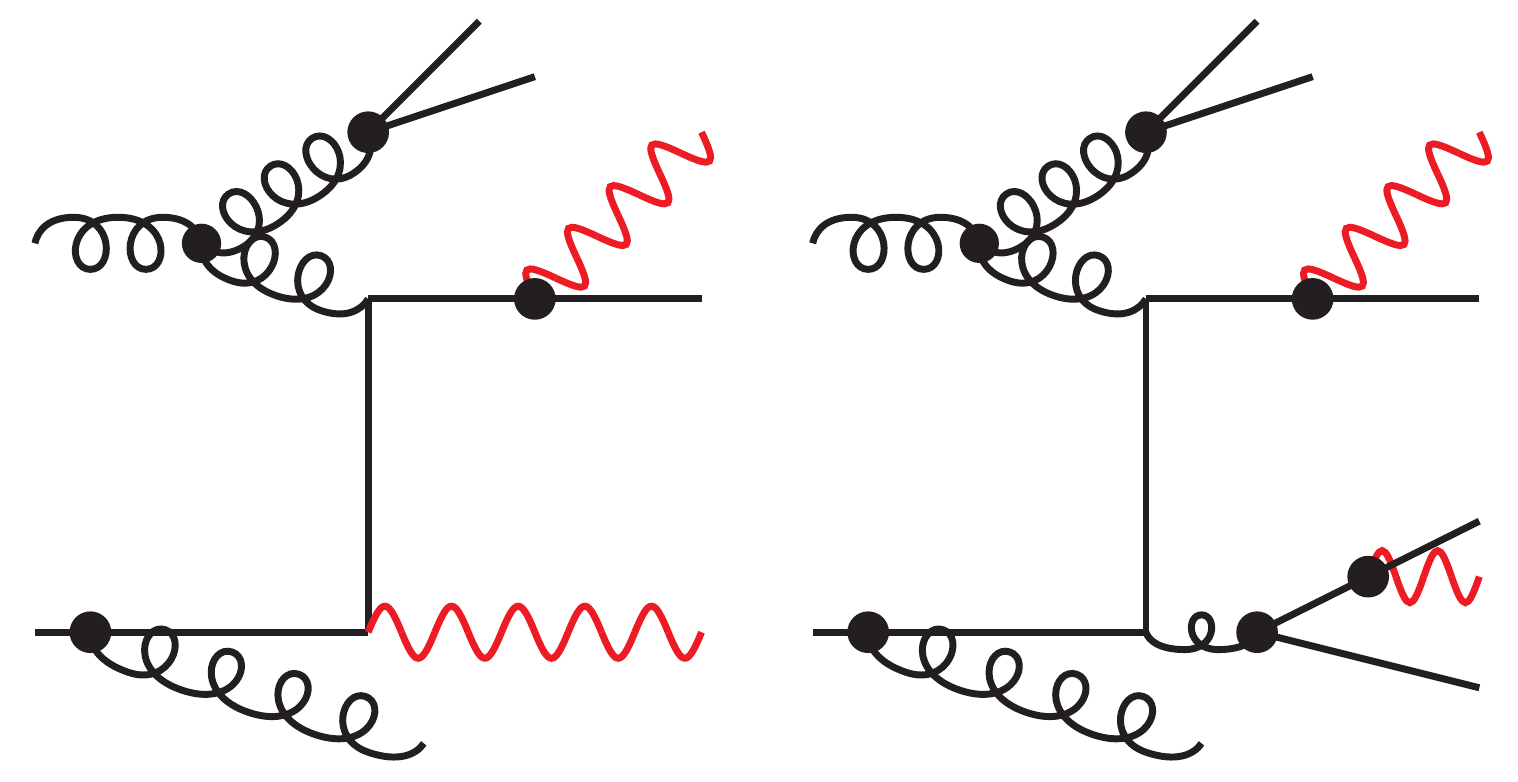}
\caption{Single- and double-fragmentation photons}
\label{fig:psfrag}
\end{subfigure}
\begin{subfigure}[b]{.3\linewidth}
\centering
\includegraphics[width=0.8\textwidth]{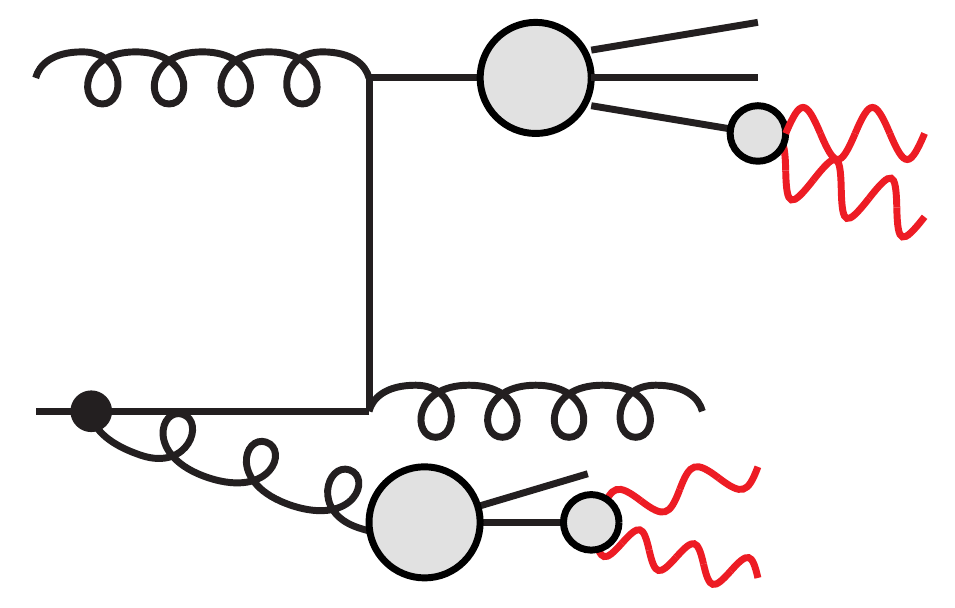}
\vspace*{6mm}
\caption{Non-prompt photons}
\label{fig:psnonprompt}
\end{subfigure}
\caption{Schematics of theoretical photon pair production mechanisms (a, b) and an example of experimental background (c) from non-prompt multiphoton production. Quarks are represented by straight lines, gluons by curly lines, and photons by red wavy lines. The black dots represent resummed bremsstrahlung, e.g.\ from a fragmentation function or parton shower, while the grey circles represent hadronisation and hadron decay processes. The presence of jets is not required in the following and only included for illustration purposes.}
\label{fig:ps}
\end{figure}

Non-resonant photons can be produced by various mechanisms in $pp$ collisions at the LHC.
In theoretical calculations, prompt photons, i.e.\ those not produced in hadron decays, are typically subdivided into two different production mechanisms: the \textit{direct} production mode, represented by a $q\bar{q}\to\gamma\gamma$ $t$-channel diagram at leading order; and the single- and double-\textit{fragmentation} component with a hard $\gamma$+jet or dijet configuration and sub-leading photon emissions in a resummed approach. A schematic representation exemplifies these photon production mechanisms in Figures~\ref{fig:psdirect} and \ref{fig:psfrag}, respectively.
Hadron decays such as $\pi^0\to\gamma\gamma$, cf.~Figure~\ref{fig:psnonprompt}, are the most abundant source of photons in $pp$ collisions, but are considered as background in this publication. To suppress this non-prompt contribution, photons are required to be isolated from hadronic activity in their vicinity.
 
In this publication, prompt diphoton production is measured in $\sqrt{s}=13$~\TeV\ $pp$ collisions at the LHC using the full Run-2 dataset.
Previous diphoton cross-section measurements were performed in $p\bar{p}$ collisions at $\sqrt{s}=1.96$~\TeV\ by CDF~\cite{Aaltonen:2012jd} and D0~\cite{Abazov:2013pua}, in $pp$ collisions at $\sqrt{s}=7$~\TeV\ by CMS~\cite{CMS-QCD-10-035,CMS-SMP-13-001} and ATLAS~\cite{STDM-2011-05,STDM-2012-05}, and at $\sqrt{s}=8$~\TeV\ by ATLAS~\cite{STDM-2015-15}.
 
The main challenge and source of uncertainty in the experimental side is the estimation of the background from non-prompt photons in jet events; in this analysis, a data-driven technique is used to estimate this background. The background-subtracted yield is then corrected for detector effects using an iterative unfolding procedure. Differential cross-section measurements are presented in a fiducial region as functions of several kinematic observables of the photon pair. State-of-the-art theoretical predictions from fixed-order calculations and from Monte Carlo (MC) event generators are compared with the data.

\section{ATLAS detector}
\newcommand{\AtlasCoordFootnote}{
ATLAS uses a right-handed coordinate system with its origin at the nominal interaction point (IP)
in the centre of the detector and the \(z\)-axis along the beam pipe.
The \(x\)-axis points from the IP to the centre of the LHC ring,
and the \(y\)-axis points upwards.
Cylindrical coordinates \((r,\phi)\) are used in the transverse plane,
\(\phi\) being the azimuthal angle around the \(z\)-axis.
The pseudorapidity is defined in terms of the polar angle \(\theta\) as \(\eta = -\ln \tan(\theta/2)\).
Angular distance is measured in units of \(\Delta R \equiv \sqrt{(\Delta\eta)^{2} + (\Delta\phi)^{2}}\).}
 
The ATLAS experiment~\cite{PERF-2007-01} at the LHC is a multipurpose particle detector
with a forward--backward symmetric cylindrical geometry and a near \(4\pi\) coverage in
solid angle.\footnote{\AtlasCoordFootnote}
It consists of an inner tracking detector surrounded by a thin superconducting solenoid
providing a \SI{2}{\tesla} axial magnetic field, electromagnetic and hadronic calorimeters, and a muon spectrometer.
 
The inner tracking detector covers the pseudorapidity range \(|\eta| < 2.5\).
It consists of silicon pixel, silicon microstrip, and transition radiation tracking detectors.
 
The calorimeter system covers the pseudorapidity range \(|\eta| < 4.9\).
Within the region \(|\eta|< 3.2\), electromagnetic (EM) calorimetry is provided by a lead/liquid-argon (LAr) sampling calorimeter with accordion geometry.
It is divided into a barrel section covering $|\eta|<1.475$ and two endcap sections covering $1.375<|\eta|<3.2$.
For $|\eta|<2.5$, it is divided into three layers in depth, which are finely segmented in $\eta$ and $\phi$.
An additional thin LAr presampler layer covering \(|\eta| < 1.8\) is used to correct for energy loss in material upstream of the calorimeters.
Hadronic calorimetry is provided by a steel/scintillator-tile calorimeter, segmented into three barrel structures within \(|\eta| < 1.7\), and two copper/LAr hadronic endcap calorimeters. The solid angle coverage is completed with forward copper/LAr and tungsten/LAr calorimeter modules optimised for electromagnetic and hadronic measurements respectively.
 
The muon spectrometer surrounds the calorimeters and is based on
three large air-core toroidal superconducting magnets with eight coils each.
The field integral of the toroids ranges between \num{2.0} and \SI{6.0}{\tesla\metre}
across most of the detector.
The muon spectrometer includes a system of precision tracking chambers and fast detectors for triggering.
 
A two-level trigger system is used to select events.
The first-level trigger is implemented in hardware and uses a subset of the detector information
to keep the rate of accepted events below \SI{100}{\kHz}.
This is followed by a software-based trigger that
reduces the accepted event rate to \SI{1}{\kHz} on average
depending on the data-taking conditions~\cite{TRIG-2016-01}.
 
An extensive software suite~\cite{ATL-SOFT-PUB-2021-001} is used for the reconstruction and analysis of real and simulated data, for detector operations, and in the trigger and data acquisition systems of the experiment.

\section{Data and simulated event samples}
\subsection{Data samples}
 
The data used in this measurement were recorded by the ATLAS detector in $pp$ collisions at $\sqrt{s}=13$~\TeV{} during the LHC Run-2 data taking period. Events were selected using a diphoton trigger requiring the presence of at least two clusters of energy depositions with $|\eta|<2.47$ in the EM calorimeter, with transverse momentum ($\pt$) above 35~\GeV\ and 25~\GeV\ for the \pt-ordered leading and sub-leading clusters, respectively~\cite{TRIG-2018-05}. Only events taken during stable beam conditions and satisfying detector and data-quality requirements~\cite{DAPR-2018-01} are considered. These requirements ensure that the calorimeters and inner tracking detectors are operating normally.
 
The total integrated luminosity of the collected sample after trigger and data-quality requirements amounts to 139~fb$^{-1}$. Multiple $pp$ interactions (pile-up) can occur in the same bunch crossing, with 34 simultaneous interactions produced on average.
The efficiency of the diphoton triggers relative to the offline selection presented in the next section ranges between 96\% and 100\%~\cite{TRIG-2018-05} as a function of the sub-leading photon~$\pt$.
 
\subsection{Simulated event samples for signal and background processes}
\label{sec:sample:mc}
 
MC simulation samples are used for the subtraction of subdominant background as well as for cross-checks of the subtraction of the dominant background, for the unfolding of detector effects, and as theoretical predictions in the final comparison with the measured data.
 
In all samples, pile-up effects from both the same bunch crossing and previous/subsequent crossings were simulated by overlaying additional generated minimum-bias events on the hard-scatter event prior to reconstruction.
The MC samples were reweighted to match the distribution of the number of pile-up interactions in data events.
All samples were processed through the ATLAS detector simulation~\cite{SOFT-2010-01} based on \GEANT~\cite{Agostinelli:2002hh} and through the same reconstruction algorithms as used for data.
 
\paragraph*{Diphoton signal samples.}
Simulated samples of prompt diphoton production were generated with two different programs, \Sherpa~\cite{Bothmann:2019yzt} and \PYTHIA[8]~\cite{Sjostrand:2007gs}. The nominal samples used in the analysis were simulated with the
\Sherpa[2.2] generator. In this set-up,
matrix elements for the $pp\to \gamma\gamma+0,1j$ process\footnote{Here and in the following, $j$ denotes a `jet', which at matrix-element level corresponds to a quark or gluon.} at next-to-leading-order (NLO) accuracy in the strong coupling constant \alphas and $pp\to \gamma\gamma+2,3j$ at leading-order (LO) accuracy are matched and
merged with the \Sherpa parton shower
based on Catani--Seymour dipoles~\cite{Gleisberg:2008fv,Schumann:2007mg}
using the \MEPSatNLO prescription~\cite{Hoeche:2011fd,Hoeche:2012yf}.
The virtual QCD correction for matrix elements at NLO accuracy is
provided by the \OPENLOOPS library~\cite{Cascioli:2011va,Denner:2016kdg}.
Samples were generated using the
NNPDF3.0 next-to-next-to-leading-order (NNLO) set~\cite{Ball:2014uwa} of parton distribution functions (PDF), along with the dedicated set of
tuned parton-shower parameters developed by the \Sherpa authors.
Both the direct and fragmentation components of isolated photon production are included by setting the merging scale dynamically in the scheme of Ref.~\cite{Siegert:2016bre}, and a photon isolation requirement based on a smooth cone~\cite{Frixione:1998jh} is used with $n=2$, $\delta_0=0.1$ and $\varepsilon=0.1$. In addition, the loop-induced $gg\to \gamma\gamma$ box process is included in these samples at LO accuracy.
 
An alternative signal sample was generated with \PYTHIA[8]. Direct production was simulated using LO matrix elements for $pp \to \gamma\gamma$ and showered in \PYTHIA[8.186]~\cite{Sjostrand:2007gs} using the NNPDF2.3 LO~\cite{Ball:2012cx} PDF set with the A14 set of tuned parameters~\cite{ATL-PHYS-PUB-2014-021}. The same framework was used to generate the single-fragmentation component based on LO matrix elements for $pp \to \gamma j$ with one additional photon produced in the parton shower, and the double-fragmentation component based on LO matrix elements for $pp \to jj$ with both additional photons produced in the parton shower. No photon isolation requirement is applied in these samples at the generator level.
 
\paragraph*{Electron background sources.}
Since electrons can be misidentified as photons, background processes with prompt electron production from Drell--Yan processes, $Z/\gamma^*(\to ee)$ and $W(\to e\nu)$, are considered and estimated from MC simulation. The samples used to estimate these backgrounds were generated using \POWHEGBOX[v2]~\cite{Alioli:2010xd,Nason:2004rx} interfaced to the \PYTHIA[8.186]~\cite{Sjostrand:2007gs} parton shower model with the \AZNLO set of tuned parameters~\cite{STDM-2012-23}. The effect of QED final-state radiation was simulated with \PHOTOSpp[3.52]~\cite{Davidson:2010ew}. The $W$/$Z$ samples are normalised to NNLO cross sections calculated with FEWZ~\cite{Anastasiou:2003ds}. Even though the impact is negligible, poor modelling of the $p_\perp(Z)$ spectrum in the $Z(\to ee)$ sample is corrected for using data/MC correction factors from the measurement in Ref.~\cite{STDM-2018-14}.
The $e\to \gamma$ misidentification rate is studied using $Z\to ee$ events, and the simulation sample is corrected to match the data; the correction value per electron varies as a function of the pseudorapidity, rising to 10\% in some regions.
 
\paragraph*{$\yj$ simulation.}
While the non-prompt background is determined in a data-driven approach, \Sherpa[2.2]~\cite{Bothmann:2019yzt} simulations of the $pp\to \gamma + $jets process are used for cross-checks. The set-up is the same as for the diphoton signal samples described above, including matrix elements for $pp\to \gamma+1,2j$ at NLO and $pp\to \gamma+3,4j$ at LO accuracy in \alphas.

\section{Event selection and definition of observables}
The photon and event selections are explained in Section \ref{sec:definition:photons}, the particle-level selection to which the data are unfolded is described in Section~\ref{sec:definition:events:particlelevel}, and the definition of
the observables is included in Section \ref{sec:definition:observables}.
 
\subsection{Photon reconstruction and event selection}
\label{sec:definition:photons}
 
Photon candidates are reconstructed from clusters of energy depositions in the EM calorimeter~\cite{EGAM-2018-01};
the clusters are formed using a dynamical topological cell-clustering algorithm~\cite{PERF-2014-07}.
Clusters matched to a track consistent with originating from an electron produced in the beam interaction region are considered electron candidates, otherwise they are considered photon candidates.
Photons which convert to electron--positron pairs in the inner detector are distinguished from unconverted photons by matching reconstructed conversion vertices or tracks consistent with originating from photon conversion to the clusters.
The photon energy is calibrated in several steps as described in Refs.~\cite{PERF-2013-05,PERF-2017-03,EGAM-2018-01}.
The energy resolution is optimised using a multivariate regression algorithm based on properties of the shower development in the EM calorimeter and trained on simulated events.
The absolute energy scale is adjusted using $Z\to ee$ events, and validated using radiative $Z$ boson decays.
The energy scale uncertainty depends on the $\pt$ and $\eta$ of the photon, and varies between 0.25\%--1\% for photons with transverse momentum in the range 30--100~\GeV\ dominating this analysis.
 
For this study, at least two photon candidates are required per event, with transverse momentum $p_\mathrm{T,\gamma}$ above 40~$\GeV$ and 30~$\GeV$ for the $p_\mathrm{T,\gamma}$-ordered leading and sub-leading photons, respectively.
These $p_\mathrm{T,\gamma}$ thresholds are each 5~$\GeV$ above the corresponding trigger thresholds, at the plateau of the trigger efficiency versus $p_\mathrm{T,\gamma}$~\cite{TRIG-2018-05}.
The photon pseudorapidity is required to be $|\eta_{\gamma}| < 2.37$, excluding the transition region between the barrel and endcaps ($1.37 < |\eta_{\gamma}| < 1.52$).
In this acceptance region, the high granularity of the EM calorimeter allows efficient identification of photons.
 
The momenta of the photon candidates are corrected for the position of their associated reconstructed primary interaction vertex. The associated vertex is chosen among all reconstructed vertices, which are each required to have at least two inner detector tracks with $\pt > \SI{0.5}{\GeV}$. The measured trajectories of the two photons, along with the reconstructed vertex information in the event, are used as inputs to a neural-network algorithm trained on simulated events to select the most probable primary vertex~\cite{HIGG-2013-08} for the photon pair.
The photon trajectories are reconstructed using the shower-depth (longitudinal) segmentation of the EM calorimeter;
for converted photons, the position of the conversion vertex is also used if the conversion tracks have hits in the silicon detectors.
 
The identification of photons is based on multiple variables that quantify the longitudinal and lateral shape of the EM shower produced by the photon. Requirements on these shower variables are used to identify photons. A detailed description of all variables and selections for the \textit{tight} photon identification used in this analysis is given in Ref.~\cite{EGAM-2018-01}.
The identification variables $f_\mathrm{side}$ (energy fraction outside of core cells) and $w_{s\mathrm{3}}$ (lateral shower width) are used to define control regions for the background estimation presented in Section~\ref{sec:background:jetfake}.
These variables use information from the finely segmented first layer of the electromagnetic calorimeter, and are sensitive to differences between candidates from prompt photons and background candidates, e.g.\ from the two collimated photons of a neutral hadron decay.
 
To further reject background candidates from hadron decays, which are in general surrounded by hadronic activity, the photons are required to be isolated.
The isolation measurement is based on calorimeter information \cite{EGAM-2018-01}. The initial isolation variable is defined as the scalar sum of the transverse momenta computed from topological clusters of calorimeter cells~\cite{PERF-2014-07} in a cone of $\Delta R = 0.2$ around the photon candidate direction. The contribution of the photon to the transverse isolation energy is subtracted, based on the energy depositions in a rectangular window of $\Delta\eta \times \Delta\phi = 0.125 \times 0.175$ around the photon candidate and the energy expected to leak out of this window.
Effects of pile-up and from the underlying event (UE) are corrected for in calculating the isolation variable $\etisoptwo{,\gamma_{1(2)}}$ used for the analysis. The pile-up and UE correction is estimated on an event-by-event basis, using the ambient transverse momentum density~\cite{Cacciari:2008gn}. This density is determined as the median transverse momentum density of jets reconstructed with the $k_t$-algorithm~\cite{Ellis:1993tq} with a radius parameter $R=0.5$;
it is computed separately in two pseudorapidity regions, $|\eta|<1.5$ and $1.5<|\eta|<3$, and used accordingly depending on the photon pseudorapidity.
The photon candidates are required to have $\etisoptwo{,\gamma_{1(2)}}<\recoIsoFactor\cdot p_{\mathrm{T},\gamma_{1(2)}}$. This isolation requirement is optimised to minimise the overall uncertainties, achieving high signal purities while retaining good signal efficiency. More stringent requirements and use of a larger isolation area would increase the purity at the cost of higher uncertainties from pile-up modelling in the simulation.
 
The angular separation of the photons is required to be $\Delta R_{\gamma\gamma} > 0.4$, thus avoiding overlap between their isolation cones, and minimising the correlation between the transverse isolation energy of the two photons.
The isolation criterion is also used for the background estimation (cf.~Section~\ref{sec:background}), and such a correlation between the isolation criteria of the two photons would have an impact on the background estimation results.
 
The event selection described above is summarised in Table~\ref{tab:event_selection}, where the particle-level event selection is also presented, which is described in more detail in the next section.
With these requirements \num{4708978} events are selected in the data.
 
\begin{table}[htb]
\newcommand{\fixverticalspacing}{\begin{minipage}[c][7mm][t]{0.1mm}\end{minipage}}
\centering
\caption{Overview of the event selection at the detector level and at the particle level, i.e.\ definition of the fiducial phase space.}
\label{tab:event_selection}
\begin{tabular}{|c | c | c|}\hline
Selection & Detector level & Particle level
\fixverticalspacing\\\hline\hline
Photon kinematics
& \multicolumn{2}{c|}{$p_\mathrm{T,\gamma_{1(2)}} > 40\,(30)\:\GeV$,\quad $|\eta_{\gamma}| < 2.37$ excluding $1.37 < |\eta_{\gamma}| < 1.52$}
\fixverticalspacing\\\hline
Photon identification
& tight
& stable, not from hadron decay
\fixverticalspacing\\\hline
Photon isolation
& $\etisoptwo{,\gamma}< \recoIsoFactor\cdot p_\mathrm{T, \gamma}$
& $\etisoptwo{,\gamma}<\truthIsoFactor\cdot p_\mathrm{T, \gamma}$
\fixverticalspacing\\\hline
Diphoton topology
& \multicolumn{2}{c|}{$N_\gamma\geq 2$,\quad $\Delta R_{\gamma\gamma} > 0.4$}
\fixverticalspacing\\\hline
\end{tabular}
\end{table}
 
\subsection{Particle-level event selection}
\label{sec:definition:events:particlelevel}
 
The particle-level selection, to which the data are unfolded (cf.~Section~\ref{sec:unfolding}), closely follows the detector-level event requirements, and is also shown in Table~\ref{tab:event_selection}. Some aspects are clarified in detail in the remainder of this section.
 
The particle-level event selection in MC samples takes the leading two prompt photons into account, and vetoes the event if one of them fails the fiducial cuts.
 
The transverse isolation energy at particle level is determined as the scalar sum of the transverse momenta of all stable particles,\footnote{Stable particles are defined as those with a decay length $c\tau>10$~mm.} excluding particles from pile-up, muons, neutrinos and the photon itself. Furthermore, to ensure that the side effects of the detector-level UE subtraction are matched at the particle level,\footnote{While pile-up particles are excluded at particle level, there are still other effects, e.g.\ multiple parton interactions, that can cause such a subtraction procedure to have an influence.} a final-state-based subtraction using the same method as described in Section~\ref{sec:definition:photons} is applied. The value of the isolation requirement at the particle level (0.09) is chosen to be different from the one at the detector level (0.05). This difference accounts for the non-compensating nature of the ATLAS calorimeter. The particle-level value is optimised to minimise the number of events where photons pass a detector-level isolation requirement but fail the particle-level one, or vice versa.
 
\subsection{Observables}
\label{sec:definition:observables}
 
Cross sections are measured differentially as functions of several observables defined for the diphoton system:
\begin{itemize}
\item Transverse momentum \phonept (\phtwopt) of the leading (sub-leading) photon.
\item Invariant mass \yym and transverse momentum \yypT of the photon pair.
\item The absolute value of the cosine of the scattering angle with respect to the $z$-axis in the Collins--Soper frame~\cite{CollinsSoper:1977}:
\begin{equation*}
\yycosTS = \left | \frac{\sinh(\Delta \eta_{\gamma\gamma})}{\sqrt{1+(\yypT/\yym)^2}}
\cdot \frac{2\phonept\phtwopt}{\yym^2} \right|
= \left|\frac{p^+_{\gamma_1}p^-_{\gamma_2}\,-\,p^-_{\gamma_1}p^+_{\gamma_2}}{\yym \,\cdot \sqrt{\yym^2+\yypT^2}}\right|,
\end{equation*}
with $\Delta \eta_{\gamma\gamma}=\eta_{\gamma_1}-\eta_{\gamma_2}$ and $p^\pm_\gamma = E_\gamma \pm p_{z,\gamma}$, where $E_\gamma$ is the energy of the photon and $p_{z,\gamma}$ is the $z$-component of the momentum of the photon. The definition in the Collins--Soper frame simplifies an interpretation of the scattering angle in the presence of initial-state radiation.
\item An angular variable sensitive to $\yypT$ defined as:
\begin{equation*}
\yyphiStar=\tan\frac{\yypiMDphi}{2}\,\sin\theta_{\eta}^*=\tan\frac{\yypiMDphi}{2}\,\sqrt{1-\left(\tanh\frac{\Delta \eta_{\gamma\gamma}}{2}\right)^2}.
\end{equation*}
Angular variables are typically measured with better resolution than the photon energy. Therefore, a particular reference frame that allows $\yyphiStar$ to be expressed in terms of angular variables only, denoted by the subscript $\eta$ and described in Ref.~\cite{Banfi:2010cf}, is used in order to optimise the resolution. In particular, $\yyphiStar$ is an ideal probe of QCD effects because it is sensitive to similar dynamics to $\yypT$, but with a significantly better resolution in the low-$\yypT$ region. A similar argument holds for $\yypTt$ below.
\item Acoplanarity \yypiMDphi, where $\Delta\phi_{\gamma\gamma}$ is the azimuthal angular separation of the two photons.
\item Transverse component of \yypT with respect to the thrust axis:\footnote{The thrust axis is defined as $\hat{t}=(\vec{p}_{\mathrm{T},\gamma_1}-\vec{p}_{\mathrm{T},\gamma_2})/|\vec{p}_{\mathrm{T},\gamma_1}-\vec{p}_{\mathrm{T},\gamma_2}|$.}
\begin{equation*}
\yypTt = 2\cdot\frac{\left| p_{x,\gamma_1}p_{y,\gamma_2}-p_{y,\gamma_1}p_{x,\gamma_2}\right|}{\left|\vec{p}_{\mathrm{T},\gamma_1}-\vec{p}_{\mathrm{T},\gamma_2}\right|},
\end{equation*}
where $p_{x(y),\gamma_{1(2)}}$ are the $x(y)$-component of the leading (sub-leading) photon momenta, and $\vec{p}_{\mathrm{T},\gamma_1}-\vec{p}_{\mathrm{T},\gamma_2}$ is the vectorial difference of the transverse momenta of the two photons.
\end{itemize}
 
A fine observable binning is used, taking advantage of the detector resolution and large sample size.
In general, the bin size is set to be roughly four times the detector resolution; this requirement ensures large purities in the response matrices used for unfolding the measurement to particle level.
To avoid very large statistical uncertainties in the regions with low event counts, the minimum bin size is limited based on the expected number of signal events, keeping the statistical uncertainty below 10\%.
The two previous requirements would still allow an extremely fine binning in some regions for $\yycosTS$, $\yyphiStar$ and $\yypiMDphi$.
In those regions the bin size is arbitrarily enlarged to make the plot easily readable.
 

\section{Background estimation}
\label{sec:background}
 
\newcommand*{\yyFracRnd}{\ensuremath{(60.4\pm3.0)\%}\xspace}
\newcommand*{\yjFracRnd}{\ensuremath{(20.0\pm1.3)\%}\xspace}
\newcommand*{\jyFracRnd}{\ensuremath{(10.1\pm1.1)\%}\xspace}
\newcommand*{\jjFracRnd}{\ensuremath{(6.3\pm1.2)\%}\xspace}
\newcommand*{\eFakeFracRnd}{\ensuremath{(2.6\pm 0.1)\%}\xspace}
\newcommand*{\PUFracRnd}{\ensuremath{(0.6\pm0.4)\%}\xspace}
 
Reconstructed photon pairs can appear in the detector from a variety of sources beyond prompt production processes. These are summarised in this section along with details about how they are estimated and subtracted from the selected events to give the actual $\yy$ signal yield.
 
Most of the background arises from jets misidentified as photons;
these are mainly jets containing a neutral hadron which carries most of the jet energy and decays into a collimated pair of photons with overlapping showers in the calorimeter.
This background consists of $\yj$, $\jy$, or $\jj$ events with one or two jets misidentified as photons, respectively;
the notation $\jy$ ($\yj$) here and in the following refers to the component where the leading (sub-leading) photon candidate stems from a misidentified jet, while in $\jj$ events both photon candidates do.
The data-driven determination is described in detail in Section~\ref{sec:background:jetfake}.
 
A few percent of the sample corresponds to the \textit{electron background (ee)}, with one or two electrons\footnote{An `electron' here and in the following always refers to an electron or a positron.} being misidentified as photons. The electromagnetic showers created by electrons in the calorimeter are very similar to the ones from photons, and therefore electrons can be misidentified as photons. The main process contributing to this background is Drell--Yan production. It is estimated using the MC simulation described in Section~\ref{sec:sample:mc}. It makes its largest contribution to the background in the analysis region at $\yym \sim m_\textrm{Z}$.
 
Finally, less than a percent of the sample is associated with \textit{pile-up background} (PU), i.e.\ pairs of $\yj$ events from different $pp$ collisions in the same bunch crossing. It is determined with a data-driven approach discussed in Section~\ref{sec:background:pileup}.
 
The selected diphoton data sample also contains signal events in which the two photon candidates come from the decay of a Higgs boson. Their contribution to the signal region is predicted to be $0.2 \%$ of the integrated signal and increases to a few percent in the relevant $\yym$ bin. This contribution is not subtracted from the data as background, since it is a small genuine part of prompt-photon pair production.
 
\subsection{Jet background}
\label{sec:background:jetfake}
 
The background from jets misidentified as prompt photons is estimated using a data-driven method, extrapolating the amount from multiple control regions enriched in background to the \textit{signal region}, i.e.\ the region defined by the nominal event selection.
This extrapolation is done performing a Poisson likelihood fit. The method is an extension of the `two-dimensional sideband' technique used in the single-photon analyses \cite{STDM-2010-08,STDM-2011-04,STDM-2011-28,STDM-2012-18,STDM-2012-16,STDM-2014-09,STDM-2015-11,STDM-2016-08,STDM-2017-01,STDM-2017-29} to the two-photon case.
It is similar to the method used in previous diphoton analyses \cite{STDM-2011-05,STDM-2012-05}.
The fit is performed separately in each bin of each observable.
 
\begin{figure}
\begin{center}
\includegraphics[width=0.6\textwidth]{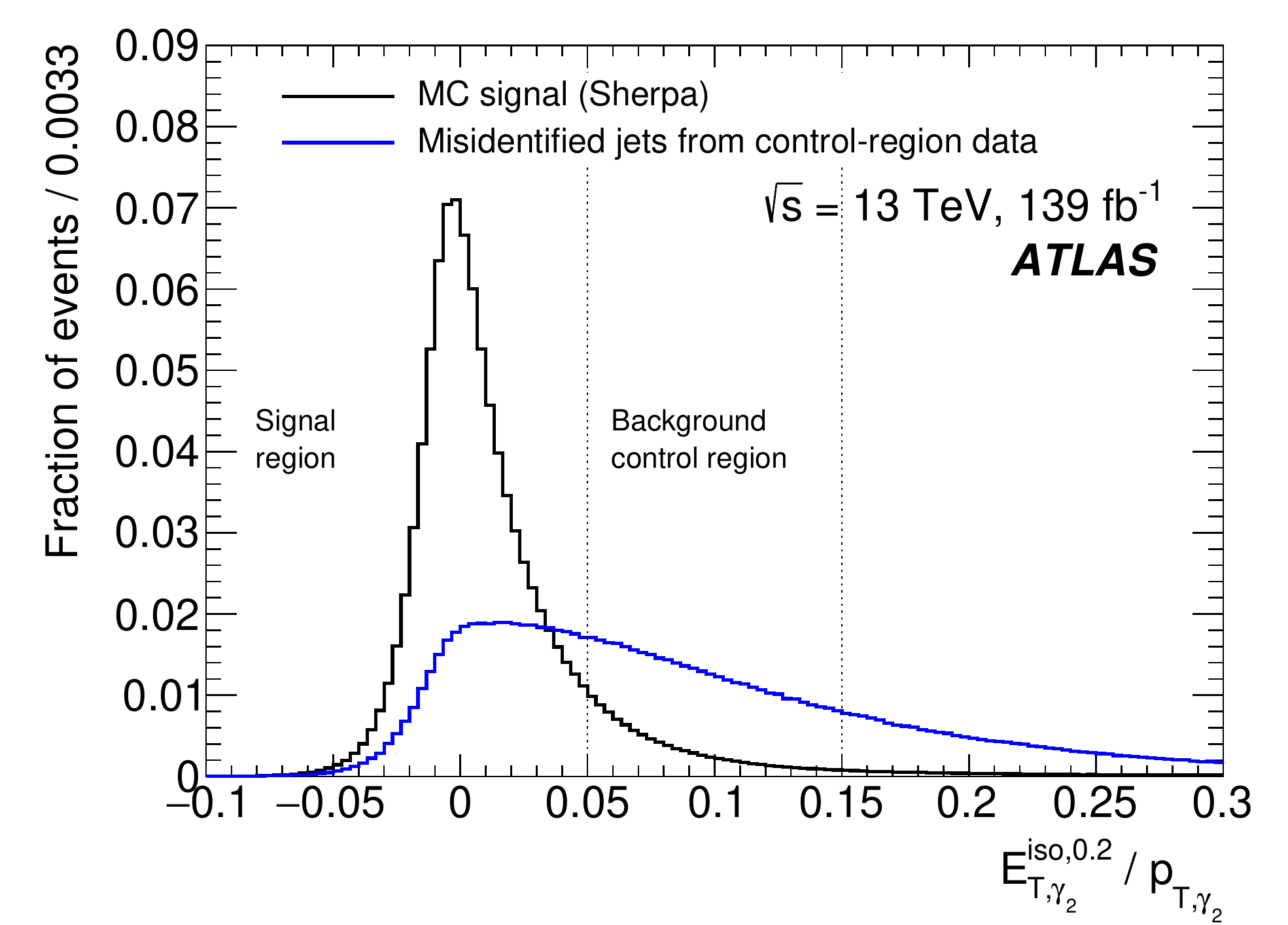}
\caption{
Relative photon isolation distribution for the sub-leading photon candidate as extracted from the \Sherpa diphoton sample (black histogram), and from misidentified jets, estimated from data events with sub-leading photon candidates failing the tight identification requirement (blue histogram). A small prompt-photon contribution to the non-tight control region is subtracted from the blue histogram, according to the \Sherpa sample predictions.}
\label{fig:background:ph2_iso}
\end{center}
\end{figure}

To define the background control regions, two almost independent jet--photon discriminant criteria are used: the lateral shape of the shower at the first layer of the electromagnetic calorimeter and the photon isolation.
Specifically, if a photon candidate fails the tight identification selection, by failing the requirements on either $f_\mathrm{side}$, $w_{s\mathrm{3}}$, or both shower shape variables (defined in Section~\ref{sec:definition:photons}), but satisfies the rest of the identification criteria, the event enters into one of the control regions.
If a photon fails the nominal isolation requirement, but falls into $\recoIsoFactor\cdot p_{\mathrm{T},\gamma_{1(2)}}<\etisoptwo{,\gamma_{1(2)}}<0.15\cdot p_{\mathrm{T},\gamma_{1(2)}}$, the event enters in a control region.
The isolation signal and control regions are indicated in Figure~\ref{fig:background:ph2_iso}, which shows the $\etisoptwo{,\gamma_{2}}/\phtwopt$ relative isolation distribution for prompt photons and for misidentified jets.
 
\begin{figure}
\begin{center}
\includegraphics[width=0.4\textwidth]{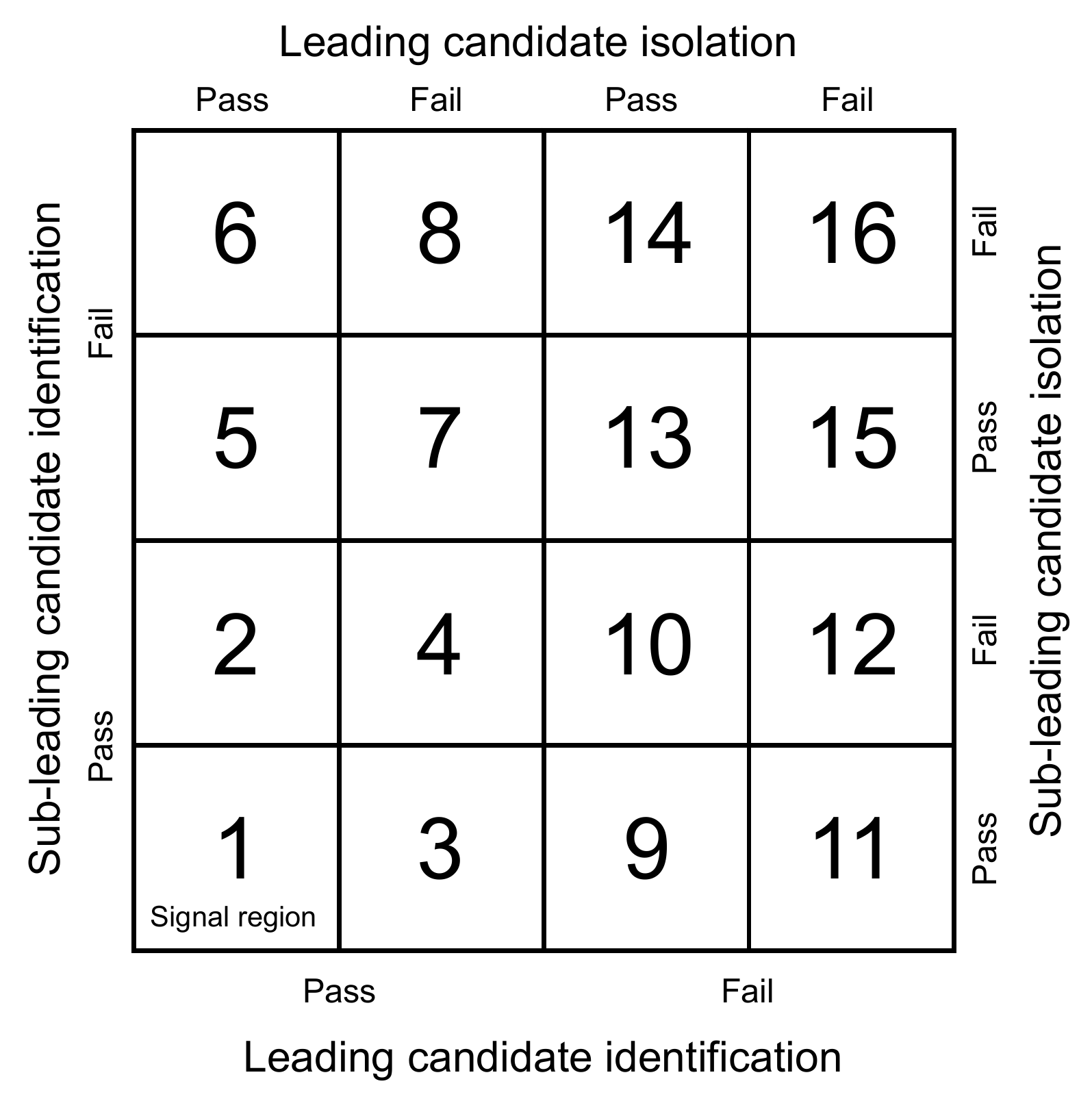}
\caption{
Definition of background control regions, and their corresponding label index.}
\label{fig:background:4DBinning}
\end{center}
\end{figure}

In addition to the signal region, where both photons satisfy both the isolation and identification criteria, 15 background control regions are defined representing all possible combinations where at least one photon fails to meet at least one of the criteria, as illustrated in Figure~\ref{fig:background:4DBinning}.
These regions are labelled by the index $i=1\dots 16$, with $i=1$ being the signal region.
 
The fit model is built of probability functions $f_{p,i}$, corresponding to the probability of an event from process $p \in \{\yy,\yj,\jy,\jj,\ee,\PU\}$ to fall in the diphoton region $i$. They are constructed as a product of the four probabilities: the probabilities for each of the two photon candidates to fulfil both the isolation and identification requirements of the given region, according to
\begin{equation}
\label{eq:fpi}
\begin{array}{ll}
f_{p,i} & = f_{p,i}(\eiso{1},\, \eiso{2},\, \riso{p},\, \eid{1},\, \eid{2},\, \rid{p},\, \rii{p}{1},\, \rii{p}{2}) \\
\\
& = \left\{
\setlength{\tabcolsep}{10mm}
\renewcommand{\arraystretch}{1.2}
\begin{array}{l@{\,\,}l@{\,\,}l@{\,\,}l@{\,\,\,\,\,\,\,\,}l}
\eiso{1} & \eiso{2} & \eid{1} & \eid{2} & \mbox{for }i=1 \\
\eiso{1} & \fiso{2} & \eid{1} & \eid{2} & \mbox{for }i=2 \\
\fiso{1} & \eriso & \eid{1} & \eid{2} & \mbox{for }i=3 \\
\fiso{1} & \friso & \eid{1} & \eid{2} & \mbox{for }i=4 \\ [3mm]
 
\eiso{1} & \eriio{2} & \eid{1} & \fid{2} & \mbox{for }i=5 \\
\eiso{1} & \friio{2} & \eid{1} & \fid{2} & \mbox{for }i=6 \\
\fiso{1} & \erisoriio & \eid{1} & \fid{2} & \mbox{for }i=7 \\
\fiso{1} & \frisoriio & \eid{1} & \fid{2} & \mbox{for }i=8 \\ [3mm]
 
\eriio{1} & \eiso{2} & \fid{1} & \erid & \mbox{for }i=9 \\
\eriio{1} & \fiso{2} & \fid{1} & \erid & \mbox{for }i=10 \\
\friio{1} & \eriso & \fid{1} & \erid & \mbox{for }i=11 \\
\friio{1} & \friso & \fid{1} & \erid & \mbox{for }i=12 \\ [3mm]
 
\eriio{1} & \eriio{2} & \fid{1} & \frid & \mbox{for }i=13 \\
\eriio{1} & \friio{2} & \fid{1} & \frid & \mbox{for }i=14 \\
\friio{1} & \erisoriio & \fid{1} & \frid & \mbox{for }i=15 \\
\friio{1} & \frisoriio & \fid{1} & \frid & \mbox{for }i=16. \\
\end{array}
\right.\\
\end{array}
\end{equation}
 
Here, \eiso{n} (\eid{n}) is the efficiency of the leading ($n=1$) or sub-leading ($n=2$) photon candidate to fulfil the signal region isolation (identification) requirement, relative to the looser requirements that include the control regions.
The correlation correction factors $R$ account either for correlations between the isolation and identification of one photon candidate, $\rii{p}{n}$, or for correlations between the isolation (identification) of both candidates, \riso{p} (\rid{p}).
A value $R=1$ corresponds to no correlation between the relevant variables, while values below or above unity indicate a certain degree of positive or negative correlation, respectively.
The $R$ values are typically close to one, slightly below one, with the lowest values reaching $0.9$; more details are given in Section~\ref{sec:background:jetfake:inputs}.
 
The expected number of events $n^{\mathrm{exp}}_i$ in each region $i$ can then be written as:
\begin{equation}\nonumber
\begin{array}{rl}
n^{\mathrm{exp}}_i =& \dfrac{n_{\yy}}{\eff{id}{\yy}{1}\, \eff{iso}{\yy}{1}\, \eff{id}{\yy}{2}\, \eff{iso}{\yy}{2}} \, f_{\yy,i}
+\,
N_{\yj}f_{\yj, i} \,+\,
N_{\jy}f_{\jy, i}
+\,
N_{\jj}f_{\jj, i} \,+\,
N_{\ee}f_{\ee, i} \,+\,
N_{\PU}f_{\PU, i}. \\
\end{array}
\end{equation}
Here, $N_{\yy}$, $N_{\yj}$, $N_{\jy}$, $N_{\jj}$, $N_{\ee}$ and $N_{\PU}$ are the total number of events integrated in all 16 regions, attributed to the signal or background processes, respectively.
The parameter of interest, i.e.\ the number of $\yy$ events in the signal region, is given by $n_{\yy} \equiv N_{\yy}\cdot \eff{id}{\yy}{1}\, \eff{iso}{\yy}{1}\, \eff{id}{\yy}{2}\, \eff{iso}{\yy}{2}$.
 
Most of the parameters associated with misidentified jets are estimated from the data, by allowing them to float in the fit, because their MC predictions are not accurate, which motivates this sophisticated approach.
The correlation correction factors $R$ are fixed in the fit after being estimated either from MC simulation or from data in validation regions, except for $\rid{\jj}$ and $\riso{\jj}$, which are allowed to float in the fit.
In total, a set of 13 nuisance parameters are allowed to float in the fit in addition to $n_{\yy}$:
\begin{equation}\nonumber
\theta = \{
N_{\yj},\,
\eff{id}{\yj}{2},\, \eff{iso}{\yj}{2},\,
N_{\jy},\, \eff{id}{\jy}{1},\, \eff{iso}{\jy}{1},\,
N_{\jj},\, \eff{id}{\jj}{1},\, \eff{iso}{\jj}{1},\, \eff{id}{\jj}{2},\, \eff{iso}{\jj}{2},\, \rid{\jj},\, \riso{\jj}
\}.
\end{equation}
 
The Poisson likelihood function is given by
\begin{equation}\nonumber
\mathcal{L}(n_{\yy},\, \theta | \,\vec{n}^{\mathrm{obs}}) = \prod_{i} \frac{\left[\nexp\right]^{\nobs}}{\nobs!} \mathrm{e}^{-\nexp},
\end{equation}
where $\nobs$ is the number of diphoton candidates observed in each region $i$.
 
The efficiency parameters $\varepsilon$ corresponding to prompt photons are determined from MC simulation and fixed in the fit.
Details about the predetermined input parameters are given in Section~\ref{sec:background:jetfake:inputs}.
The parameters of the electron and pile-up background probability functions, $f_{\ee, i}$ and $f_{\PU, i}$, as well as $N_{\ee}$ and $N_{\PU}$, are also predetermined and fixed in the fit;
the normalisation of the pile-up background $N_{\PU}$ is estimated in a data-driven way, as described in Section~\ref{sec:background:pileup}, while the values for the other parameters are extracted from MC simulation.
 
\begin{figure}[t]
\begin{center}
\includegraphics[width=0.8\textwidth]{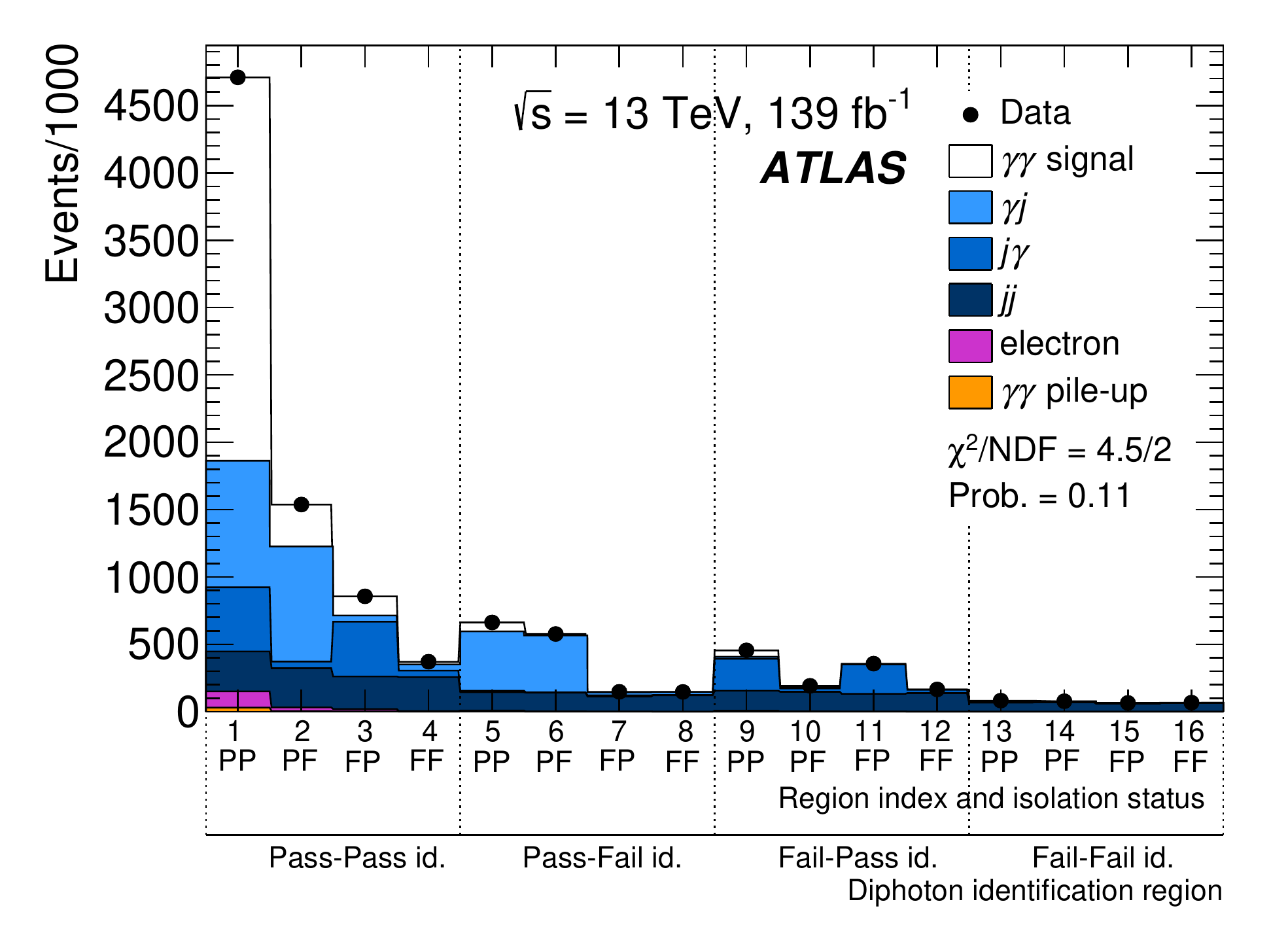}
\caption{Event yields of the inclusive data sample in the 16 diphoton regions. Both the observed counts and the fit result, including the estimated contributions from the signal and each background component, are shown. The quoted $\chi^2$ value is computed taking into account the systematic uncertainties of the fit model, and their correlation between bins. Labels `PP/PF/FP/FF' are used to denote the isolation status (`P' = pass, `F' = fail) of the leading and sub-leading photon candidates. Similarly, the labels `Pass-Pass id.', `Pass-Fail id.', etc., denote the identification status of the two photon candidates.}
\label{fig:background:fit_inclusive}
\end{center}
\end{figure}

\begin{figure}[tbp]
\centering
\vspace{-3mm}
\includegraphics[width=0.49\textwidth]{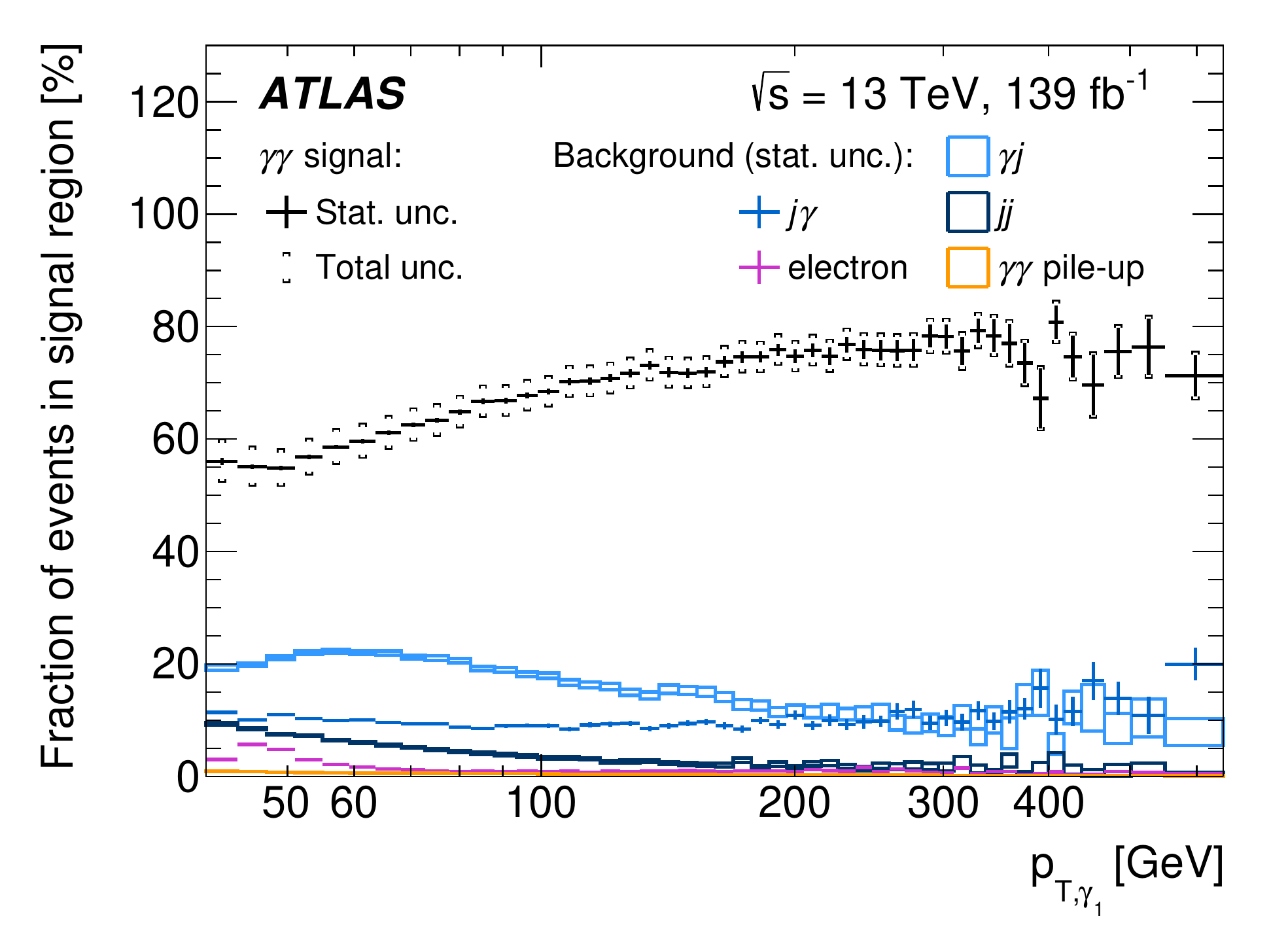}
\includegraphics[width=0.49\textwidth]{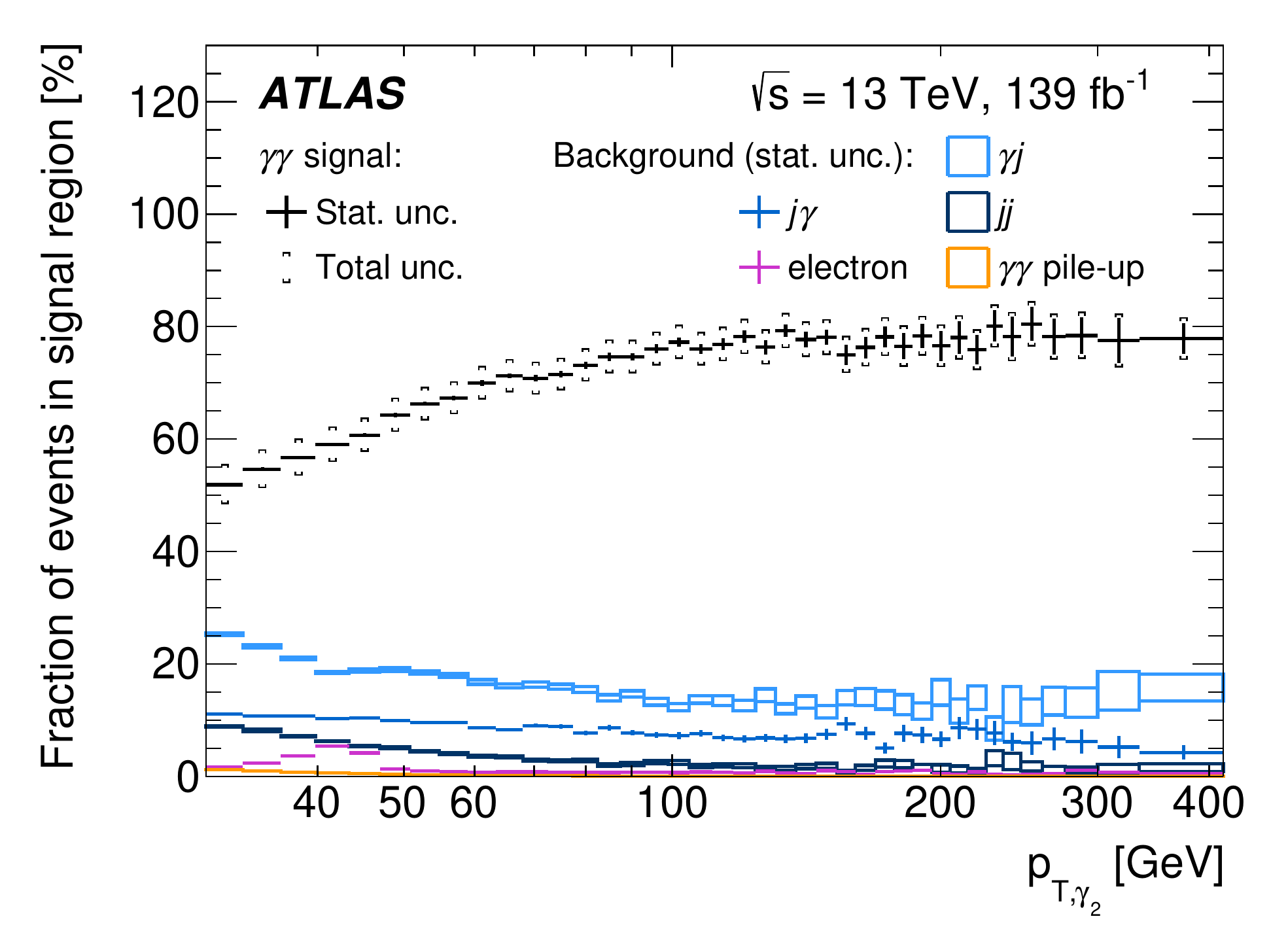}
\includegraphics[width=0.49\textwidth]{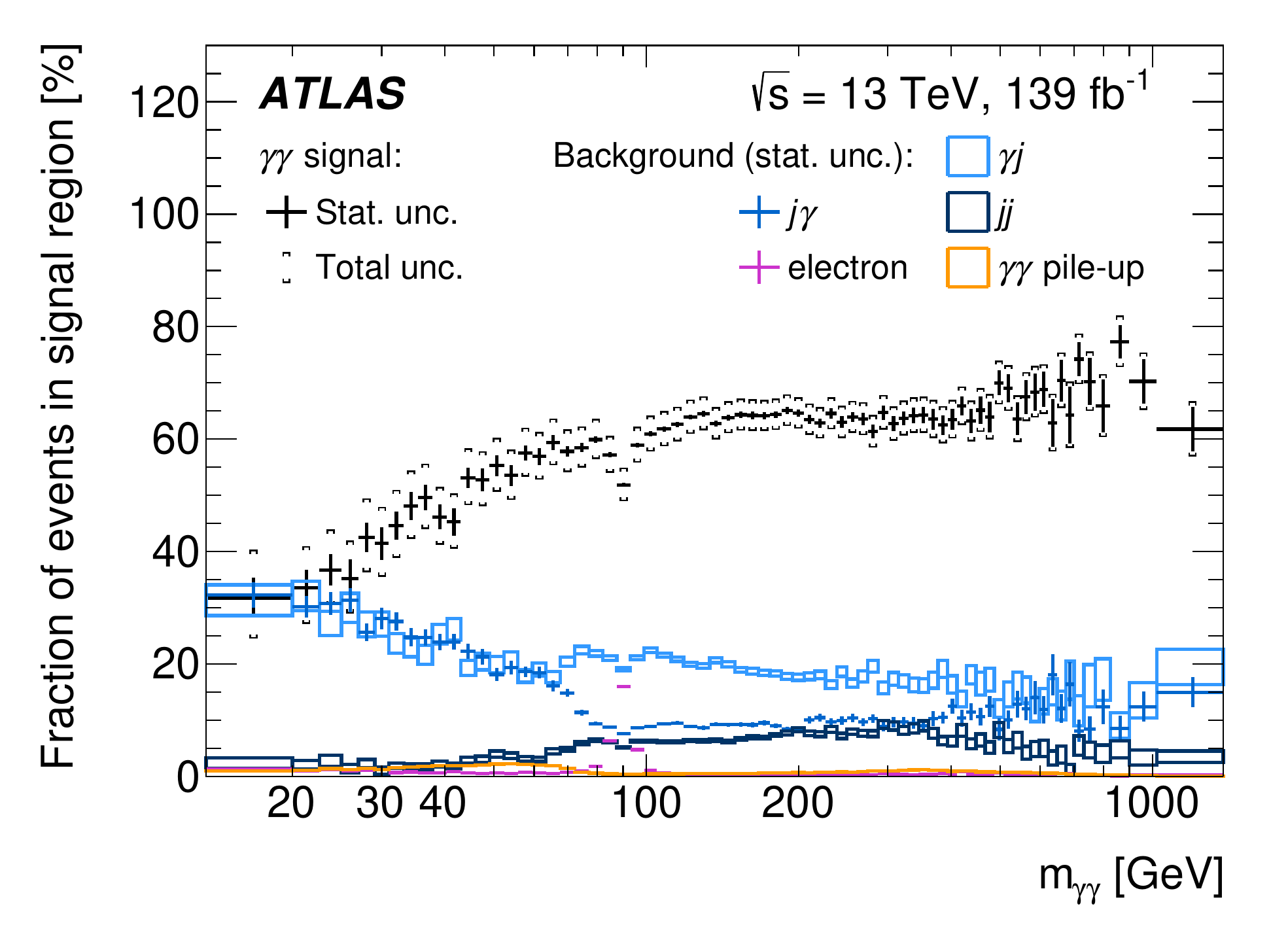}
\includegraphics[width=0.49\textwidth]{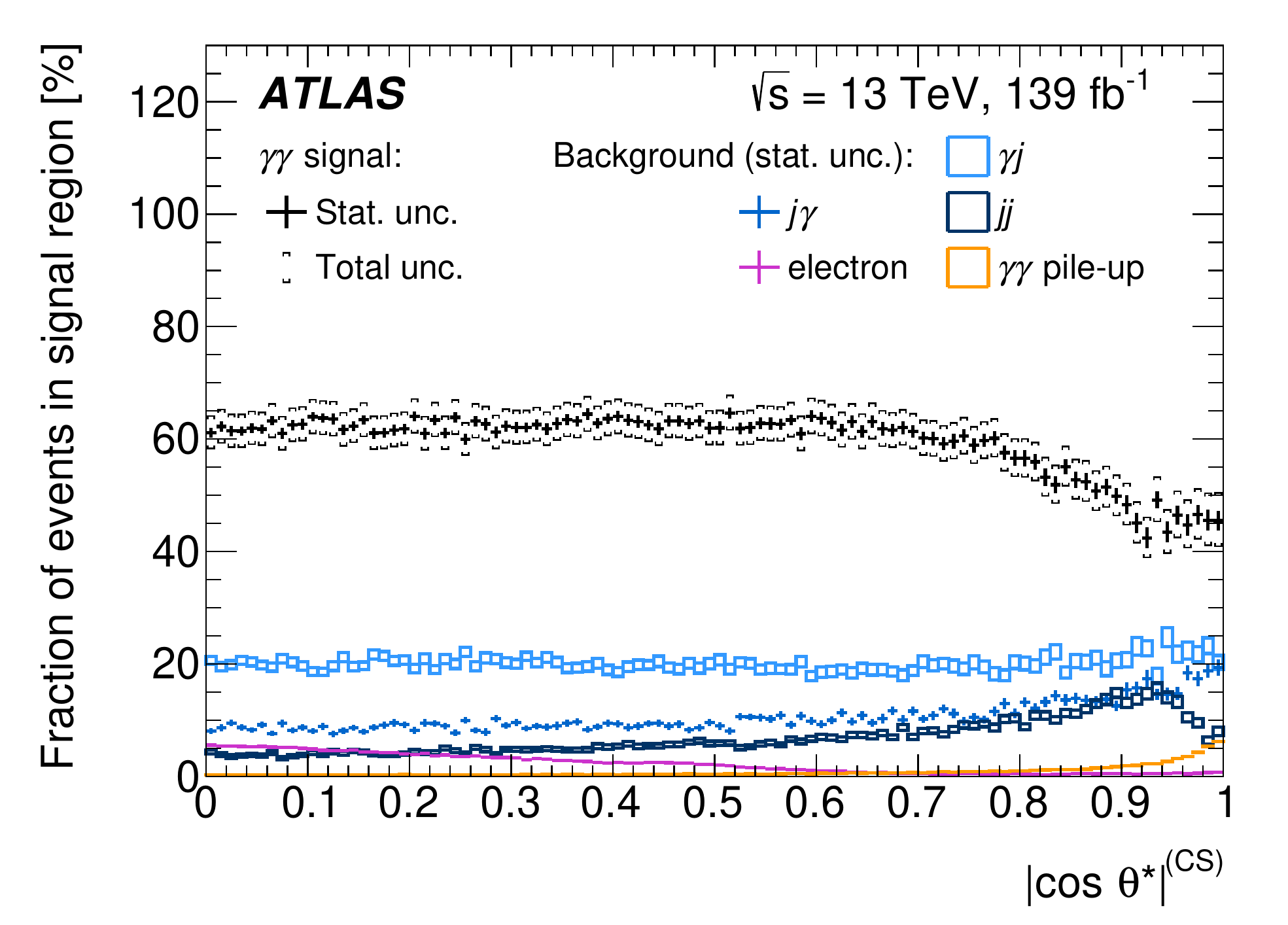}
\includegraphics[width=0.49\textwidth]{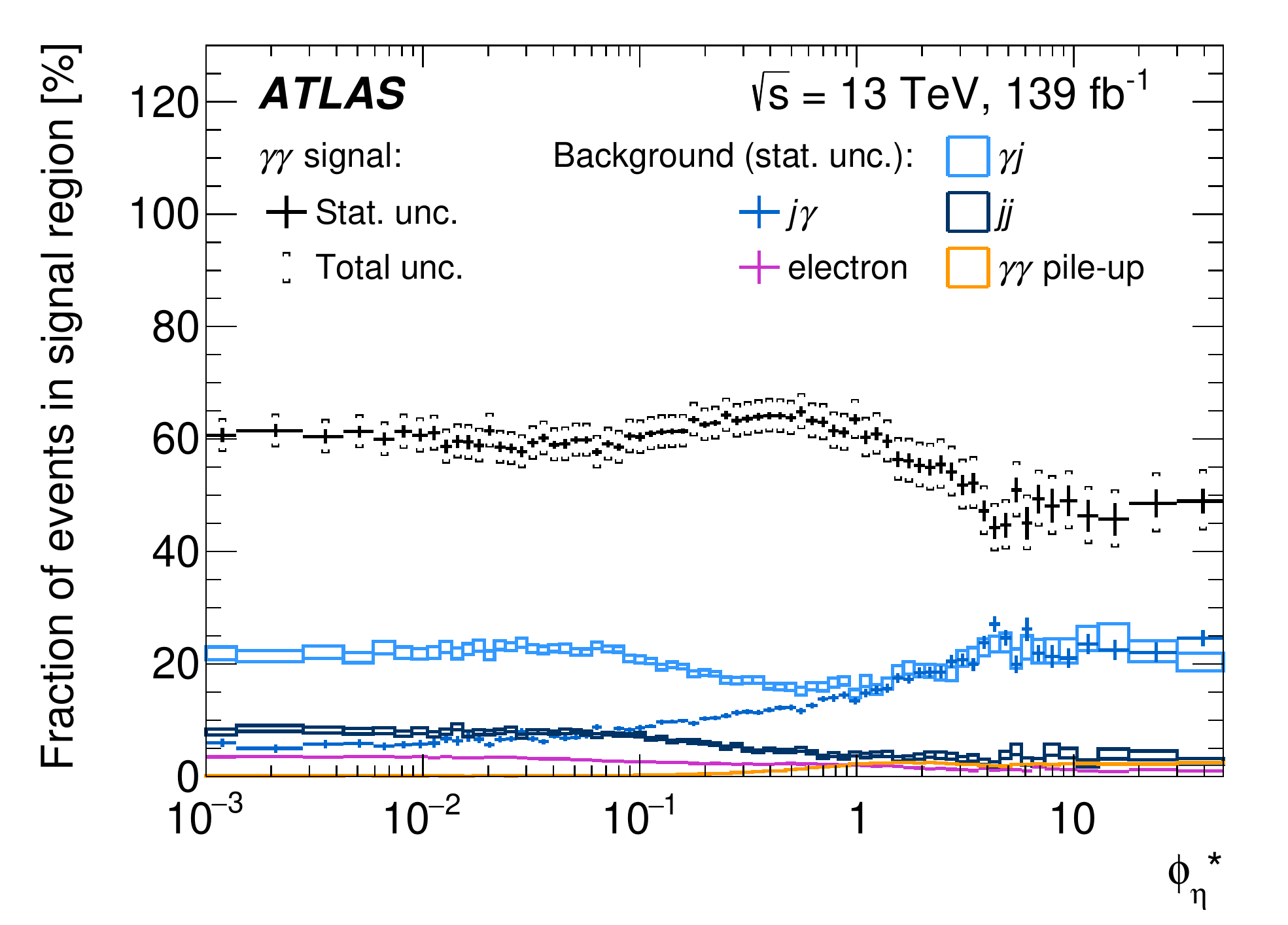}
\includegraphics[width=0.49\textwidth]{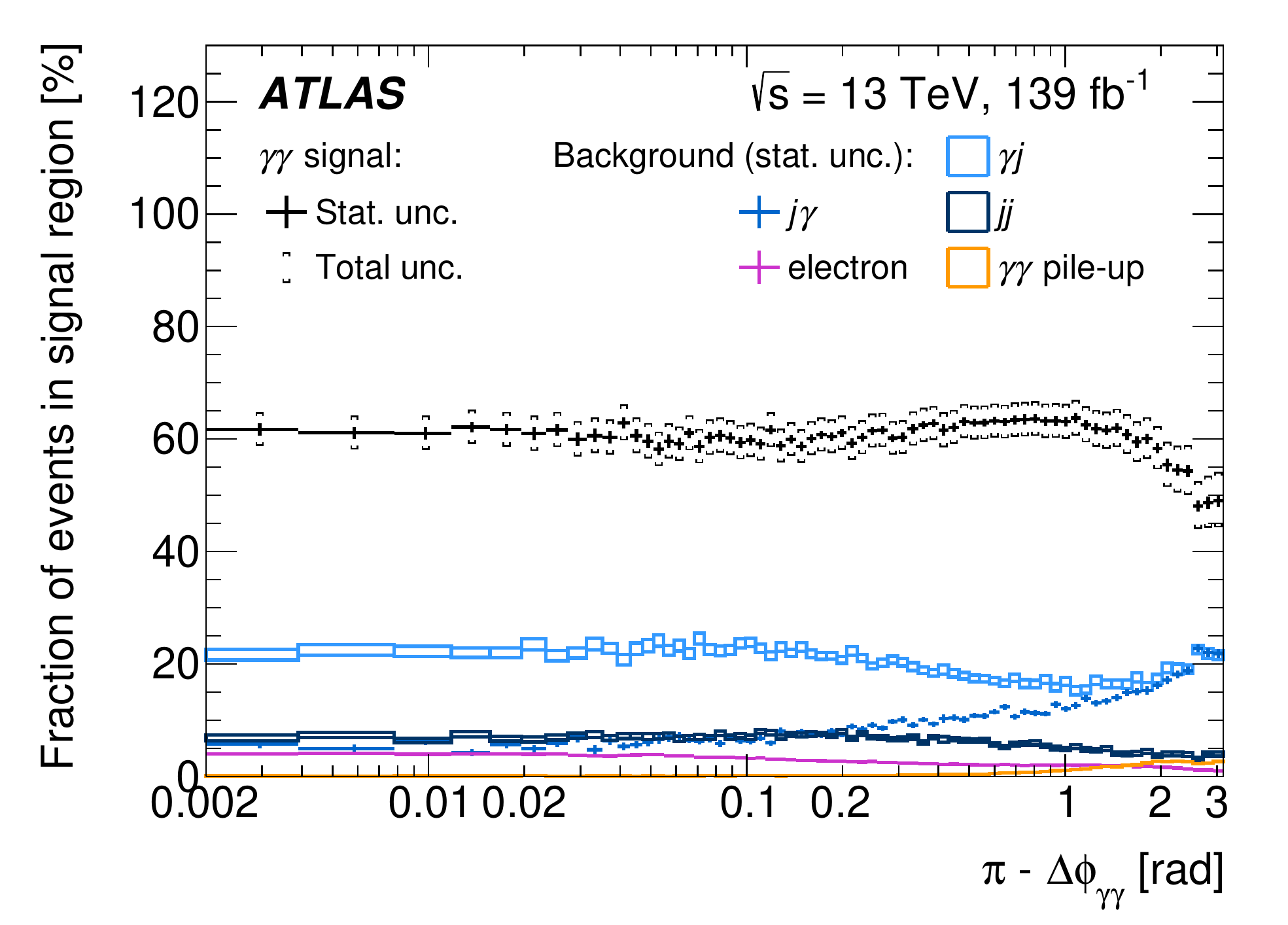}
\includegraphics[width=0.49\textwidth]{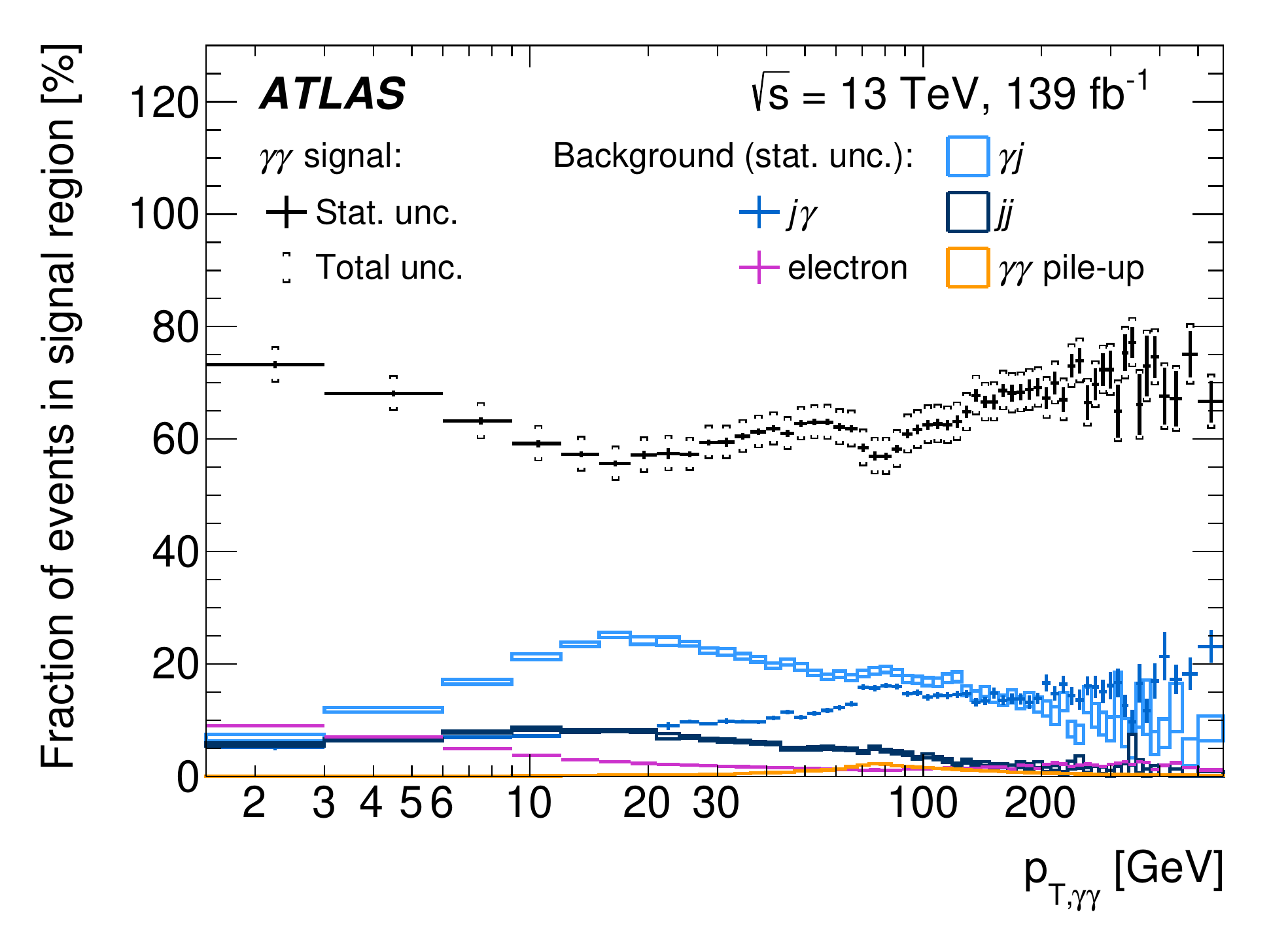}
\includegraphics[width=0.49\textwidth]{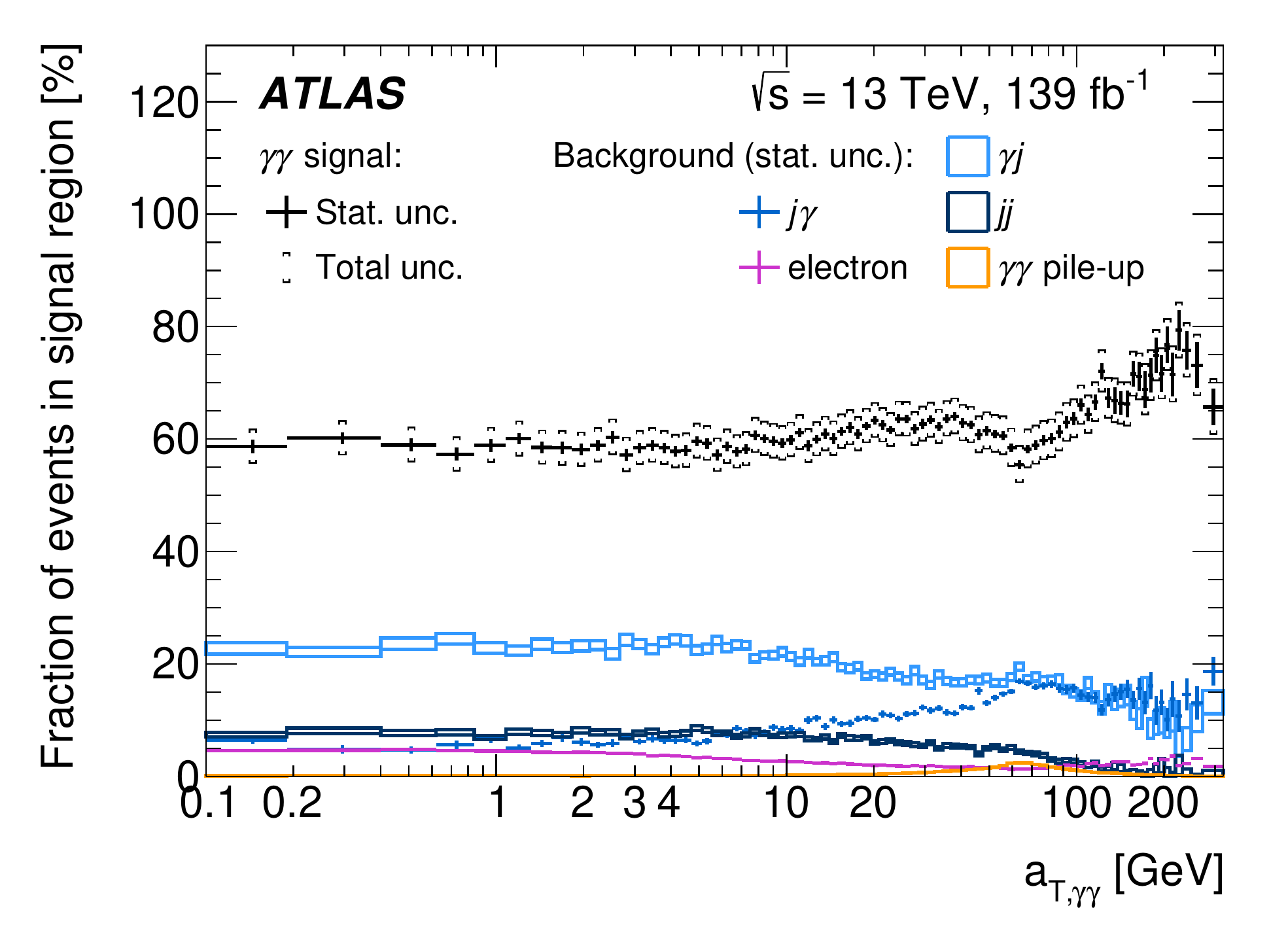}
\vspace{-3mm}
\caption{Sample decomposition as a function of each observable. For the $\yy$ signal fraction, the total uncertainty is shown, while for the background fractions only the data and Monte Carlo statistical uncertainty component is shown.}
\label{fig:background:purity}
\end{figure}

As an example, the distribution of the inclusive data sample in the 16 regions is shown in Figure~\ref{fig:background:fit_inclusive}, together with the result of the fit to this inclusive sample;
the fractional composition of the data in the signal region resulting from this fit is given in Table~\ref{tab:bkgresult}.
For the nominal measurement, as mentioned above, the fit is performed separately in each bin of each observable, and the equivalent sample decomposition in the signal region is displayed differentially in Figure~\ref{fig:background:purity}. The purity of the $\yy$ signal process ranges from $35\%$ at low invariant mass to $80\%$ at high $p_\mathrm{T,\gamma}$.
 
\begin{table}
\begin{center}
\caption{
Estimated composition of the signal region diphoton candidate samples, resulting from the inclusive sample fit.
The uncertainties are fully dominated by the systematic uncertainties, which are discussed in Section~\ref{sec:uncertainties}.}
\label{tab:bkgresult}
\begin{tabular}{lcrclc}
\hline\hline
Process & \,\,\, & & \hspace{-17mm} Event fraction [\%] \hspace{-17mm} & & \,\,\, \\
\hline
\yy signal & & 60.4 & \hspace{-4mm} $\pm$ \hspace{-4mm} & 3.0 &\\
\yj & & 20.0 & \hspace{-4mm} $\pm$ \hspace{-4mm} & 1.3 &\\
\jy & & 10.1 & \hspace{-4mm} $\pm$ \hspace{-4mm} & 1.1 &\\
\jj & & 6.3 & \hspace{-4mm} $\pm$ \hspace{-4mm} & 1.2 &\\
Electron & & 2.6 & \hspace{-4mm} $\pm$ \hspace{-4mm} & 0.1 &\\
\yy pile-up & & 0.6 & \hspace{-4mm} $\pm$ \hspace{-4mm} & 0.4 &\\
\hline\hline
\end{tabular}
\end{center}
\end{table}
 
This background and signal estimation method is tested using MC simulation.
A pseudo-dataset is prepared including $\yy$ signal and jet background events, with proportions similar to the ones observed in the data.
The results obtained with the fit to the pseudo-dataset are compatible with the fractional composition of the MC events included in the sample.
 
\subsubsection{Fit model input parameters}
\label{sec:background:jetfake:inputs}
 
This section describes the parameters of the fit model that are predetermined and fixed in the fit, and how their values and systematic uncertainties are determined.
The propagation of the systematic uncertainties to the estimated $n_{\yy}$ is done by rerunning the fit using systematic variations of the input parameters, keeping the uncertainty from each source fully correlated between the bins of the observables.
 
\begin{figure}[t]
\centering
\includegraphics[width=0.6\textwidth]{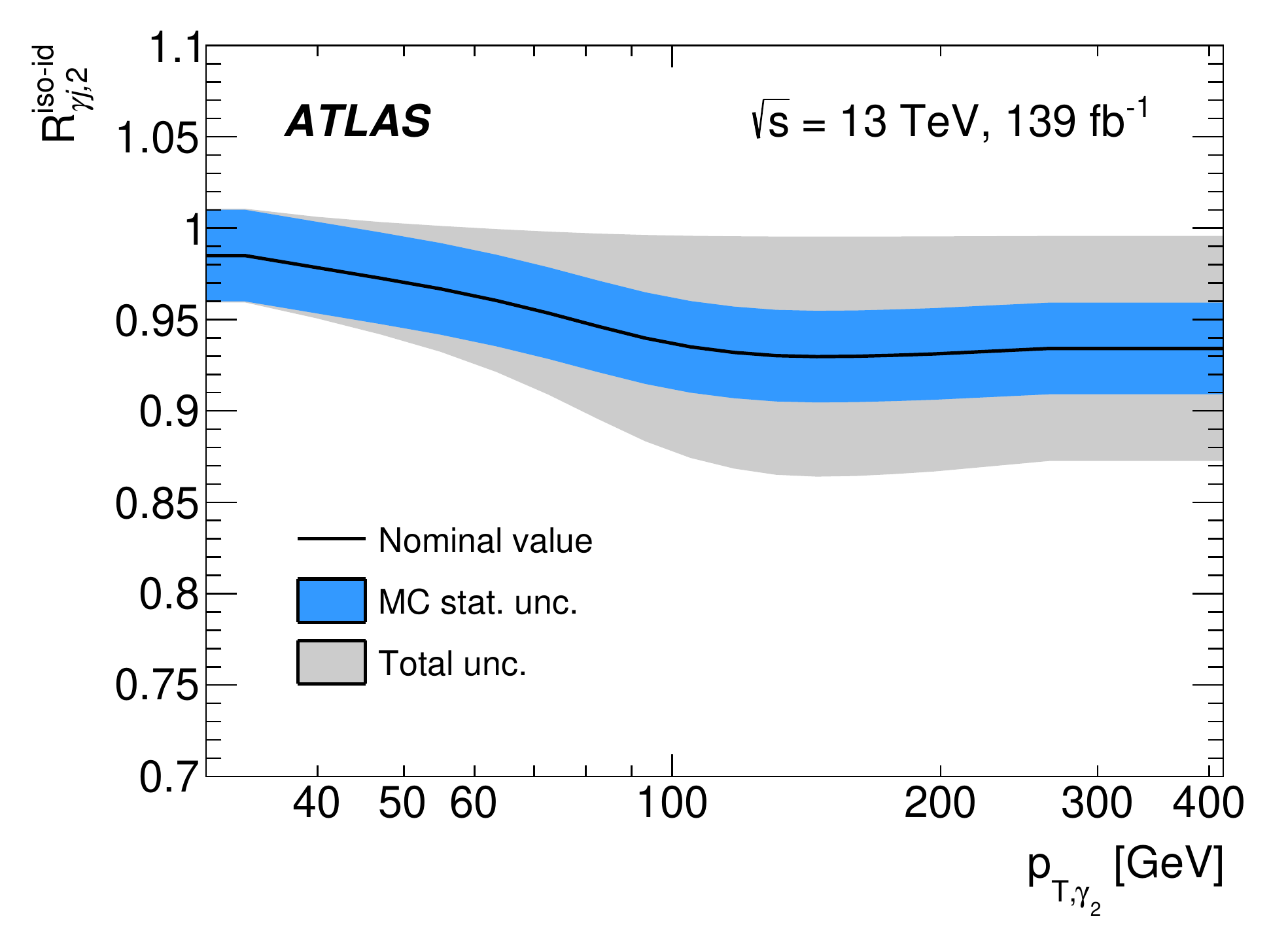}
\caption{Correction factor \rii{\yj}{2} for the isolation--identification correlation, for jets misidentified as photons. The nominal value is indicated in black, and is the average of the values obtained from a data validation region and a $\yj$ MC sample. The blue band corresponds to the simulation statistical uncertainty, and the grey band shows the total uncertainty, taking into account half of the difference between the MC and data results.}
\label{fig:background:risoidj}
\end{figure}

\begin{description}
\item[Signal parameters: \eff{id}{\yy}{1}, \eff{id}{\yy}{2}, \rid{\yy}, \eff{iso}{\yy}{1}, \eff{iso}{\yy}{2}, \riso{\yy}, \rii{\yy}{1} and \rii{\yy}{2}]\hfill\\
The values of the eight \yy signal parameters are extracted from the \Sherpa MC signal sample, corrected to match the isolation and identification variable distributions in data~\cite{EGAM-2018-01}. These parameters are extracted as a function of each observable for which the differential cross section is measured. Uncertainties in these parameter values due to the detector simulation and theory modelling, in particular for the isolation distribution, are taken into account as described in Section~\ref{sec:uncertainties:photon}.
\item[Prompt-photon isolation and identification efficiencies in \yj and \jy events: \eff{iso}{\yj}{1}, \eff{iso}{\jy}{2}, \eff{id}{\yj}{1}, \eff{id}{\jy}{2}]\hfill\\
For the \yj and \jy background components, isolation and identification efficiencies for the prompt photons are derived from the \yy signal MC sample. The underlying assumption is cross-checked by comparing these efficiencies with the ones derived from the \yj \sherpa MC sample. The efficiencies are compatible within statistical uncertainties. In addition, a change of 2\% in those efficiencies is tested, which is roughly the size of the statistical uncertainty of the comparison ratio, and the impact on the estimated signal yield result is found to be negligible.
\item[Isolation--identification correlation for misidentified jets: \rii{\yj}{2}, \rii{\jy}{1}, \rii{\jj}{1} and \rii{\jj}{2}]\hfill\\
The correlation between isolation and identification for jets misidentified as photons is studied with $\yj$ events as a function of \phtwopt.
A twofold approach is used to estimate it.
First, a $\yj$ MC simulation sample is used, yielding $R_{\yj,2}^\mathrm{iso-id,MC}=0.990\pm 0.025$ (stat) without a significant trend as a function of \phtwopt given the limited MC sample size.
In addition to the MC study, a $\yj$-rich validation region in data is constructed by requiring the sub-leading photon to fail a track isolation requirement. The track isolation variable $p_{\mathrm{T},\gamma_{2}}^\mathrm{iso,0.2}$ is computed as the sum of the transverse momenta of charged-particle tracks coming from the primary vertex, in a cone of $\Delta R = 0.2$ around the photon candidate direction. In the validation region, the sub-leading photon candidates are required to have $p_{\mathrm{T},\gamma_{2}}^\mathrm{iso,0.2} > 0.02\cdot p_{\mathrm{T},\gamma_{2}} + 1.5\,\gev$. The contamination from $\yy$ events in that validation region is subtracted using the signal MC simulation. With the higher statistical power of the data, an enhancement of the positive correlation as a function of \phtwopt is observed, i.e.\ values of $R_{\yj,2}^\mathrm{iso-id}$ significantly below one. This dependency is extrapolated to the signal region using the MC simulation.\\
An average of the MC and data results is used for the central values of this correction factor versus \phtwopt; the values vary in a range of $0.93<R^\mathrm{iso-id}<0.99$ as shown in Figure~\ref{fig:background:risoidj}.
The MC simulation is susceptible to mismodelling of the correlation, while the data estimation relies on the MC-based subtraction of $\yy$ events in the validation region, and on the low correlation between the calorimetric and track isolation.
Half of the difference between the MC and data results is assigned as a systematic uncertainty, and combined in quadrature with the MC statistical uncertainty.
For each bin of the observables and each of the four $R^\mathrm{iso-id}$ parameters for misidentified jets, an average is calculated taking into account the corresponding photon candidate $p_\mathrm{T,\gamma}$ distribution, estimated from data events in the control regions dominated by each background process.\\
In addition to \phtwopt, the dependency of $R_{\yj,2}^\mathrm{iso-id}$ on the photon candidate pseudorapidity is also studied; however, its impact is found to be negligible.
\item[Correlation between isolation in \yj and \jy events: \riso{\yj} and \riso{\jy}]\hfill\\
The correction factor $\riso{p}$ parameterises the correlation between the isolation variables of the two photon candidates. For the signal probability function, this parameter is extracted from MC simulation as \riso{\yy}. A similar extraction for misidentified jets from \yj MC samples alone is not possible because of the limited statistical precision available.
A slight correlation is expected, as found for the signal process, where $0.95 < \riso{\yy} < 1$. It is also extracted from \jj background events, as \riso{\jj} is left to float in the fit and yields values in the range $0.9 < \riso{\jj} < 1$. For the \yj and \jy backgrounds, $\riso{\yj}=\riso{\jy}=0.95\pm 0.05$ is chosen in all regions of phase space, thus the uncertainties cover the extreme cases.
This uncertainty is treated as fully correlated between all the bins of the observables, in the same way as all the uncertainties discussed in this section. \\
In order to have the common correction factor $\riso{\yj}=\riso{\jy}$ consistently affecting the isolation efficiency of the photon candidate associated with the misidentified jet in both processes, $\yj$ and $\jy$, the probability function for the $\jy$ background $f_{\jy, i}$ is slightly modified relative to the Eq.~(\ref{eq:fpi}) function.
In Eq.~(\ref{eq:fpi}), the $\riso{p}$ parameter corrects the isolation efficiency of the sub-leading photon candidate \eff{iso}{p}{2} in the regions where the leading candidate fails the nominal isolation requirement, i.e.\ in the regions 3, 4, 7, 8, 11, 12, 15 and 16.
In the \jy case, the function is modified, and $\riso{\jy}$ corrects the leading candidate efficiency $\eff{iso}{\jy}{1}$ in the regions where the sub-leading candidate fails the nominal isolation criterion, i.e.\ in the regions 2, 4, 6, 8, 10, 12, 14 and 16.
For example, for regions 3 and 4:
\begin{equation*}
f_{\jy,i} = \left\{
\setlength{\tabcolsep}{10mm}
\renewcommand{\arraystretch}{1.2}
\begin{array}{l@{\,\,}l@{\,\,}l@{\,\,}l@{\,\,\,\,\,\,\,\,}l}
\fisojy{1} & \eisojy{2} & \eidjy{1} & \eidjy{2} & \mbox{for }i=3 \\
\frisojy & \fisojy{2} & \eidjy{1} & \eidjy{2} & \mbox{for }i=4. \\
\end{array}
\right.\\
\end{equation*}
\item[Identification correlation in \yj events: \rid{\yj} and \rid{\jy}]\hfill\\
The correlation between the identification of the two objects is in general very low. This is cross-checked with MC simulation for the \yy signal process. For the \jj process, the correlation is extracted from the data, by allowing \rid{\jj} to float in the fit. In both cases, \rid{} is close to one within 2\%. Performing the fit with the \rid{\yj} and \rid{\jy} values changed by 2\% produces negligible variations of the final result. The assumption $\rid{\yj}=\rid{\jy}=1$ is thus kept without an additional uncertainty.
\end{description}
 
\subsection{Pile-up of multiple single-photon events}
\label{sec:background:pileup}
 
Pairs of $\yj$ events from different $pp$ collisions in the same bunch crossing can form a diphoton final state in the detector. The normalisation of this background is estimated using a fit to the data. The shape of the background in the diphoton observables is based on MC simulations. Details of the procedure are given below.
 
\begin{figure}[htbp]
\centering
\includegraphics[width=0.70\textwidth]{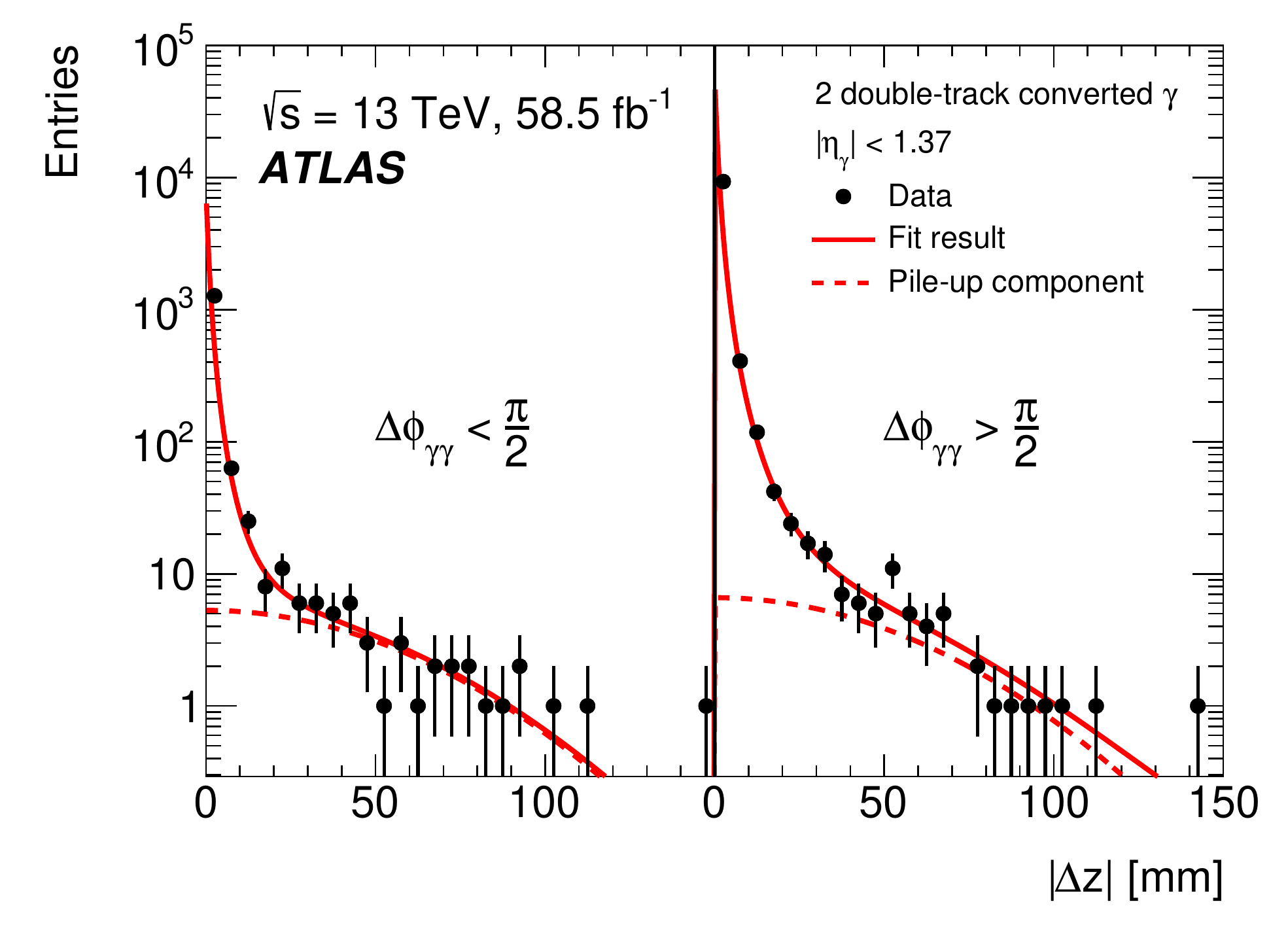}
\caption{The two-component fit to determine the global normalisation of pile-up events is shown. Data events in which each of the two photon candidates is associated with two inner detector tracks are used for the fit. The dashed lines represent the pile-up-like component of the total model marked with a solid line. The left side of the plot shows the events with azimuthal angular separation of the photon candidates $\Delta\phi_{\gamma\gamma}<\pi/2$, and the right side of the plot shows the events with $\Delta\phi_{\gamma\gamma}>\pi/2$. The fit is performed in the two regions simultaneously.}
\label{fig:background:pileupBackground}
\end{figure}

In diphoton events, in general, there is not enough information to efficiently identify and reject the pile-up background.
However, around 0.5\% of the events do have sufficiently precise information for both photons to effectively determine whether they are produced at the same primary interaction vertex or not.
This subsample of events is used to estimate the pile-up background normalisation.
It consists of events with both photons converted to electron--positron pairs within the detector material, with both conversions fully reconstructed, i.e.\ each including the reconstruction of the two charged-particle tracks, and with both photons in the central pseudorapidity region, $|\eta|<1.37$.\footnote{The fraction of converted photons is fairly constant as a function of photon $\ET$ and $\eta$ and well modelled in MC simulation. The conversion events can thus be used directly for an extrapolation to the full dataset.}
In addition, the conversion tracks are required to have hits in the silicon detectors of the inner tracking system.
For each photon, the trajectory is reconstructed using the shower-depth segmentation of the calorimeter, and the position of the conversion vertex.
The intersection of the extrapolated trajectory with the beam axis provides a precise estimate of the longitudinal position $z_\gamma$ of the vertex at which the photon is produced.
This information is used in a two-component fit to $\Delta z = |z_{\gamma_1}-z_{\gamma_2}|$, as can be seen in Figure~\ref{fig:background:pileupBackground}.
 
The first component of the model, representing the distribution of pile-up events in $\Delta z$, is considered to be a convolution of the two identical Gaussian $z_\gamma$ distributions that have a width of $\sigma_{z}=34.15\pm0.16$~mm, resulting in a Gaussian with a width of $\sigma_{\Delta z} = \sqrt{2}\times \sigma_{z}$.
The $\sigma_{z}$ value is estimated by fitting a Gaussian function to the $z_\gamma$ distribution.
Since $\sigma_{z}$ changes between data-taking periods, depending on the beam conditions, only data collected in 2018 are used for this pile-up background study; during this period $\sigma_{z}$ was relatively stable, with a root-mean-square of the $\sigma_{z}$ variations of 2.8~mm.
The integrated luminosity of the 2018 dataset is 58.5~fb$^{-1}$.
Afterwards, the estimated amount of pile-up background is extrapolated to the full Run-2 dataset, taking into account the integrated luminosity and the average number of interactions per bunch crossing $\langle \mu \rangle$ of the two datasets, $\langle \mu \rangle_{2018} = 36.1$ and $\langle \mu \rangle_\textrm{Run 2} = 33.7$.
The amount of pile-up background is proportional to $\langle \mu \rangle$.
 
The second component of the fit model, representing the non-pile-up contribution, is taken as a power-law function.
For most of the photon candidates, the $z_\gamma$ resolution is around 0.2~mm; however, some candidates have significantly larger $z_\gamma$ resolution, which induces a long tail in the $\Delta z$ distribution.
This tail is well modelled by the power-law function.
 
To extract the fraction of events due to pile-up, the power-law parameters and the normalisation of each component are fitted simultaneously to the measured $\Delta z$ distribution. The fit is performed in two regions simultaneously, $\Delta\phi_{\yy}<\pi/2$ and $\Delta\phi_{\yy}>\pi/2$.
For non-pile-up diphoton candidates, there are 10 times more events with $\Delta\phi_{\yy}>\pi/2$ than with $\Delta\phi_{\yy}<\pi/2$; this proportion is allowed to float in the fit.
On the other hand, for the pile-up background, the events are almost equally distributed between the two regions;
only the $\Delta R_{\yy} > 0.4$ selection requirement unbalances the $\Delta\phi_{\yy}$ distribution, slightly reducing the fraction of pile-up events with $\Delta\phi_{\yy}<\pi/2$.
In this case, this fraction is fixed in the fit, according to the MC simulation prediction.
The difference between the proportions of pile-up and non-pile-up events in the two $\Delta\phi_{\yy}$ regions helps to constrain the fit parameters.
The fit yields an uncertainty of $\delta_\textrm{fit}^\textrm{PU}=\pm15\%$ for the pile-up background fraction.
 
The event yield resulting from this fit still contains contributions from misidentified jets, which are subtracted in the main background fit also for pile-up events. To avoid a double subtraction, the nominal background fit is performed for the events with $\Delta z>48$~mm to determine the fraction of prompt $\yy$ events within the pile-up contribution. The large uncertainty of $\delta_{\yy}^\textrm{PU}=\pm 50\%$ in this fraction is of statistical nature due to the low number of input events, and is added in quadrature with the $\delta_\textrm{fit}^\textrm{PU}$ to form the total uncertainty in the pile-up background propagated to the final results, as described in Section~\ref{sec:uncertainties}.
 
With the normalisation of the pile-up background fixed, the shape of this background in the fiducial phase space and as a function of each observable is derived using a sample of pseudo-events. Events in this sample are constructed by overlaying two separate reconstructed events from a $\yj$ MC simulation. This shape is combined with the normalisation determined as above and fed into the background estimation fit as a predetermined component.
Given the large relative uncertainty of the normalisation and the small contribution from this background, the uncertainties of the MC sample are negligible.
 
In the inclusive sample, the pile-up background is found to be $(0.6\pm0.4)\%$.
Differentially, the largest contribution from this background is around 6\% at \yycosTS close to one, cf.\ Figure~\ref{fig:background:purity}.

\section{Correction to particle level}
\label{sec:unfolding}
Effects such as the finite resolution of the detector and the event selection efficiency have to be corrected for, in order to be able to compare results from different experiments and to allow comparisons with theory predictions without detector simulation. The particle-level selection to which the data are unfolded is presented in Table~\ref{tab:event_selection}. The unfolding procedure is applied to the background-subtracted distribution at the detector level in three steps.
 
First, the measured distribution is corrected for the fraction of MC events reconstructed and selected at detector level that do not pass the particle-level selection. The contribution of these \textit{non-fiducial} reconstructed events is estimated using the signal MC samples. The main source of non-fiducial photons is failure to satisfy the transverse momentum and isolation requirements. The particle-level selection is chosen to be close to the detector-level selection to reduce the influence of this correction.
 
The second step is to correct for detector resolution effects which cause migrations from one particle-level bin to multiple detector-level bins. For this step, an iterative unfolding method~\cite{DAgostini:1994fjx,Adye:2011gmfix} based on Bayes' theorem is used. Response matrices parameterise these migrations and are constructed from simulated events passing both the detector-level and particle-level selections. A first iteration of the migration correction uses the particle-level prediction from \Sherpa as the prior input. The prior then gets updated with the obtained result, and the procedure is repeated; three iterations are done in this analysis. The chosen number of iterations is high enough to minimise the dependence on the prior, but low enough to not be sensitive to statistical fluctuations.
 
\begin{figure}[t]
\centering
\begin{subfigure}[b]{.49\textwidth}
\centering
\includegraphics[width=\textwidth]{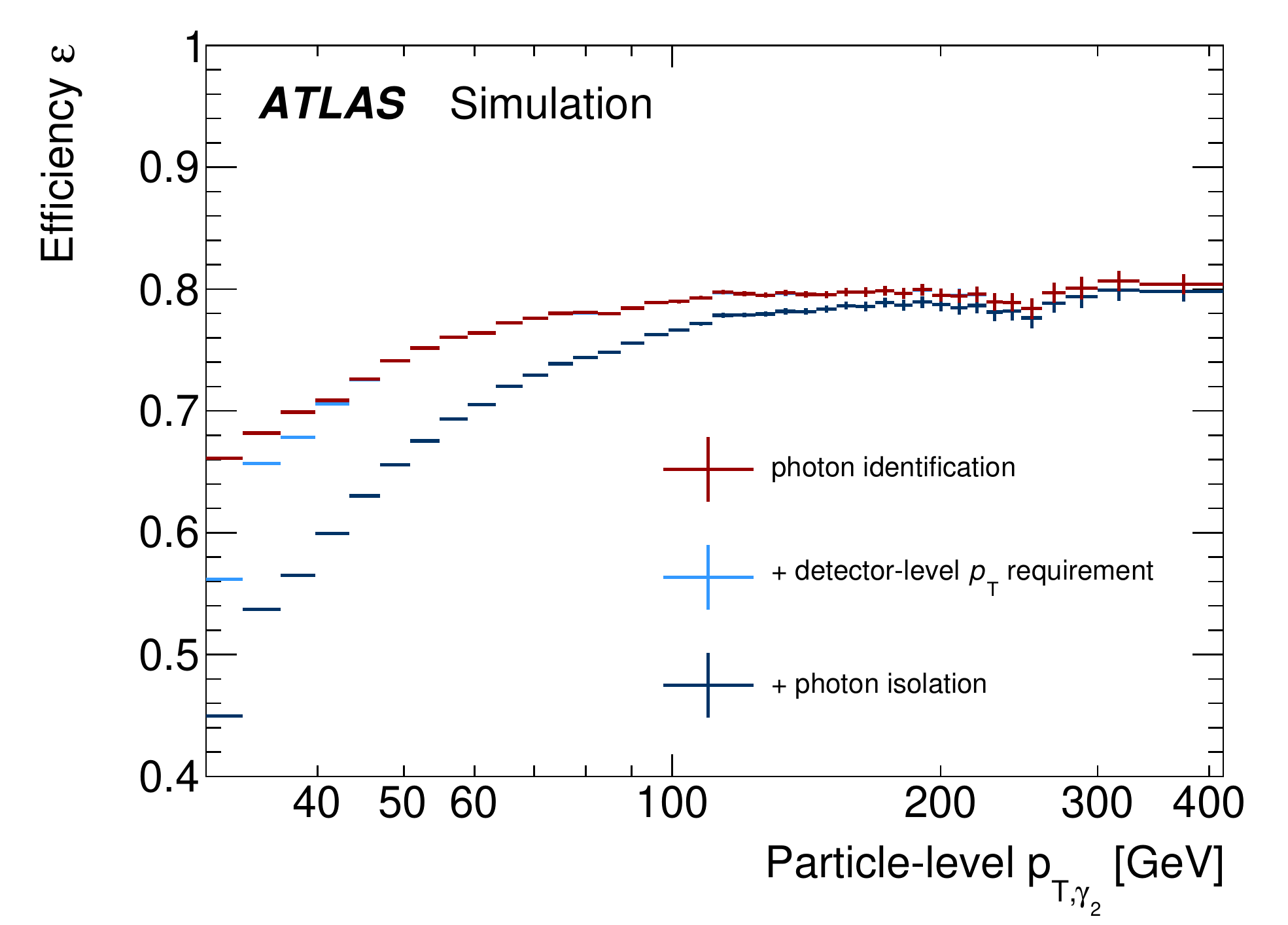}
\caption{}
\label{fig:unfolding:efficiency}
\end{subfigure}
\begin{subfigure}[b]{.49\textwidth}
\centering
\includegraphics[width=\textwidth]{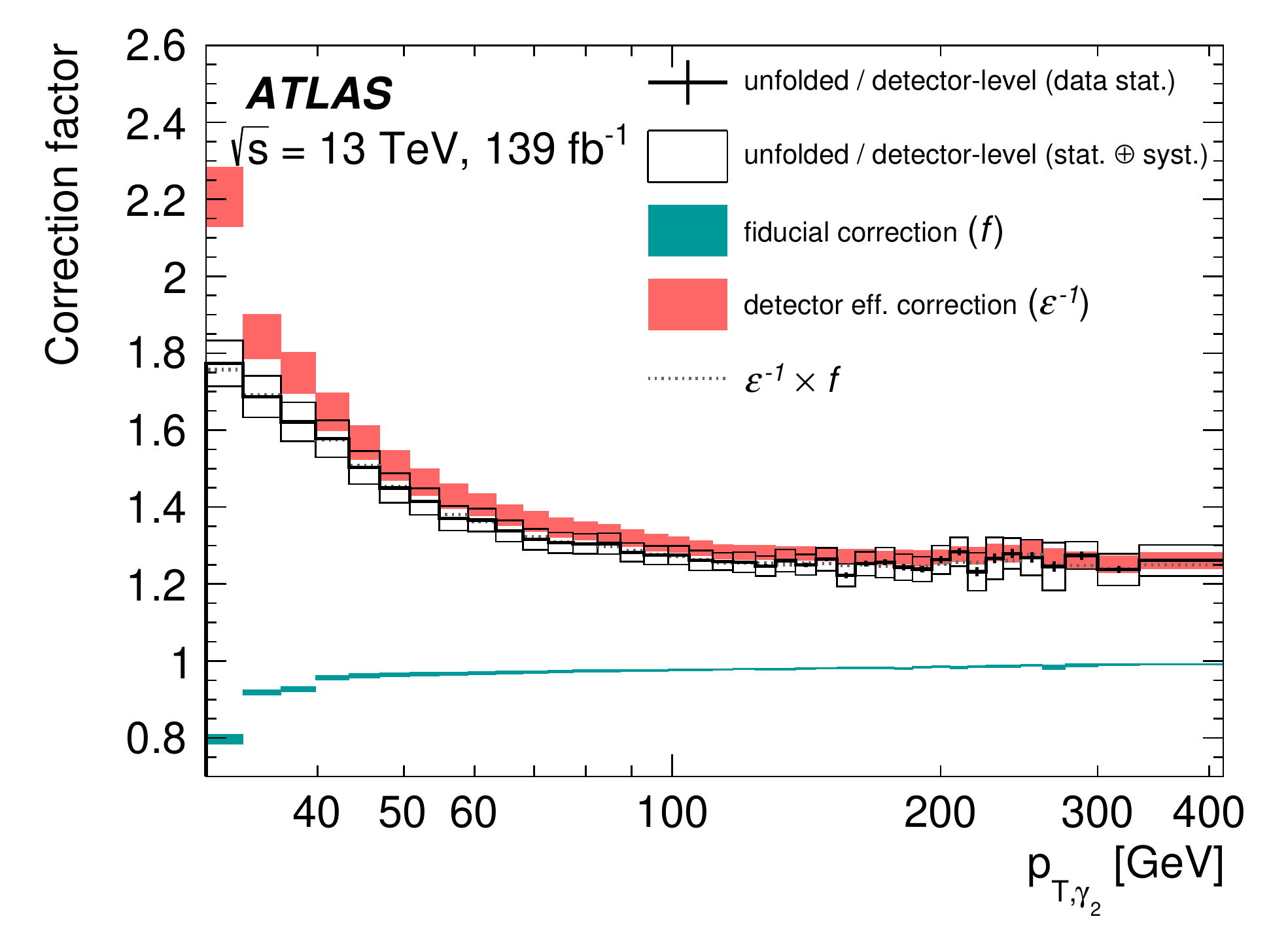}
\caption{}
\label{fig:unfolding:corrections}
\end{subfigure}
\caption{
(a) Efficiency of consecutive sets of detector-level selections for diphoton events passing the particle-level selection as a function of the particle-level \phtwopt.
(b) Impact of the unfolding on the detector-level yields after background subtraction as a function of \phtwopt. The correction for the fraction of non-fiducial events not passing the particle-level cuts is depicted in green. The correction for the detector efficiency is shown in red. The black line is the ratio of the detector-level yields to the result after the three corrections (i.e.\ for non-fiducial events, the iterative migration unfolding and the correction for detector inefficiencies) are applied. The impact of the iterative unfolding can be seen by comparison with the grey dotted line, which is the product of the efficiency and fiducial correction.}
\end{figure}

In the last step, the unfolded results are corrected for detector inefficiencies. This accounts for the fraction of particle-level selected events in the signal region which do not pass the signal selections at detector level. Their contribution is also estimated using the signal MC samples. The largest corrections stem from the photon identification and isolation selection, as shown in Figure~\ref{fig:unfolding:efficiency} as a function of \phtwopt.
 
Figure~\ref{fig:unfolding:corrections} illustrates individual steps of the unfolding correction, as well as the total correction, for the \phtwopt observable.
The efficiency and fiducial corrections partially cancel each other out due to the symmetric shape of the detector response, e.g.\ for the photon transverse momentum criteria.
 
\section{Uncertainties}
\label{sec:uncertainties}
\begin{figure}[tbp]
\centering
\includegraphics[width=0.45\textwidth]{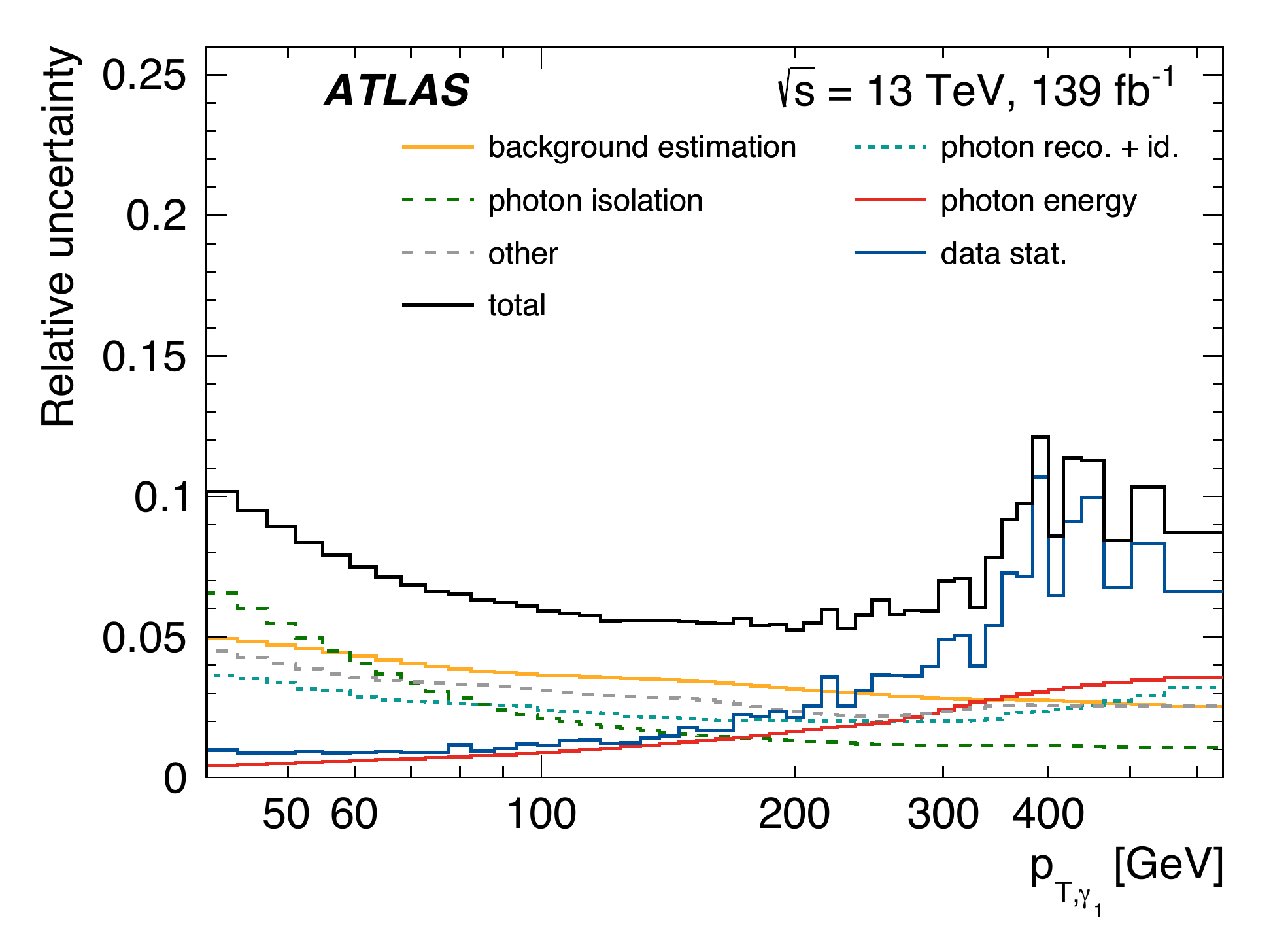}
\includegraphics[width=0.45\textwidth]{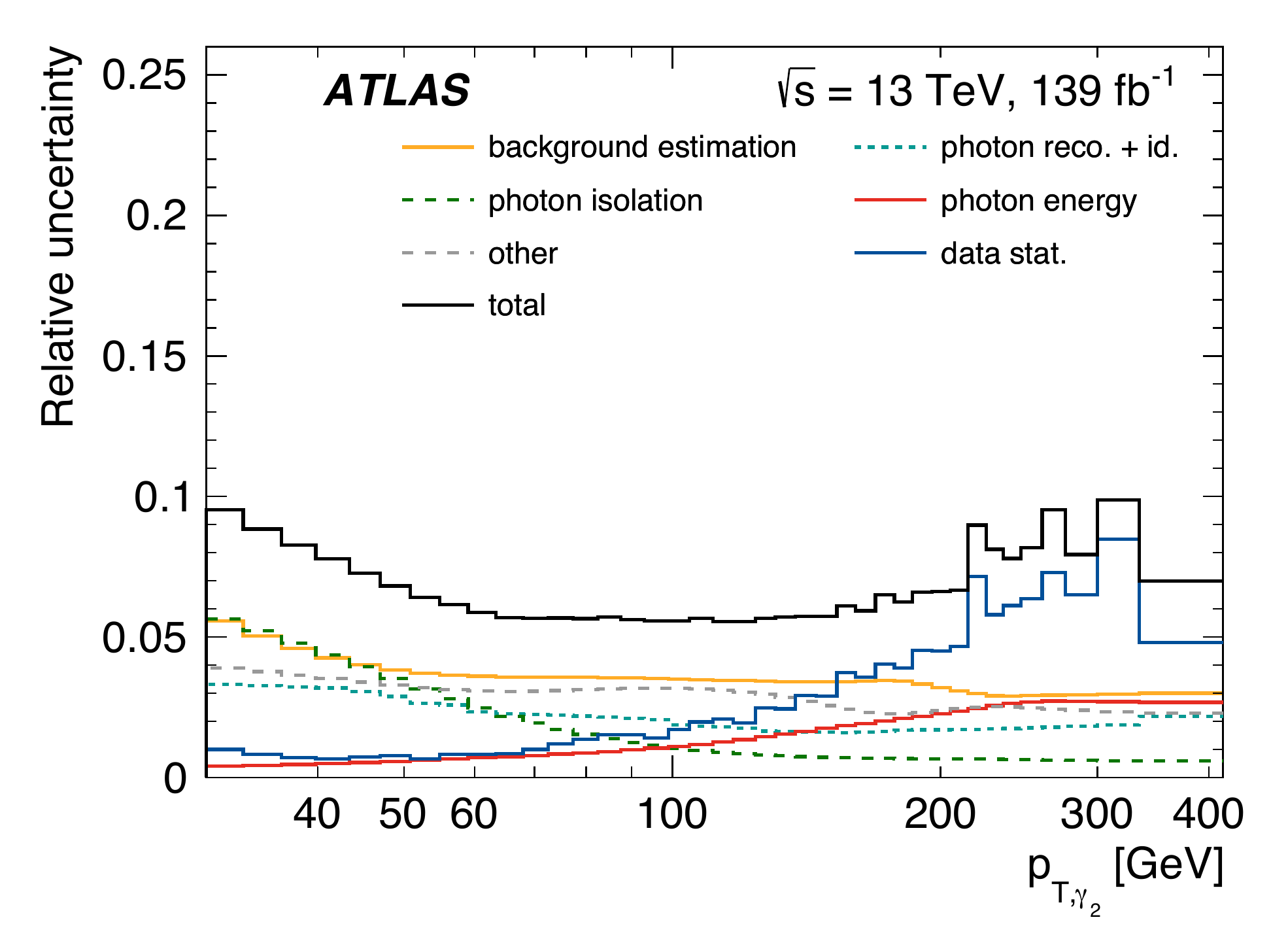}
\includegraphics[width=0.45\textwidth]{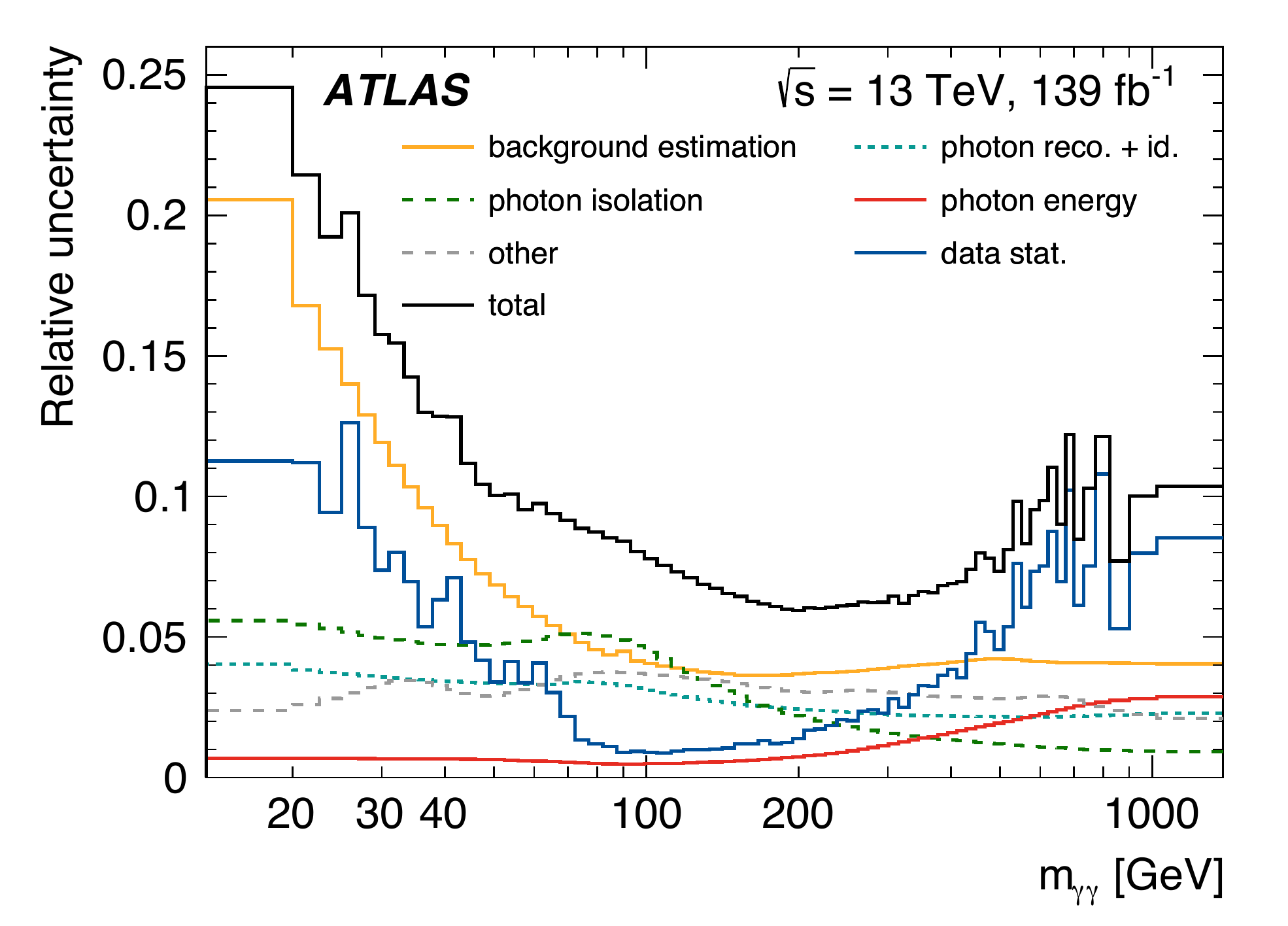}
\includegraphics[width=0.45\textwidth]{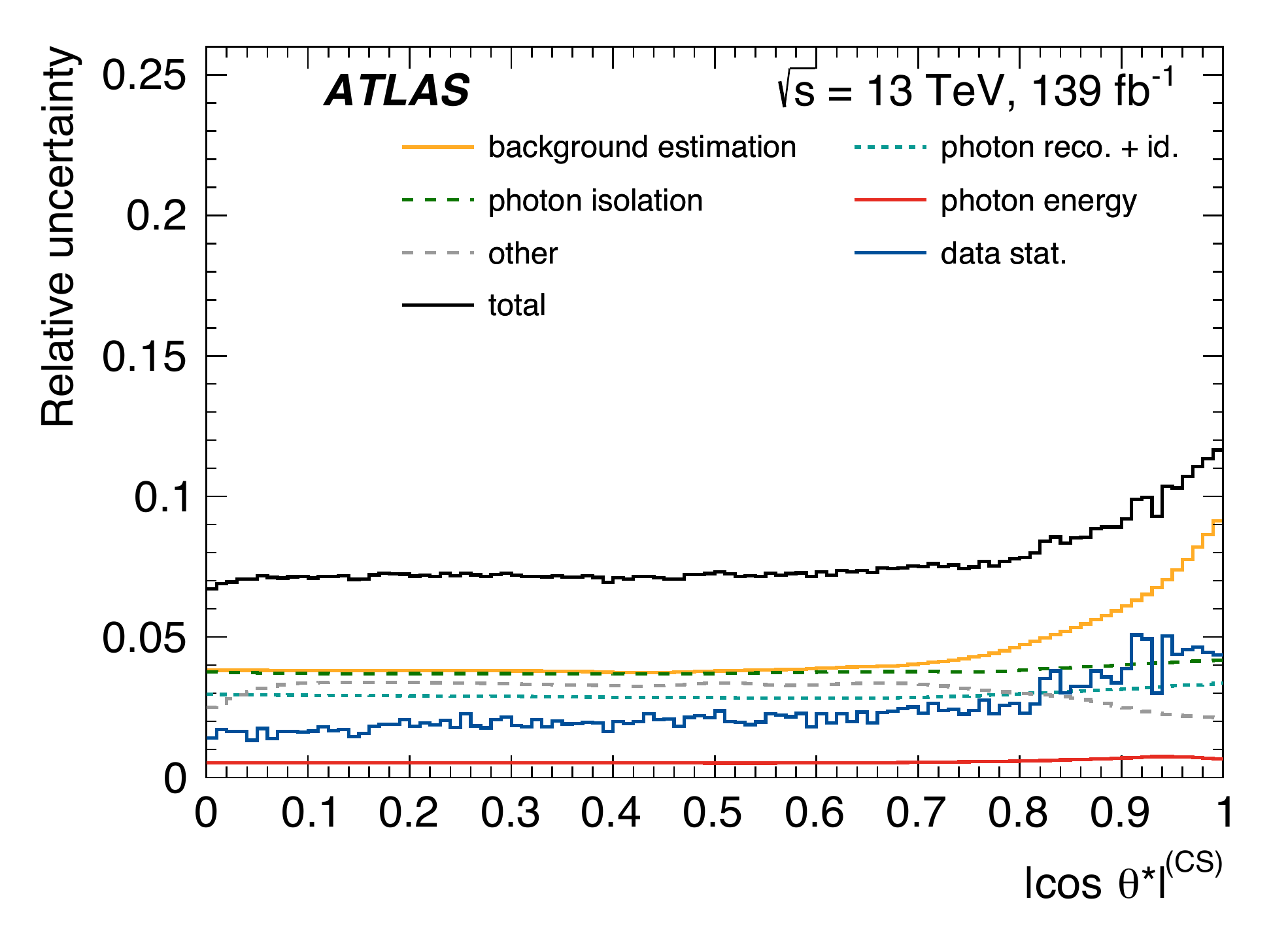}
\includegraphics[width=0.45\textwidth]{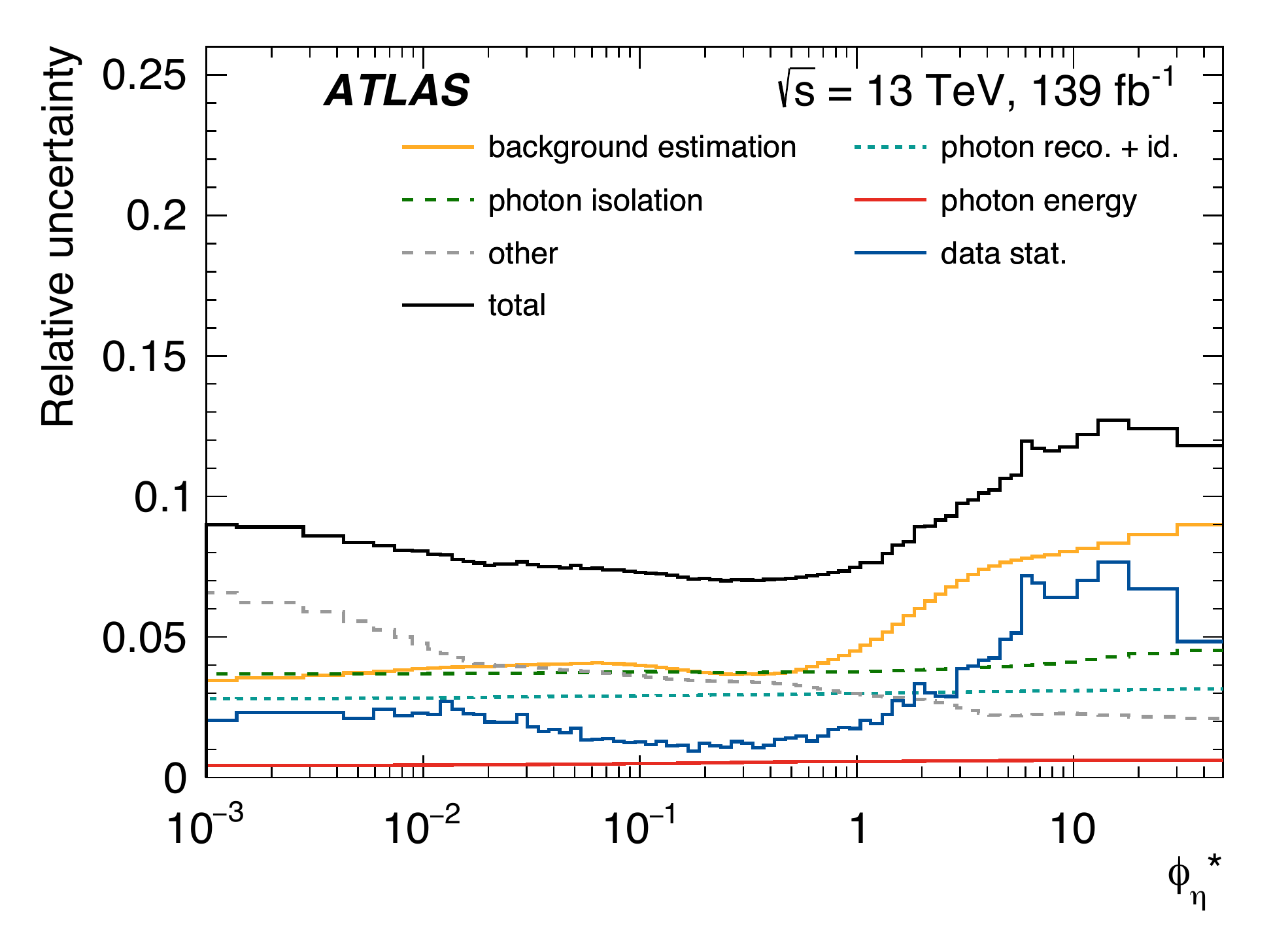}
\includegraphics[width=0.45\textwidth]{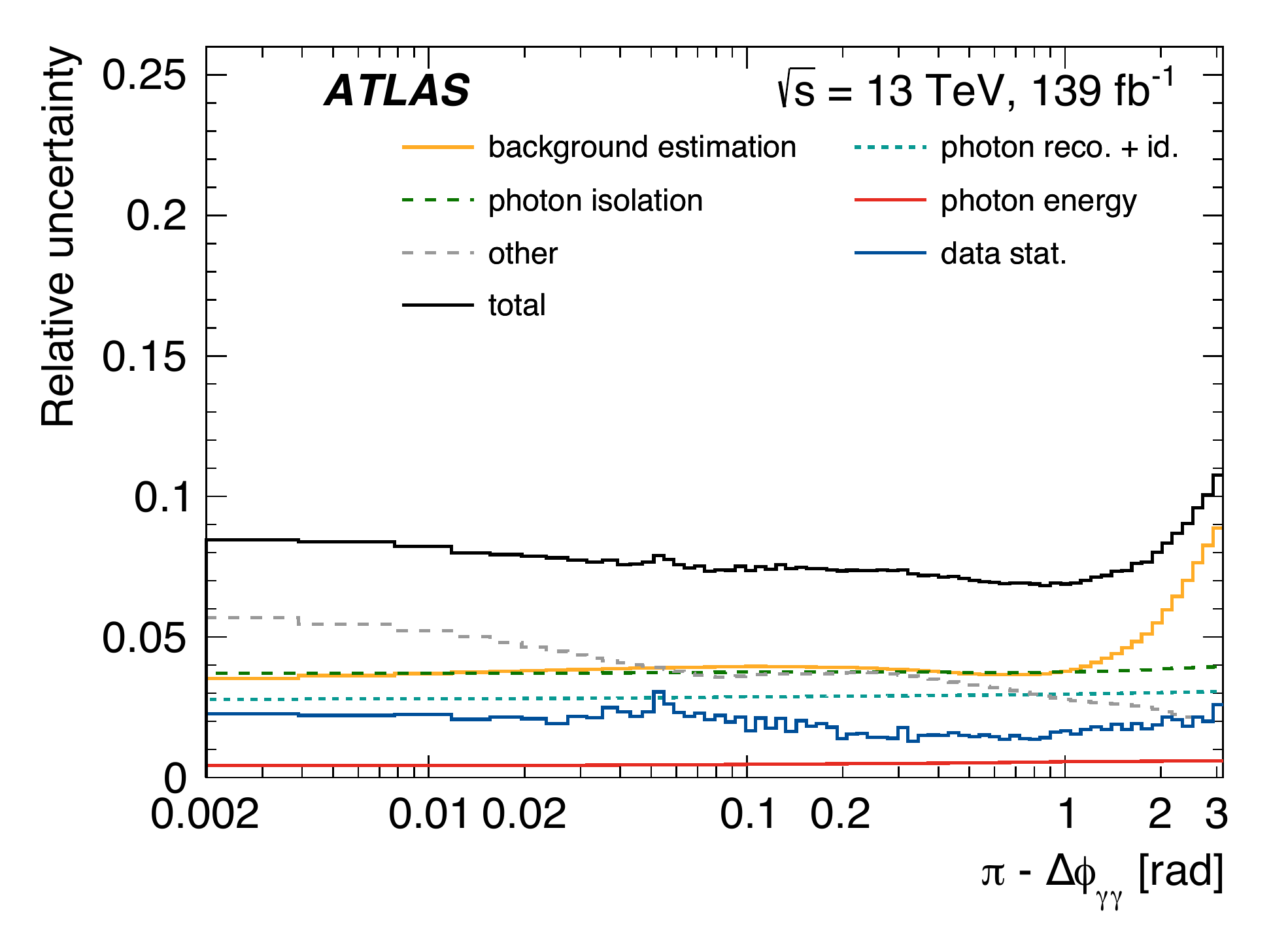}
\includegraphics[width=0.45\textwidth]{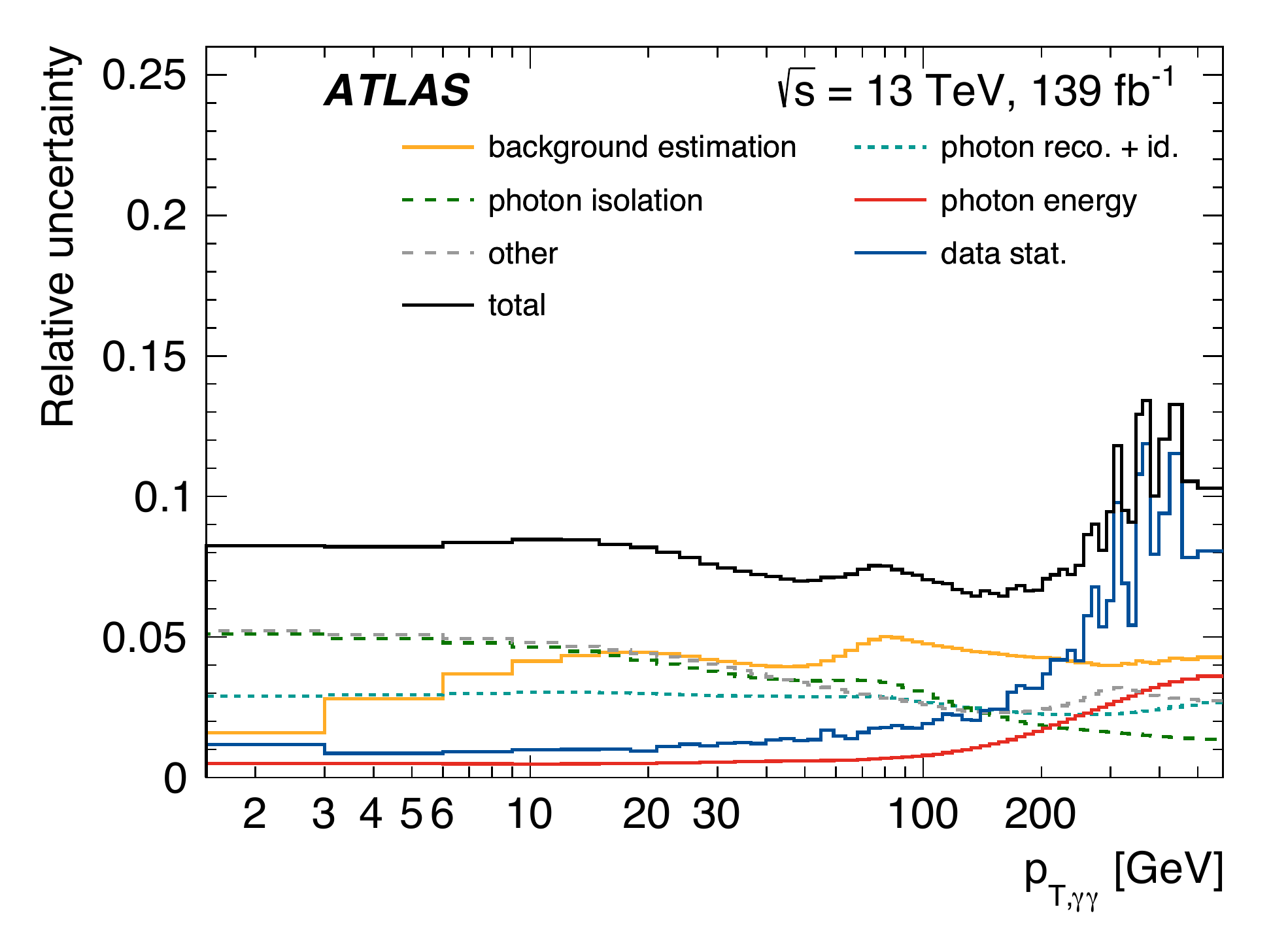}
\includegraphics[width=0.45\textwidth]{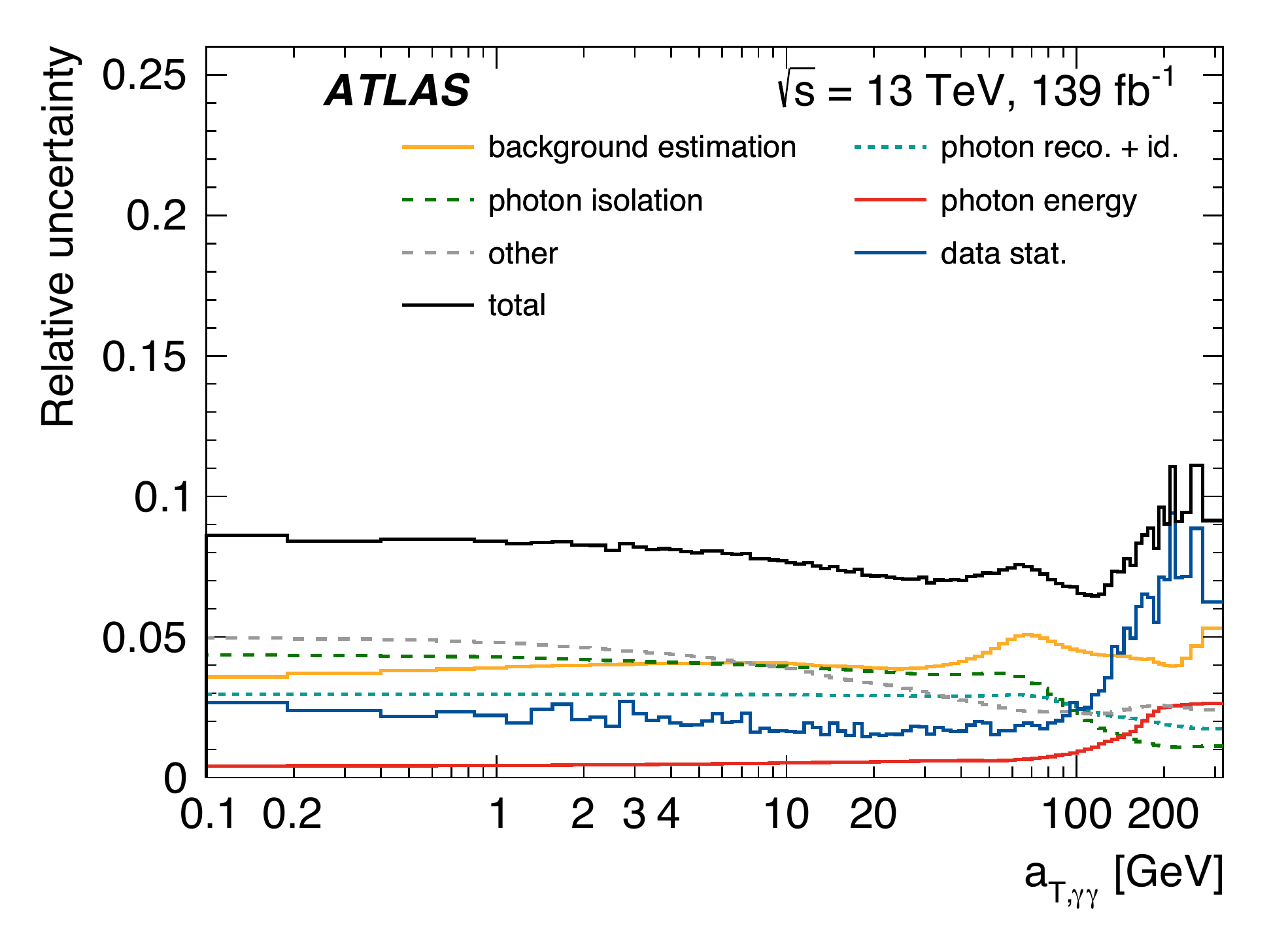}
\caption{The relative uncertainties in the unfolded result associated with different sources are summarised. The uncertainty components and the total uncertainty are shown as a function of each of the observables.}
\label{fig:uncertainties:summary}
\end{figure}

Several uncertainties affect the measurement. They are evaluated independently and then added in quadrature as uncorrelated components into a total uncertainty, with its breakdown shown in Table \ref{ref:total_fiducial_breakdown} for the integrated fiducial cross section. Figure~\ref{fig:uncertainties:summary} shows the breakdown of the systematic uncertainties in the differential cross section.
 
\begin{table}[tb]
\centering
\normalsize
\caption{Breakdown of the relative uncertainties in the integrated fiducial cross-section measurement. The uncertainties are grouped into different categories and the first row of each category shows the sum in quadrature of its components.}
\label{ref:total_fiducial_breakdown}
\begin{tabular}{l p{1.0cm} r@{.}l}
\hline
\hline
Source& \multicolumn{3}{l}{Relative uncertainty {[}\%{]}}\\%
\hline
\hline
Background estimation& & 4&3\\%
\qquad $R^\mathrm{iso-id}_{j}$ & & 4&2\\%
\qquad $\gamma\gamma$ pile-up background& & 0&6\\%
\qquad $R^\mathrm{iso}_\mathrm{\gamma j}$& & 0&5\\%
\qquad Electron background& & \ 0&2\\%
\hline
Photon isolation& & 4&1\\%
\qquad Pile-up reweighting& & 3&5\\%
\qquad Data-driven correction& & 1&9\\%
\qquad Theory modelling of isolation& & 0&6\\%
\hline
Photon reconstruction and identification& & 3&0\\%
\qquad Photon identification& & 2&9\\%
\qquad Photon reconstruction& & 0&4\\%
\hline
Other& & 3&5\\%
\qquad Data-period stability& & 2&9\\%
\qquad Luminosity& & 1&7\\%
\qquad Trigger efficiency& & 0&7\\%
\qquad Theory modelling of $\phonept$& & 0&2\\%
\qquad MC statistical uncertainty& & 0&1\\%
\qquad Unfolding method& & $<$0&1\\%
\hline
Photon energy& & 0&5\\%
\hline
\hline
Total systematic uncertainty& & 7&5\\%
Data statistical uncertainty& & 0&3\\%
\hline
\hline
\end{tabular}
\end{table}
 
The individual sources of uncertainties and the methods to propagate them to the final result are described in more detail in the following sections.
 
\subsection{Background estimation}
\begin{description}
\item[Jets misidentified as photons.] The correlation between isolation and identification for jets misidentified as photons, \risoidj, is determined both from MC simulations and from a data-driven approach, as described in \cref{sec:background:jetfake:inputs}. While the average of the two methods is used for the central value, half of the difference is assigned as an uncertainty to cover both approaches and is combined in quadrature with the statistical uncertainty of the MC result. This is by far the dominant contribution to the total uncertainty for the background, which amounts to $4.2\%$ in general and rises to $17\%$ in the low-\yym region.\\
The correlation between the photon and jet isolation in \yj and \jy events is varied around its central value by $\pm 5\%$, as determined from cross-checks in the \yy MC sample and in the control regions dominated by \jj events in data. This uncertainty is at the sub-percent level almost everywhere, except at low \yym, where it rises to $6\%$.\\
Uncertainties in the identification and isolation efficiencies of the prompt photons also affect the background estimation. They are discussed in the Section~\ref{sec:uncertainties:photon}.
\item[Electrons misidentified as photons.] Three components are taken into account as uncertainties for the electron-related background sources. The physics modelling of the $Z\to ee$ MC sample is varied by a $5\%$ normalisation uncertainty, and a $p_\perp^Z$-dependent shape uncertainty covers the difference between MC simulation and data~\cite{STDM-2018-14}. The uncertainty related to the modelling of $e\to \gamma$ fake rates is estimated from a dedicated measurement using $Z\to ee$ events. All of these uncertainties have a completely negligible contribution almost everywhere, rising to $2\%$ only in the $\yym \sim m_Z$ bin.
\item[Pile-up background.] The uncertainty in the normalisation of the pile-up background stems from a combination of two uncertainties arising from the fits, namely the fit to extract the total pile-up yield, and the $\gamma\gamma$ extraction in the pile-up sample. The main impact of this background contribution is located at $\yycosTS \to 1$ (see Figure~\ref{fig:background:purity}) and its uncertainty rises to $6\%$ in this region, whereas it is small in all other regions of phase space.
\end{description}
The uncertainties described above are propagated to the final cross-section results by using different background-subtracted signal yields as input to the unfolding procedure, with all bins varied in a correlated way, and response matrices unaffected.
 
\subsection{Photon selection}
\label{sec:uncertainties:photon}
The detector simulation and reconstruction of photons within the MC samples play significant roles in the background estimation and unfolding. This results in uncertainties from various sources.
\begin{description}
\item[Photon isolation.] The modelling of the photon isolation variable in simulated events is corrected using a data-driven technique when deriving the nominal results~\cite{EGAM-2018-01}. An uncertainty due to a potential isolation mismodelling is estimated by disabling these data-driven corrections. The corrections modify the position of the peak of the isolation variable.\\
Additionally, the width of the isolation transverse energy distribution is affected significantly by the modelling of pile-up in the simulation.
In the MC samples, the noise induced by the pile-up in the calorimeters is found to be overestimated, increasing the width of the distribution of the used calorimetric isolation.
This simulation feature is observed in the electron isolation distribution in $Z\to ee$ events.
Therefore, to extract the efficiency of the photon isolation requirement, the simulation pile-up profile is reweighted, scaling down the number of pile-up interactions by 17\%, based on the observations from the $Z\to ee$ studies.
Systematic uncertainties are estimated by further reducing and increasing the amount of pile-up similarly by 17\%.
The uncertainty in the modelling of pile-up dominates the isolation uncertainty and constitutes one of the leading sources of uncertainty in the analysis, with a typical size of $4\%$.\\
The isolation distribution is also affected by a theory modelling uncertainty; the estimation of this uncertainty is discussed in Section~\ref{sec:uncertainties:other}.
\item[Photon trigger, reconstruction and identification  efficiency.] The uncertainties in the modelling of the detector efficiency of the photon trigger, reconstruction and identification are addressed using scale factors that depend on the transverse momentum $p_{\mathrm{T},\gamma}$ and pseudorapidity $\eta$ of the photon.
The trigger and identification scale factors and their uncertainties are derived to match the detector efficiency in simulation to data as described in Refs.~\cite{TRIG-2018-05,EGAM-2018-01}.
The trigger efficiency uncertainty is around 0.7\%, while the uncertainty in the measured diphoton cross section related to the photon identification efficiency ranges between 2\% and 4\%.
The photon reconstruction efficiency is studied with $Z\rightarrow\mu\mu\gamma$ events by looking for events in which the reconstructed cluster is misclassified as an electron. The fraction of photons misreconstructed as electrons is around 4\%. To match the data, the simulated reconstruction efficiency of the diphoton systems is corrected on average by $-1.8\%$, with an uncertainty of 0.4\%. The reconstruction efficiency corrections are derived and applied independently for data taken in the years 2015+2016 and 2017+2018.
\item[Photon energy.] Uncertainties in the photon energy scale and resolution are taken into account by modifying the MC simulation used to determine response matrices, efficiencies and non-fiducial-photon rates~\cite{PERF-2017-03}. This is done separately for independent sources of the uncertainty, e.g.\ from the overall energy scale determined in $Z\to ee$ decays, the non-linearity of the energy measurement at the calorimeter cell level, the relative calibration of the different calorimeter layers, the amount of material in front of the calorimeter, and the simulation modelling of the photon conversion reconstruction, lateral shower shape, and energy resolution. They are propagated separately to account for correlations and their $\eta$ dependence, and are combined in quadrature afterwards. This uncertainty is at the sub-percent level in most regions of the analysis, rising to 3--4\% only at high $p_{\mathrm{T},\gamma}$, $\yym$, $\yypT$ and $\yypTt$.
\end{description}
For the photon isolation and identification variations described above, modified MC simulations are used coherently in the background fit and unfolding, to propagate the uncertainty for each source to the final result and maintain the correlations. The photon energy and photon trigger efficiency uncertainty affect only the unfolding.
 
\subsection{Other uncertainties}
\label{sec:uncertainties:other}
 
\begin{description}
\item[Signal theory modelling.] The theory modelling in the signal MC samples can affect the correction to the particle level as well as the signal model parameters in the background fit. The resulting uncertainty is estimated by reweighting the nominal \sherpa sample in two different ways and repeating the whole analysis chain. The first reweighting is based on the photon isolation distribution taken from a \PYTHIA[8] sample at particle level.
The uncertainty in the final results is relatively constant at the $1\%$ level but rises to $4\%$ in the low-\yym region. Additionally, a reweighting of the $\phonept$ distribution is performed such that the detector-level distribution matches the background-subtracted result. The effect of this variation is below 1\%.
The differences between the nominal and each of the two reweighted results are treated as independent uncertainties and applied coherently in the background fit and the unfolding.
\item[Unfolding method uncertainties.] The results of the unfolding procedure can depend on the choice of prior and the number of iterations. To estimate an uncertainty, the MC simulation is reweighted such that it matches the data at the detector level, and this emulated data is then unfolded using the nominal sample. The non-closure with respect to the reweighted particle-level simulation then yields an additional uncertainty orthogonal to the one from signal theory modelling and is negligible everywhere.
\item[Data period stability]
A test is performed comparing the results obtained by splitting the data into 2015+2016, 2017 and 2018 periods, and the MC sample in their corresponding representative subsamples. The whole analysis is redone for these three individual datasets. An additional uncertainty is assigned as a function of the different observables to cover the non-closure among the different years. It is typically around 3\%.
\item[MC statistics.] The influence of the limited size of the MC samples on the final result is evaluated using pseudo-experiments. Each experiment is generated by assigning a random weight according to a Poisson distribution with an expectation value of one to every MC event. The MC statistical uncertainty is estimated from the standard deviation of the results of the pseudo-experiments and is negligible everywhere.
\item[Data statistics.] The statistical uncertainty of the data sample is propagated from the signal extraction to the particle level using pseudo-experiments. It is the dominant uncertainty at high $p_\mathrm{T}$ and high $\yym$, but sub-leading in general.
\item[Luminosity.] The uncertainty in the combined 2015--2018 integrated luminosity is \SI{1.7}{\percent} \cite{ATLAS-CONF-2019-021}, obtained using the LUCID-2 detector \cite{Avoni:2018iuv} for the primary luminosity measurements. This uncertainty is sub-leading in comparison with the other uncertainties.
\end{description}

\section{Theory predictions}
The unfolded data are compared with several state-of-the-art theory predictions, which are summarised in Table~\ref{tab:predictions} and detailed in this section.
 
\begin{table}[tb]
\centering
\caption{Overview of the theory predictions and their relevant features. The label `QCD res.' (`NP effects') stands for QCD resummation (non-perturbative effects).}
\label{tab:predictions}
\begin{tabular}{| c || c | c | c | c | c | c | c | c | c | c |}\hline
& \multicolumn{6}{c|}{Fixed-order accuracy} & \multicolumn{2}{c|}{Fragmentation} & \multirow{2}{1cm}{QCD res.} & \multirow{2}{1cm}{NP effects} \\
& $\yy$ & $+1j$ & $+2j$ & $+3j$ & $+\geq 4j$ & $gg\to \yy$ & single & double & & \\
\hline\hline
\diphox & NLO & LO & - & - & - & LO & \multicolumn{2}{c|}{NLO} & $-$ & $-$ \\\hline
\nnlojet & NNLO & NLO & LO & - & - & LO & $-$ & $-$ & $-$ & $-$ \\\hline
\Sherpa & \multicolumn{2}{c|}{NLO} & \multicolumn{2}{c|}{LO} & PS & LO & \multicolumn{2}{c|}{ME+PS} & PS & \checkmark \\\hline
\end{tabular}
\end{table}
 
\paragraph*{Fixed-order NNLO with \nnlojet.}
The \nnlojet framework~\cite{Gehrmann:2018szu} is used to obtain fixed-order predictions for diphoton production up to NNLO QCD accuracy. It is based on the antenna subtraction method~\cite{GehrmannDeRidder:2005cm,Currie:2013vh} and the NNLO prediction includes the loop-induced $gg\to \yy$ contribution at leading-order accuracy.
The nominal factorisation and renormalisation scales are chosen as $\mu_\mathrm{f,r}=\yym$ and varied independently to estimate the perturbative uncertainties in a 7-point variation to $\muF=c_\mathrm{f}\cdot \yym$ and $\muR=c_\mathrm{r}\cdot \yym$, where $c_\mathrm{f,r}=0.5, 1, 2$ with $0.5\leq c_\mathrm{f}/c_\mathrm{r}\leq 2$. For the nominal predictions, the NNPDF3.0 NNLO PDF set~\cite{Ball:2014uwa} is used.
NNLO predictions for the fragmentation component are not yet available. Thus, the prediction is calculated only from the direct component using an implementation of the hybrid scheme for photon isolation~\cite{Siegert:2016bre,Chen:2019zmr,Amoroso:2020lgh}:
\begin{itemize}
\item a very loose smooth cone isolation~\cite{Frixione:1998jh} with $n=2$, $\delta_0=0.1$, $\varepsilon=0.1$, and
\item the experimental cone isolation with the fiducial photon isolation given in Table~\ref{tab:event_selection}, but without the UE subtraction.
\end{itemize}
The smooth-cone isolation requirement is necessary to remove collinear photon--quark configurations from the calculation, which would be singular and would have to be factorised into a fragmentation function for a full calculation. The hybrid scheme helps to minimise the bias from not taking such configurations into account.
 
\paragraph*{Fixed-order NLO with \diphox.}
 
Fixed-order NLO predictions including the direct and single- and double-fragmentation contributions are made with \diphox~\cite{Binoth:1999qq} using CT10 NLO PDFs~\cite{Lai:2010vv}. The nominal factorisation, renormalisation, and fragmentation scales are chosen as $\muR=\muF=\mu_\textrm{frag}=\yym$. The loop-induced $gg\to\yy$ process is included in the direct component at leading order.
Since the fragmentation contributions are included, the \diphox predictions are valid for a cone-based photon isolation consistent with the one used in the measurement, $\etisoptwo{,\gamma_{1(2)}}<\truthIsoFactor\cdot p_\mathrm{T, \gamma_{1(2)}}$.
To estimate the perturbative uncertainties of the \diphox predictions, three sets of input parameters are varied. Factorisation and renormalisation scales undergo 7-point variations as defined above. This is the dominant uncertainty for this prediction. The fragmentation scale is raised and lowered by a factor of 2, yielding an uncertainty which is always significantly smaller than that from the factorisation and renormalisation scale variations. Finally, variations between predictions based on the nominal PDF set and those based on MSTW2008~\cite{Martin:2009iq} and NNPDF2.3~\cite{Ball:2012cx} are also significantly smaller than those from the scale variation in all phase-space regions.

The fixed-order predictions do not include non-perturbative effects.
The impact of the non-perturbative effects is found to be negligible, except in soft or collinear regions of the phase space, where they grow to $\mathcal{O}(10\%)$; in such regions, fixed-order predictions are not valid without resummation. Thus, no correction is applied to the fixed-order predictions.
 
\paragraph*{\Sherpa}
 
In addition to these fixed-order calculations, the unfolded data are also compared with fully hadronised particle-level predictions from the \Sherpa signal calculations described in Section~\ref{sec:sample:mc}. This sample is used extensively throughout the analysis chain, e.g.\ in the background fit and the correction for detector effects.
 
Systematic theory uncertainties for the following comparisons are obtained from three sets of variations performed using event reweighting~\cite{Bothmann:2016nao}. Factorisation and renormalisation scales are varied independently in a 7-point variation as defined above to estimate the perturbative uncertainties. A second set of uncertainties is obtained by coherently varying the strong coupling used in the simulation and in the PDF fit, $\alphas(m_Z)$ in the range 0.117--0.119. A third uncertainty set is constructed from all replicas within the NNPDF3.0 NNLO set.
These uncertainties are completely dominated by the scale uncertainties and are summed in quadrature to obtain a final uncertainty envelope for the \Sherpa predictions.
 
\section{Results}
\subsection{Integrated cross section in the fiducial phase space}
 
The measurement of the differential cross section as a function of $\phtwopt$ is used to compute the integrated fiducial cross section,
\begin{equation*}
\sigma_{\yy} = 31.4 \pm 0.1 \mathrm{\:(stat.)} \pm 2.4 \mathrm{\:(syst.)}\:\mathrm{pb}.
\end{equation*}
The integration is also performed over the other differential cross-section measurements yielding compatible results.
The fiducial phase space is defined in Table~\ref{tab:event_selection}.
 
\begin{figure}[tbp]
\centering
\includegraphics[width=0.4\textwidth]{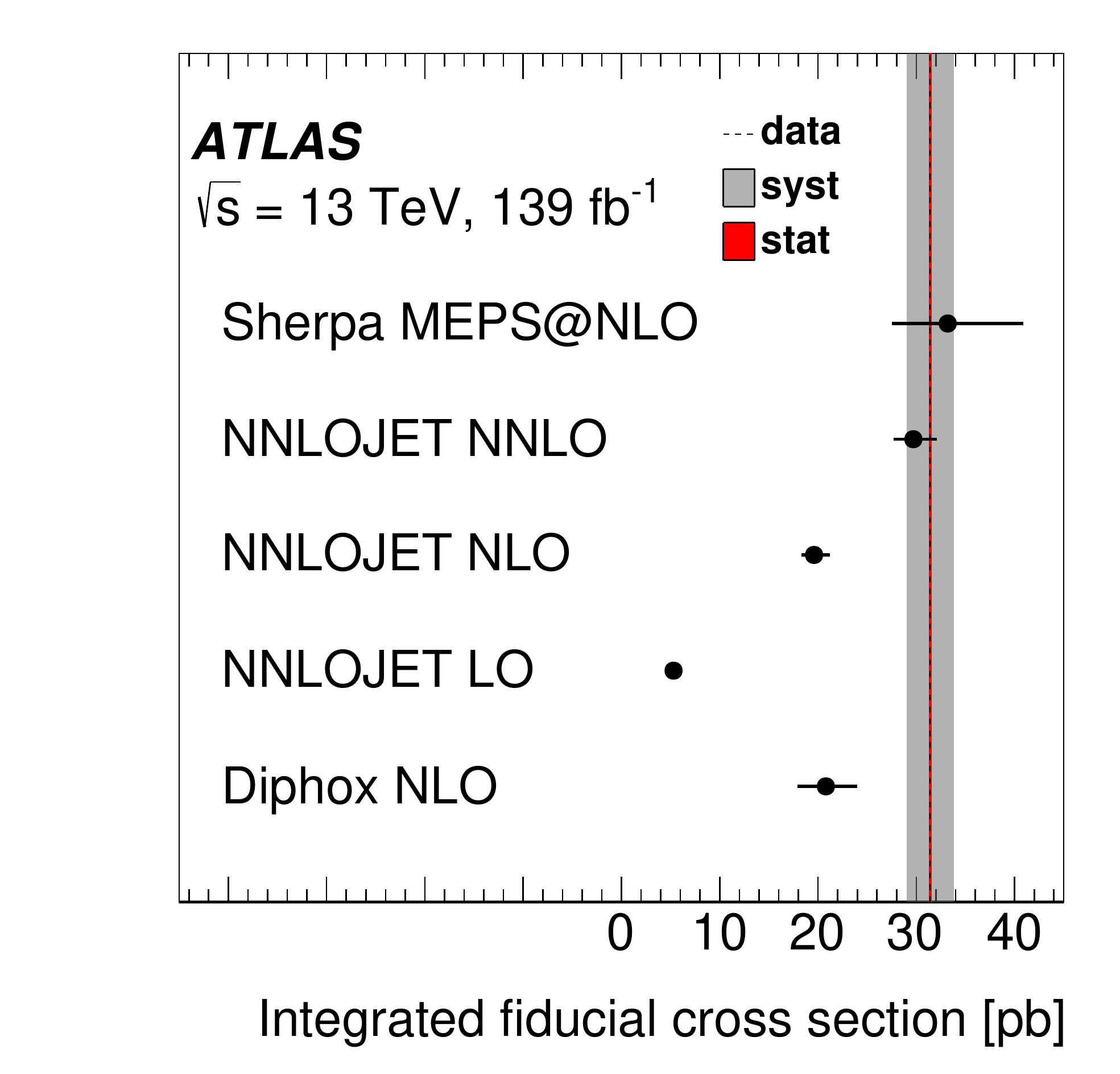}
\caption{Integrated fiducial cross-section comparison between ATLAS data and theory predictions.}
\label{fig:results:total}
\end{figure}

A comparison of the measured cross section with the theoretical predictions (see Figure~\ref{fig:results:total} and Table~\ref{tab:results:total}) shows the importance of higher-order QCD contributions even for such an inclusive diphoton measurement. The predictions from \Sherpa MEPS@NLO, which has NLO multi-leg matrix elements supplemented by a parton shower, and that from the fixed-order NNLO prediction as implemented in \nnlojet are the only predictions compatible with the data within the uncertainties. The predictions limited to LO or NLO QCD, as implemented in \nnlojet (N)LO and \diphox NLO, fail to agree with the data and also do not provide a reliable estimate of the perturbative uncertainties, as will be discussed further below.

\begin{table}[htbp]
\centering
\caption{Integrated fiducial cross section from ATLAS data compared with theory predictions with their systematic uncertainties from QCD scale variations.}
\label{tab:results:total}
\def\arraystretch{1.3}
\begin{tabular}{lrr}
\hline
Fiducial cross section {[}pb{]} & $\sigma_{\yy}$ & $\pm$ unc.\\%
\hline
\Sherpa MEPS@NLO & 33.2 & $^{+ 7.7}_{-5.6}$\\%
\nnlojet NNLO & 29.7 & $^{+ 2.4}_{-2.0}$\\%
\hphantom{\nnlojet }NLO & 19.6 & $^{+ 1.6}_{-1.3}$\\%
\hphantom{\nnlojet }LO & 5.3 & $^{+ 0.5}_{-0.5}$\\%
\diphox NLO & 20.8 & $^{+ 3.2}_{-2.9}$\\
\hline
Data & 31.4 & 2.4 \\%
\hline
\end{tabular}
\end{table}
 
\subsection{Differential cross sections}
 
\begin{figure}[tbp]
\centering
\includegraphics[width=0.47\textwidth]{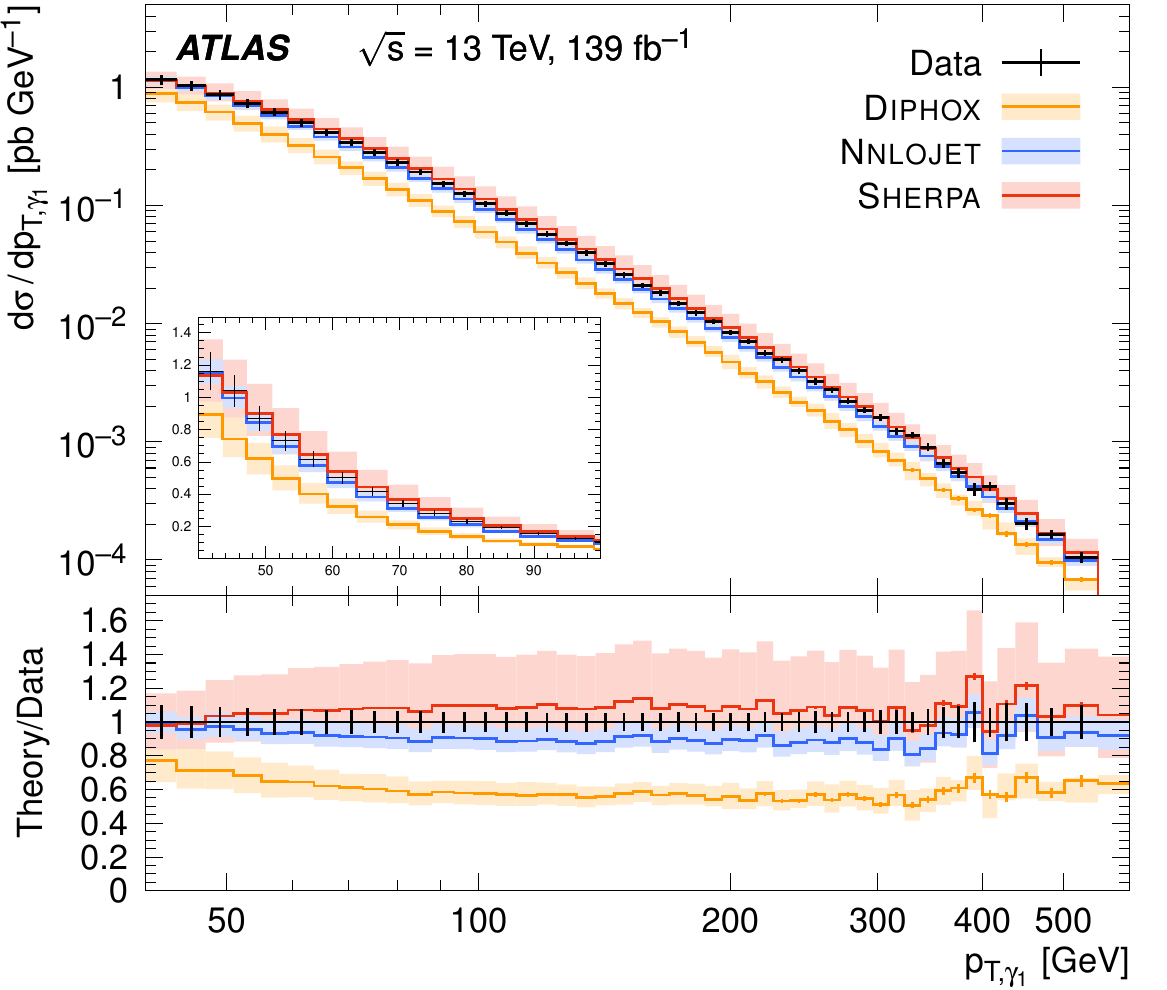}
\hspace*{2mm}
\includegraphics[width=0.47\textwidth]{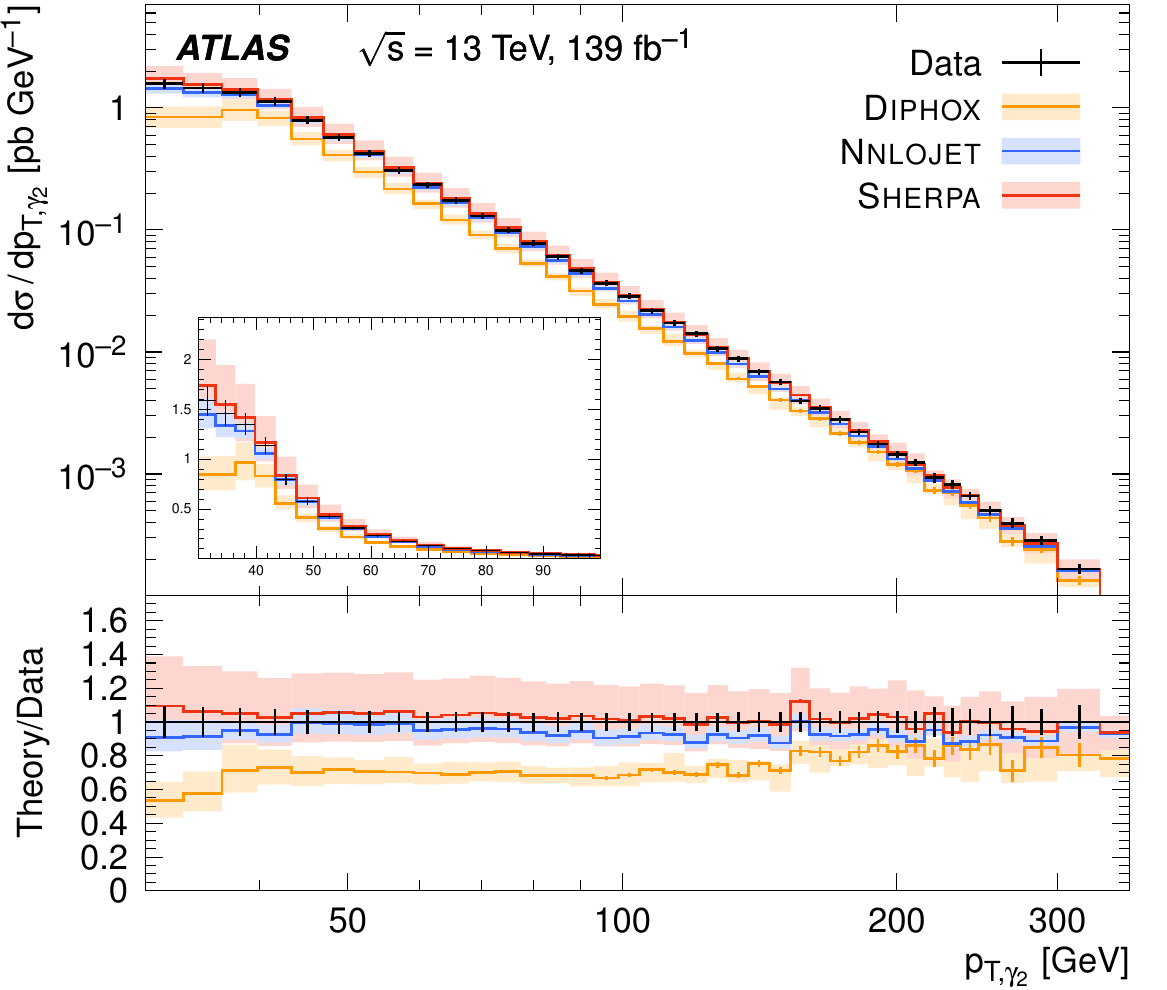}
\includegraphics[width=0.47\textwidth]{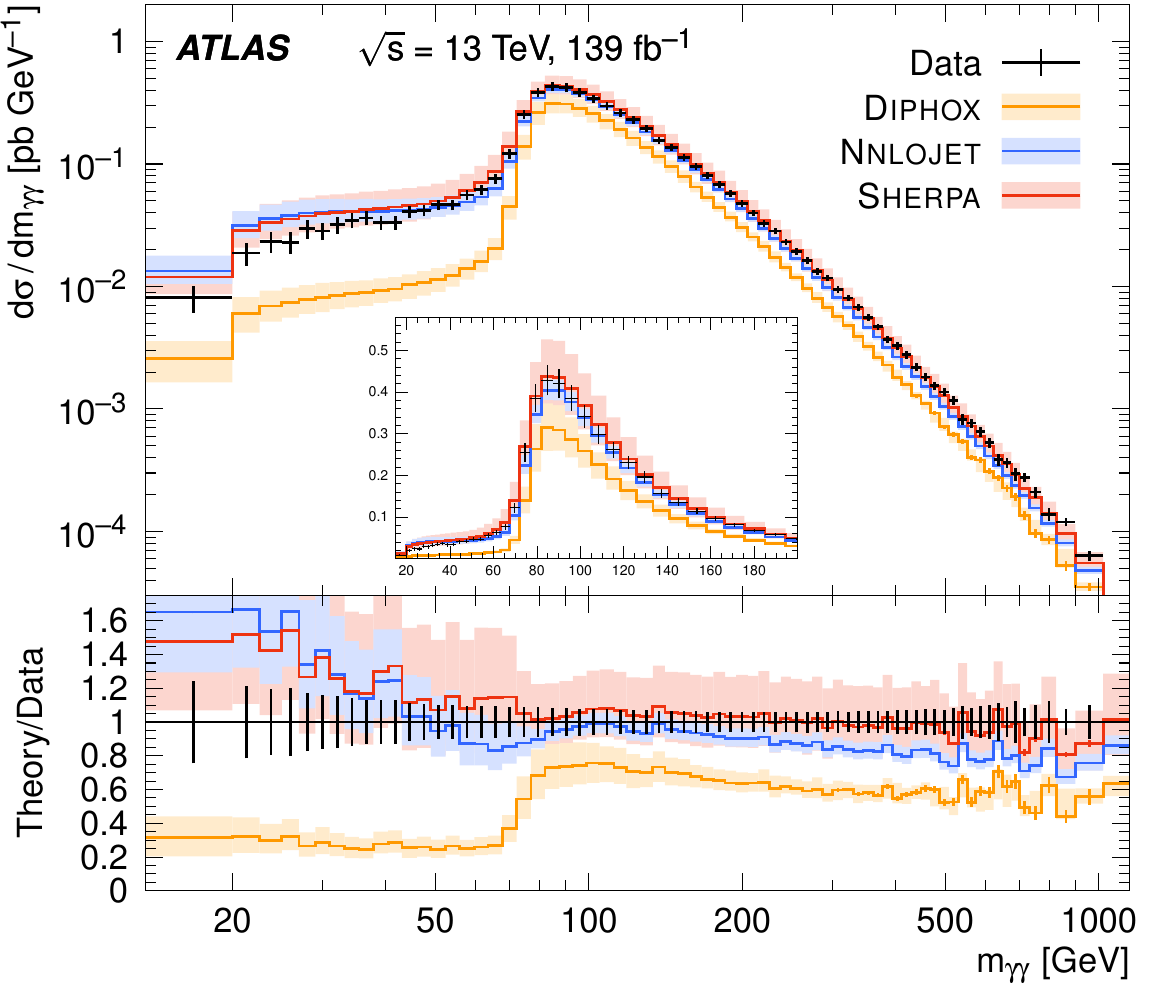}
\hspace*{2mm}
\includegraphics[width=0.47\textwidth]{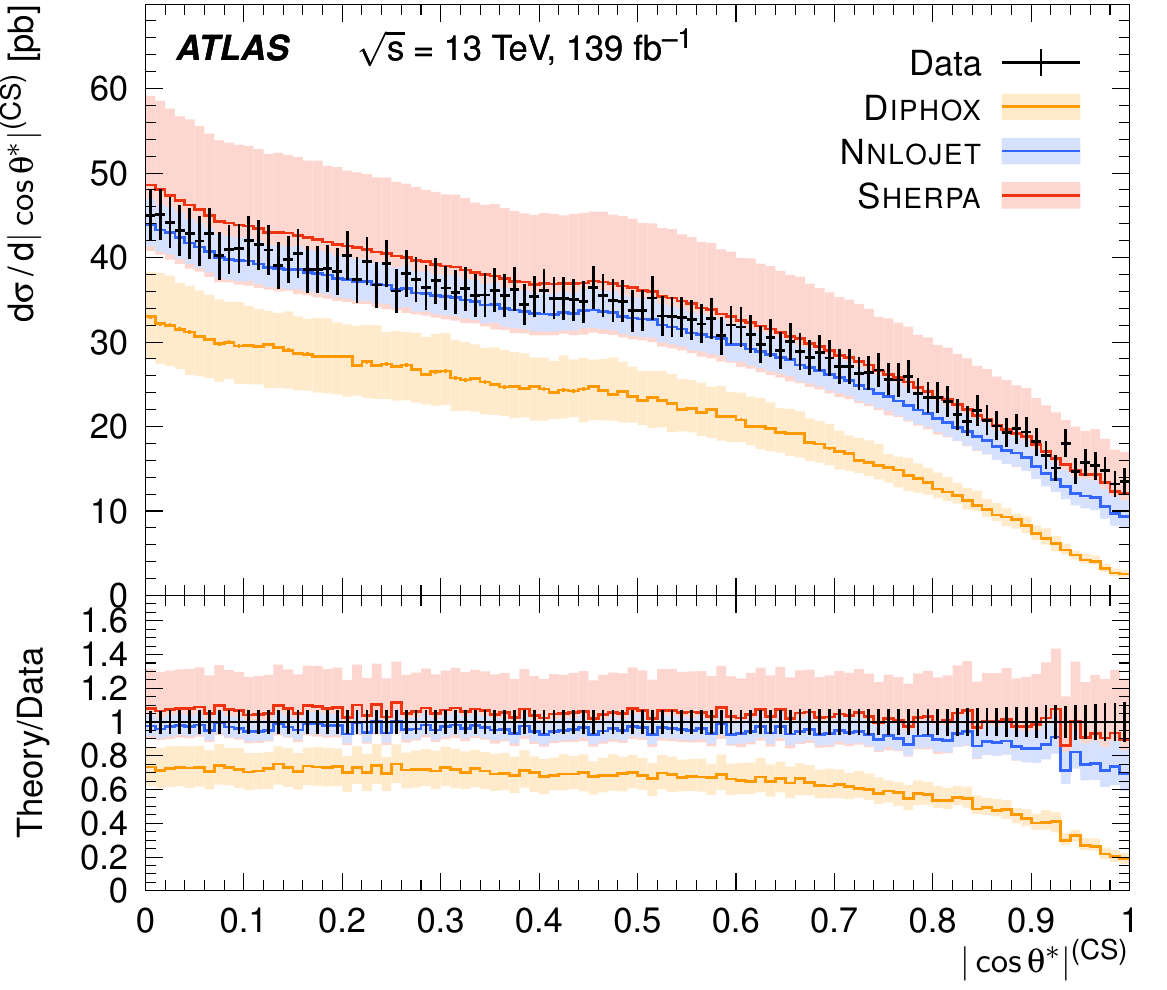}
\caption{Differential cross sections measured as functions of $\phonept$, $\phtwopt$, $\yym$ and $\yycosTS$ (black error bars) compared with the predictions from \diphox NLO, \nnlojet NNLO and \Sherpa MEPS@NLO (lines with coloured bands). At the bottom of each plot, the ratio of the prediction to the data is shown. Uncertainty bars on the data represent the total uncertainty, while uncertainty bands (bars) on the predictions represent perturbative scale (statistical) uncertainties.
Inset plots show the core of each distribution on a linear scale. For $\phonept$, $\phtwopt$ and $\yym$ the last bin is only visible in the ratio inset and includes the overflow events.}
\label{fig:results:differential:datatheory1}
\end{figure}

\begin{figure}[tbp]
\centering
\includegraphics[width=0.47\textwidth]{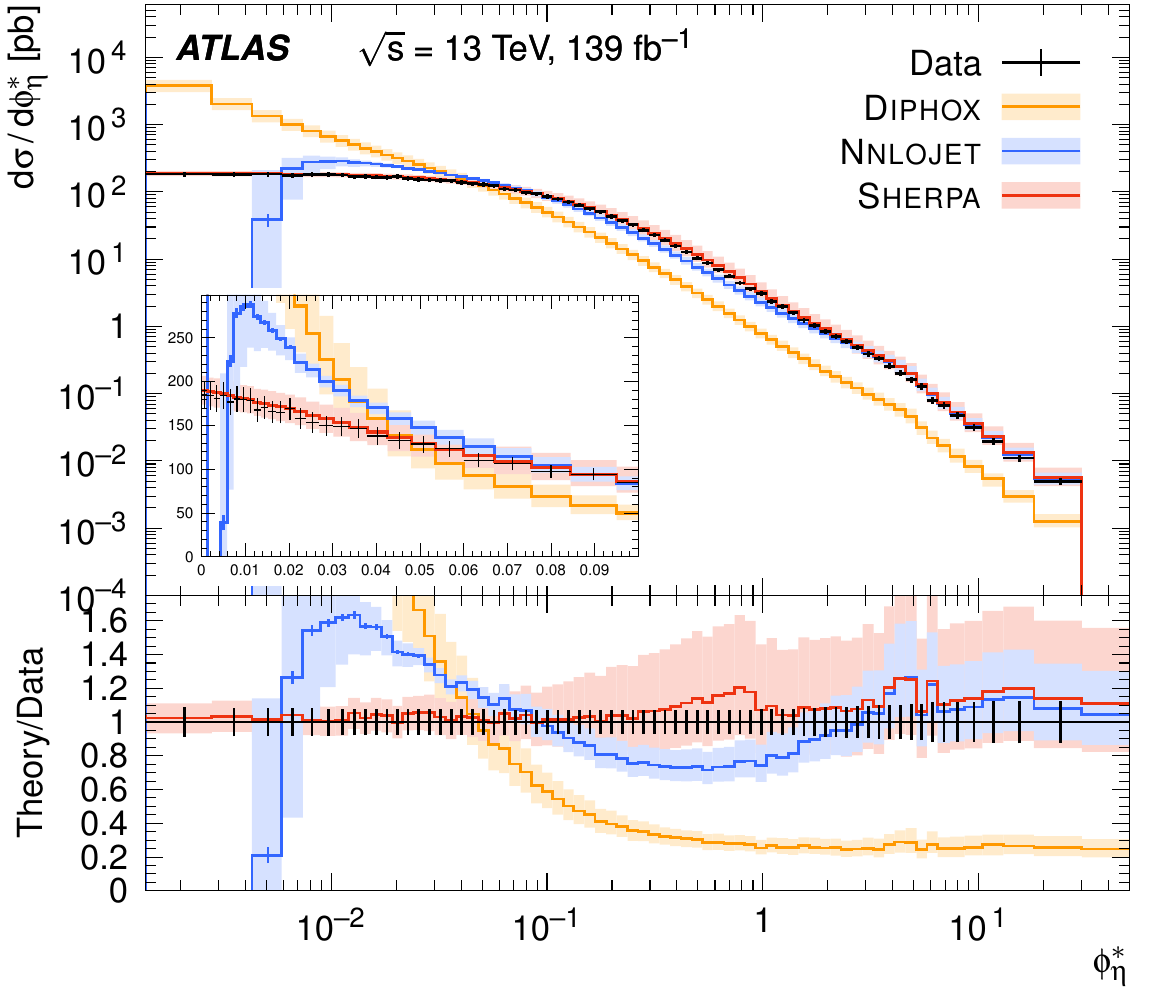}
\hspace*{2mm}
\includegraphics[width=0.47\textwidth]{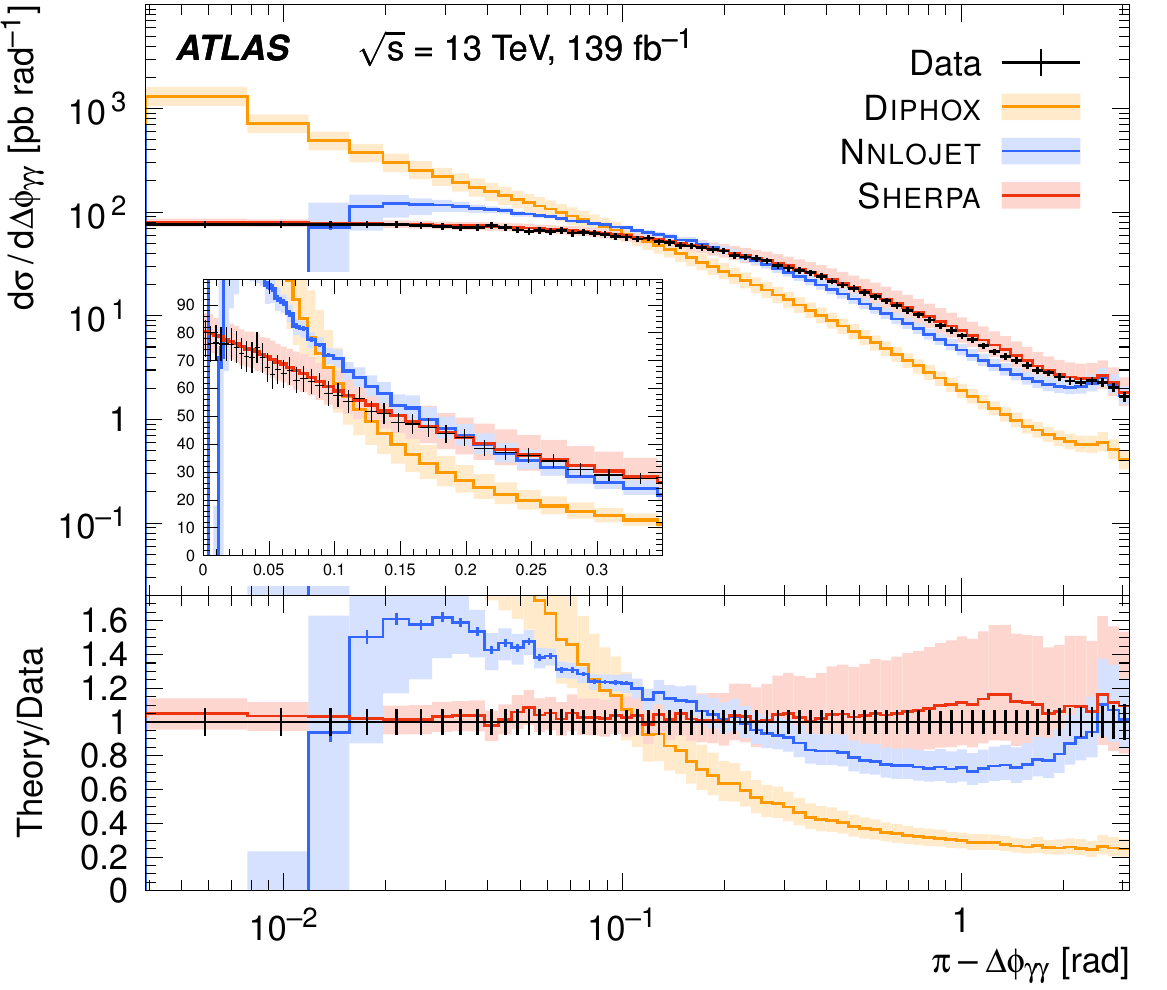}
\includegraphics[width=0.47\textwidth]{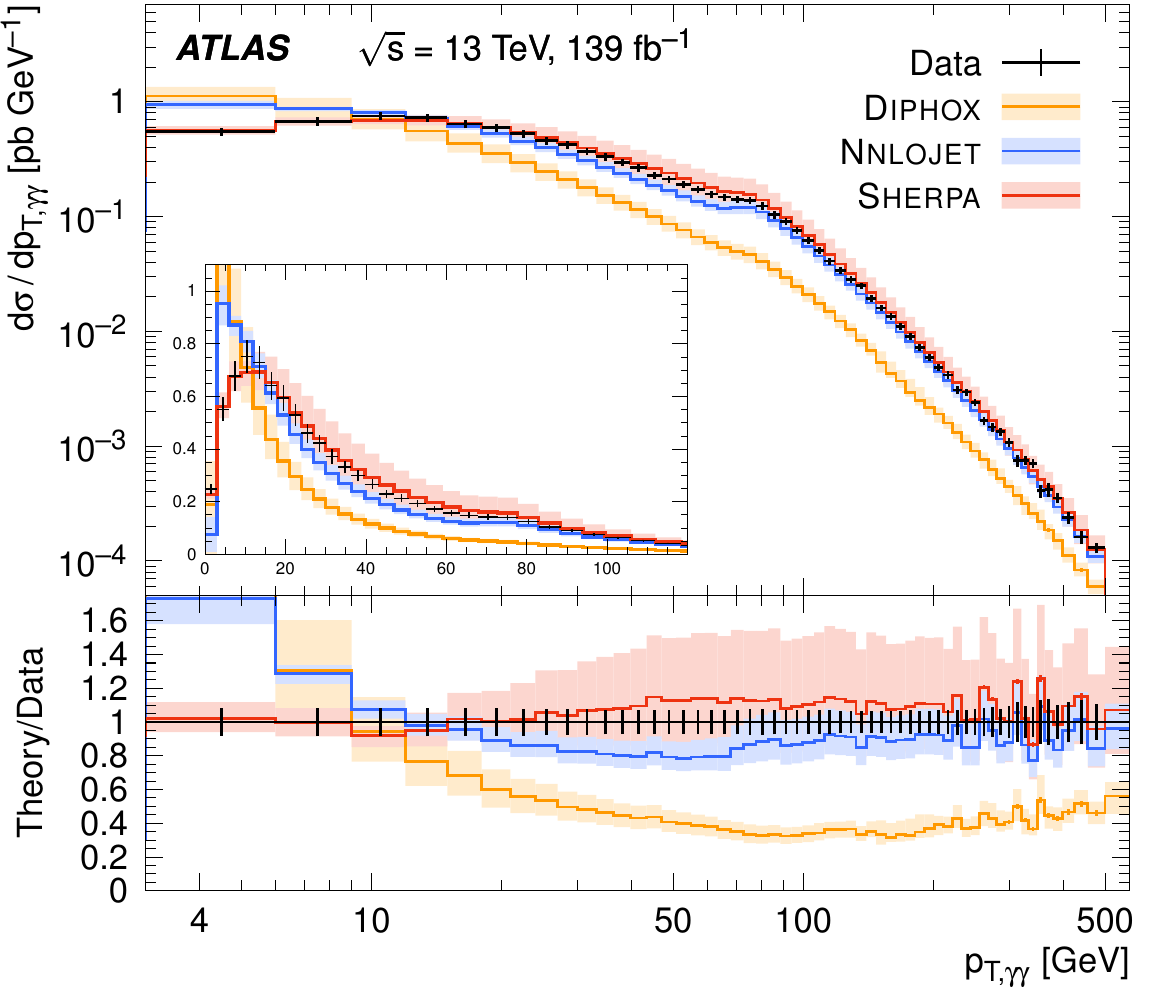}
\hspace*{2mm}
\includegraphics[width=0.47\textwidth]{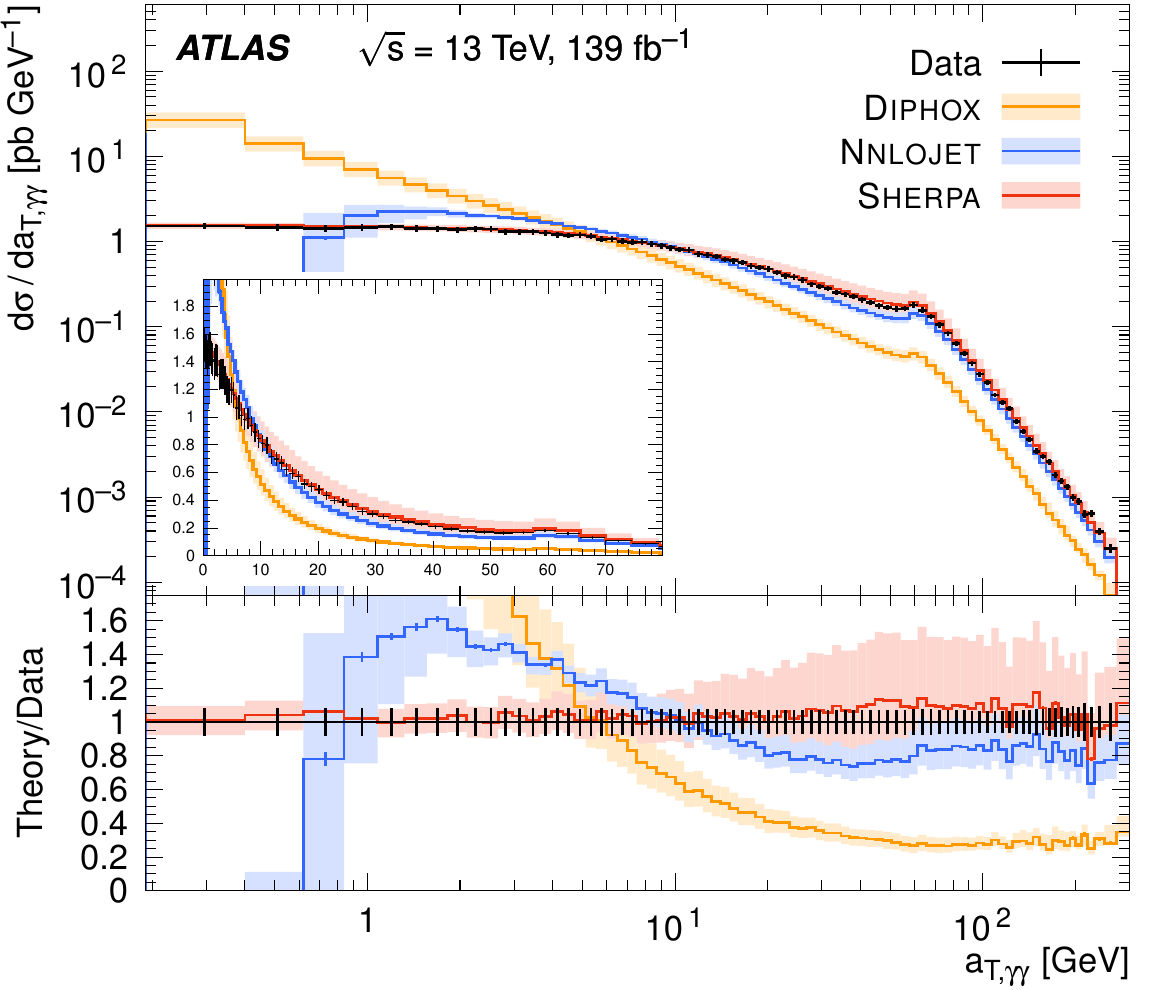}
\caption{Differential cross sections measured as functions of $\yyphiStar$, $\yypiMDphi$, $\yypT$, and $\yypTt$ (black error bars) compared with the predictions from \diphox NLO, \nnlojet NNLO and \Sherpa MEPS@NLO (lines with coloured bands). At the bottom of each plot, the ratio of the prediction to the data is shown. Uncertainty bars on the data represent the total uncertainty, while uncertainty bands (bars) on the predictions represent perturbative scale (statistical) uncertainties.
Inset plots show the core of each distribution on a linear scale. For $\yyphiStar$, $\yypT$ and $\yypTt$ the last bin is only visible in the ratio inset and includes the overflow events.}
\label{fig:results:differential:datatheory2}
\end{figure}

The measured differential cross sections as functions of the different observables are shown in Figures~\ref{fig:results:differential:datatheory1}~and~\ref{fig:results:differential:datatheory2}.
Comparisons between the data and \diphox NLO, \nnlojet NNLO and \Sherpa MEPS@NLO predictions are also included for all observables.
\nnlojet and \Sherpa provide the best agreement with the data in the regions expected to be modelled well by perturbative QCD. Since \diphox and \nnlojet are fixed-order predictions, they are not expected to be valid in regions where the effects of multiple collinear or soft QCD emissions are relevant,  while \Sherpa provides remarkably good agreement with the data in these regions also.
The systematic uncertainties of the fixed-order calculations estimated from scale variations fail to provide a reliable estimate of the true uncertainties, as indicated by the failure of the uncertainty bands to overlap with each other. This is a typical feature of the diphoton process, where significant contributions and their uncertainties only appear at higher orders, namely the single- and double-fragmentation contributions shown in Figure~\ref{fig:psfrag}. For example, the largest part of the fixed-order uncertainty bands stems from renormalisation scale variations in the real-emission topologies, which are by definition spuriously absent at LO. The multi-leg merged prediction from \Sherpa includes real-emission topologies for higher multiplicities than in the other predictions and also exhibits correspondingly large uncertainties.
 
For the transverse momentum spectrum of the leading photon, $\phonept$, \nnlojet and \Sherpa predictions are compatible with the data over the full measured range. \diphox adequately describes the shape of the measured differential cross section; however, as already seen for the integrated cross section, the normalisation of this prediction is much lower than that of the data. A similar picture holds for the transverse momentum of the sub-leading photon, $\phtwopt$, with the only notable difference being the shape of the \diphox prediction, which adequately describes the shape of the measured cross section for $\phtwopt \geq 40$~\GeV, but considerably underestimates the rate of photons at lower $\phtwopt$ values.
 
Stark features of the theory predictions can be seen in the $\yym$ spectrum. The shape of this distribution is governed by the transverse-momentum requirements placed on the individual photons, with the region $\yym < \phonept + \phtwopt$ being suppressed and only populated through $\yy$+multi-jet configurations. Such configurations are not modelled well at NLO accuracy in \diphox, but benefit significantly from the higher-order contributions included in the \nnlojet and \Sherpa predictions. The predictions from \nnlojet and \Sherpa agree with the data within uncertainties, while the uncertainties in the \diphox prediction are severely underestimated judging from the higher-order predictions. At high $\yym$, the \Sherpa predictions are compatible with the data, while \nnlojet underestimates the rate slightly.
 
For high $\yypT$, \Sherpa and \nnlojet agree well with the data, while for intermediate values the \nnlojet prediction is slightly too low to cover the data within uncertainties, and \Sherpa is slightly high but still compatible. As expected, the fixed-order predictions fail at low $\yypT < \SI{10}{\GeV}$, where soft QCD emissions are relevant, while \Sherpa is in good agreement with the data also in this phase space region. It is worth noting, that the \nnlojet prediction for $\yypT>0$ is only NLO accurate, and an even higher-order calculation would be needed to describe these phase-space regions at NNLO accuracy, e.g.\ $\gamma\gamma$+jet at NNLO~\cite{Chawdhry:2021hkp}.
 
The three observables $\yypiMDphi$, $\yypTt$, and $\yyphiStar$ are closely related and reveal very similar features in the comparison between the data and the theory predictions.
The almost back-to-back configuration of the two photons ($\yypiMDphi < 0.1$, $\yypTt < \SI{10}{\GeV}$, or $\yyphiStar < 0.1$) is difficult to model correctly with fixed-order predictions, because even almost collinear or low-energy parton emissions can decorrelate the two photons. Only the parton-shower-based \Sherpa prediction includes a resummation of these effects and is able to model the back-to-back region reliably, agreeing well with the data over the full measured range in these three observables. In the region of intermediate azimuthal separation of the photons ($\yypiMDphi \approx 1$, $\yypTt \approx \SI{40}{\GeV}$, or $\yyphiStar \approx 0.5$), \nnlojet underestimates the data, while for a large decorrelation they agree well, in contrast to the NLO predictions from \diphox, which fail to accurately model the shape of these distributions over the full measured range.
 
For scattering angles close to the limit $\yycosTS \to 1$, the fixed-order predictions underestimate the measured rate. This region is sensitive to uncorrelated photons, which can occur as a consequence of the associated production of multiple jets. The NNLO prediction is affected by larger perturbative uncertainties in this region; this is an indication of the (N)LO-accurate associated production of two jets (one jet) becoming relevant. The multi-leg merged prediction from \Sherpa provides a better description of the measured cross section in this region.
 
\FloatBarrier

\section{Conclusion}
A measurement of prompt photon pair production in proton--proton collisions at $\sqrt{s}=13$~\TeV\ is presented. The data were recorded by the ATLAS detector at the LHC with an integrated luminosity of 139~fb$^{-1}$.
 
Events with at least two photons in the well-instrumented region of the detector are selected, and the photons are required to be isolated and have $p_\mathrm{T,\gamma_{1(2)}} > 40\:(30)$~\GeV\ for the leading (sub-leading) photon ordered in transverse momentum.
The dominant background stems from non-prompt photons from hadron decays, which is very challenging to estimate using simulation. A sophisticated data-driven method is used to estimate this background using background-rich control regions based on inverted photon identification and isolation requirements. The small background from two photons being produced in separate proton--proton collisions in the same bunch crossing is estimated from data; it will become even more relevant for future LHC runs.
 
The background-subtracted signal yields are corrected for detector efficiencies and migrations, and converted into fiducial cross sections. The measured integrated fiducial cross section is $31.4\pm2.4$~pb. Differential cross sections as functions of several observables of the diphoton system are presented, including the transverse momenta of the leading (sub-leading) photon up to 550~(330)~\GeV, the invariant mass of the diphoton system from 15~\GeV\ to 1~\TeV, and the transverse momentum of the diphoton system up to 500~\GeV. Specifically, diphoton variables sensitive to different aspects of higher-order corrections in perturbative QCD are studied and compared with theory predictions from state-of-the-art MC and fixed-order calculations.
 
Good agreement is generally found with the predictions at the highest theoretical precision. An impressive improvement is observed when taking into account higher-order terms in perturbative QCD. Only the merged approach with multi-leg matrix elements at NLO, as implemented by \Sherpa, and a fixed-order NNLO calculation, as implemented by \nnlojet, give a satisfactory description of the data. The fixed-order NNLO calculation has a higher formal precision than \Sherpa and thus lower theoretical uncertainties in the regions governed by perturbative QCD. The multi-leg merged prediction yields larger uncertainties, but is in better agreement with the data in all regions. \Sherpa includes QCD resummation through the parton shower, and thus provides a more reliable prediction in regions sensitive to collinear or low-energy parton emissions. The shape of the measured cross section as a function of $\yym$ is best described by the \Sherpa predictions. The region of low $\yym$ is sensitive to multi-jet configurations, which are only modelled with low precision in all predictions.
 
\section*{Acknowledgements}
 

We thank CERN for the very successful operation of the LHC, as well as the
support staff from our institutions without whom ATLAS could not be
operated efficiently.
 
We acknowledge the support of
ANPCyT, Argentina;
YerPhI, Armenia;
ARC, Australia;
BMWFW and FWF, Austria;
ANAS, Azerbaijan;
SSTC, Belarus;
CNPq and FAPESP, Brazil;
NSERC, NRC and CFI, Canada;
CERN;
ANID, Chile;
CAS, MOST and NSFC, China;
Minciencias, Colombia;
MSMT CR, MPO CR and VSC CR, Czech Republic;
DNRF and DNSRC, Denmark;
IN2P3-CNRS and CEA-DRF/IRFU, France;
SRNSFG, Georgia;
BMBF, HGF and MPG, Germany;
GSRI, Greece;
RGC and Hong Kong SAR, China;
ISF and Benoziyo Center, Israel;
INFN, Italy;
MEXT and JSPS, Japan;
CNRST, Morocco;
NWO, Netherlands;
RCN, Norway;
MEiN, Poland;
FCT, Portugal;
MNE/IFA, Romania;
JINR;
MES of Russia and NRC KI, Russian Federation;
MESTD, Serbia;
MSSR, Slovakia;
ARRS and MIZ\v{S}, Slovenia;
DSI/NRF, South Africa;
MICINN, Spain;
SRC and Wallenberg Foundation, Sweden;
SERI, SNSF and Cantons of Bern and Geneva, Switzerland;
MOST, Taiwan;
TAEK, Turkey;
STFC, United Kingdom;
DOE and NSF, United States of America.
In addition, individual groups and members have received support from
BCKDF, CANARIE, Compute Canada and CRC, Canada;
COST, ERC, ERDF, Horizon 2020 and Marie Sk{\l}odowska-Curie Actions, European Union;
Investissements d'Avenir Labex, Investissements d'Avenir Idex and ANR, France;
DFG and AvH Foundation, Germany;
Herakleitos, Thales and Aristeia programmes co-financed by EU-ESF and the Greek NSRF, Greece;
BSF-NSF and GIF, Israel;
Norwegian Financial Mechanism 2014-2021, Norway;
NCN and NAWA, Poland;
La Caixa Banking Foundation, CERCA Programme Generalitat de Catalunya and PROMETEO and GenT Programmes Generalitat Valenciana, Spain;
G\"{o}ran Gustafssons Stiftelse, Sweden;
The Royal Society and Leverhulme Trust, United Kingdom.
 
The crucial computing support from all WLCG partners is acknowledged gratefully, in particular from CERN, the ATLAS Tier-1 facilities at TRIUMF (Canada), NDGF (Denmark, Norway, Sweden), CC-IN2P3 (France), KIT/GridKA (Germany), INFN-CNAF (Italy), NL-T1 (Netherlands), PIC (Spain), ASGC (Taiwan), RAL (UK) and BNL (USA), the Tier-2 facilities worldwide and large non-WLCG resource providers. Major contributors of computing resources are listed in Ref.~\cite{ATL-SOFT-PUB-2021-003}.
 

\printbibliography

\clearpage 
 
\begin{flushleft}
\hypersetup{urlcolor=black}
{\Large The ATLAS Collaboration}

\bigskip

\AtlasOrcid[0000-0002-6665-4934]{G.~Aad}$^\textrm{\scriptsize 99}$,    
\AtlasOrcid[0000-0002-5888-2734]{B.~Abbott}$^\textrm{\scriptsize 124}$,    
\AtlasOrcid[0000-0002-7248-3203]{D.C.~Abbott}$^\textrm{\scriptsize 100}$,    
\AtlasOrcid[0000-0002-2788-3822]{A.~Abed~Abud}$^\textrm{\scriptsize 34}$,    
\AtlasOrcid[0000-0002-1002-1652]{K.~Abeling}$^\textrm{\scriptsize 51}$,    
\AtlasOrcid[0000-0002-2987-4006]{D.K.~Abhayasinghe}$^\textrm{\scriptsize 91}$,    
\AtlasOrcid[0000-0002-8496-9294]{S.H.~Abidi}$^\textrm{\scriptsize 27}$,    
\AtlasOrcid[0000-0001-5329-6640]{H.~Abramowicz}$^\textrm{\scriptsize 157}$,    
\AtlasOrcid[0000-0002-1599-2896]{H.~Abreu}$^\textrm{\scriptsize 156}$,    
\AtlasOrcid[0000-0003-0403-3697]{Y.~Abulaiti}$^\textrm{\scriptsize 5}$
% ...
\bigskip
\\

% ...
$^{5}$High Energy Physics Division, Argonne National Laboratory, Argonne IL; United States of America.\\
% ...
$^{27}$Physics Department, Brookhaven National Laboratory, Upton NY; United States of America.\\
% ...
$^{34}$CERN, Geneva; Switzerland.\\
% ...
$^{51}$II. Physikalisches Institut, Georg-August-Universit\"{a}t G\"ottingen, G\"ottingen; Germany.\\
% ...
$^{91}$Department of Physics, Royal Holloway University of London, Egham; United Kingdom.\\
% ...
$^{99}$CPPM, Aix-Marseille Universit\'e, CNRS/IN2P3, Marseille; France.\\
$^{100}$Department of Physics, University of Massachusetts, Amherst MA; United States of America.\\
% ...
$^{124}$Homer L. Dodge Department of Physics and Astronomy, University of Oklahoma, Norman OK; United States of America.\\
% ...
$^{156}$Department of Physics, Technion, Israel Institute of Technology, Haifa; Israel.\\
$^{157}$Raymond and Beverly Sackler School of Physics and Astronomy, Tel Aviv University, Tel Aviv; Israel.\\
% ...

\end{flushleft}


\end{document}